\documentclass[prb,twocolumn,amsmath,amssymb,nofootinbib,floatfix]{revtex4}

\usepackage{graphicx,bm,epsfig,setspace}
\usepackage{color}

\makeatletter
\def\graphicscale{\twocolumn@sw{0.33}{0.4}}
\makeatother

\def\spose#1{\hbox to 0pt{#1\hss}}
\def\lesssim{\mathrel{\spose{\lower 3pt\hbox{$\mathchar"218$}}
 \raise 2.0pt\hbox{$\mathchar"13C$}}}
\def\gtrsim{\mathrel{\spose{\lower 3pt\hbox{$\mathchar"218$}}
 \raise 2.0pt\hbox{$\mathchar"13E$}}}

\def\<{\langle}
\def\>{\rangle}

\newcommand*{\beq}{\begin{eqnarray}}
\newcommand*{\eeq}{\end{eqnarray}}
\newcommand*{\bea}{\begin{eqnarray}}
\newcommand*{\eea}{\end{eqnarray}}

\def\simge{\mathrel{%
       \rlap{\raise 0.511ex \hbox{$>$}}{\lower 0.511ex \hbox{$\sim$}}}}
\def\simle{\mathrel{
       \rlap{\raise 0.511ex \hbox{$<$}}{\lower 0.511ex \hbox{$\sim$}}}}

\begin{document}

\title{Thermodynamics of star polymer solutions:
   a coarse-grained study }

\author{Roberto Menichetti}
\email{menichetti@mpip-mainz.mpg.de}
\affiliation{Max-Planck-Institut f\"ur Polymerforschung, 
Ackermannweg 10, D-55128 Mainz, Germany}
\author{Andrea Pelissetto}
\email{andrea.pelissetto@roma1.infn.it}
\affiliation{Dipartimento di Fisica, Sapienza Universit\`a di Roma and
INFN, Sezione di Roma I, P.le Aldo Moro 2, I-00185 Roma, Italy}
\author{Ferdinando Randisi}
\email{ferdinando.randisi@physics.ox.ac.uk}
\affiliation{Rudolf Peierls Centre for Theoretical Physics, 1 Keble Road,
Oxford OX1 3NP, United Kingdom}

\date{\today}

\begin{abstract}
We consider a coarse-grained (CG) model with pairwise interactions, suitable
to describe low-density solutions of star-branched polymers of functionality
$f$. Each macromolecule is represented by a CG molecule with $(f+1)$
interaction sites, which captures the star topology. Potentials are obtained
by requiring the CG model to reproduce a set of distribution functions
computed in the microscopic model in the zero-density limit. Explicit 
results are given for $f=6,12$ and 40. We use the CG model to compute the 
osmotic equation of state of the solution for concentrations $c$ such that
$\Phi_p = c/c^*\lesssim 1$, where $c^*$ is the overlap concentration.
We also investigate in detail the phase diagram for $f=40$, identifying 
the boundaries of the solid intermediate phase. 
Finally, we investigate how the 
polymer size changes with $c$. For $\Phi_p\lesssim 0.3$ polymers become harder
as $f$ increases at fixed reduced concentration $c/c^*$. 
On the other hand, for $\Phi_p\gtrsim 0.3$, polymers show the
opposite behavior: At fixed $\Phi_p$, the larger the value of $f$, the larger 
their size reduction is.
\end{abstract}

% 61.25.he Polymer solutions
% 65.20.De 
%     General theory of thermodynamic properties of liquids, including computer
%     simulation
% 82.35.Lr Physical properties of polymers

\maketitle

\section{Introduction}

Soft materials are at present the object of very active research, 
because of their peculiar
physico-chemical properties and for their technological applications. 
In particular, star polymers have attracted a considerable interest,
\cite{GFHR-96,VFPR-01} as their 
thermodynamic and rheological behavior can be tuned by varying their 
functionalities (i.e., the number of their arms), 
thereby interpolating between the 
behavior observed using linear polymers and that appropriate for hard colloids.

Star-branched molecules of functionality $f$ are obtained by tethering 
$f$ polymer chains to a central core. Their structure in very dilute solutions 
in which one can neglect polymer-polymer interactions is very well 
understood, both at a qualitative and quantitative level, at least
when the number $L$ of monomers per arm is large. For large
$f$ and $L$, the behavior of star polymers is often modelled by using the 
Daoud-Cotton model.  \cite{DC-82,BZ-84}
The scaling predictions of this model are confirmed by numerical simulations, 
although they appear to be quantitatively valid only for $f \gg 100$.
\cite{Grest-94,HNG-04,HG-04,HBL-09,RP-13} 
At finite density,
the thermodynamic behavior of dilute and 
semidilute solutions of these macromolecules is well established from 
a qualitative point of view. Scaling arguments and 
approximate calculations predict 
a qualitative change of the phase diagram as $f$ increases.
\cite{WP-86,WPC-86,LLWAKAR-98,WLL-98,WLL-99,LSLWRZ-05,MP-13} For $f$ less than
a critical value $f_c$, star polymers behave as linear chains. As the 
density increases, polymers simply overlap, so that the behavior is
qualitatively analogous to that of linear polymers. On the other hand,
for $f > f_c$, a solid phase appears, separating two distinct (one dilute 
and one dense) fluid phases. At a quantitative level, however, much less 
is known in the universal, large-$L$ regime. 
Numerical simulations, which are at present the only method that provides 
accurate results,  are quite hard.
First, it is difficult to devise efficient algorithms for these finite-density
systems and therefore simulations are very slow. Second, finite-length
corrections increase quite rapidly with the functionality $f$ for fixed 
$L$.\cite{RP-13} 
Therefore, even when using optimal models, tuned to minimize 
this type of corrections, one still needs $L\gtrsim 10^2$-$10^3$ to obtain
results in the universal (model-independent) regime. Therefore, even for
relatively small functionalities, say $f\approx 10$, one ends up with 
prohibitively large molecules.

Since we are only interested in thermodynamical and large-scale properties of
the system, a numerically affordable method to overcome 
the above-mentioned difficulties 
consists in using coarse-grained (CG) models, in which most of the internal
degrees of freedom of the polymer are treated implicitly, 
projecting the reference
system onto a set of CG molecules with a limited number of interaction 
sites (we call them blobs). Several different CG strategies have been 
proposed in the literature:
\cite{MP-02,Voth-09,Feller-09,Wilson-09,PK-09,RN-11,Noid-13,GSH-17} 
structure-based method, energy-based method, force-matching and 
relative-entropy approaches, just to mention the most popular ones.
In this work we consider a CG model with effective pairwise interactions,
which are obtained by using the structure-based
route.\cite{LBHM-00,Likos-01,BLHM-01,MP-02,PCH-07,Pelissetto-09,PK-09,%%
DPP-12-Soft,DMPP-15-SpecTopics} 
The potentials are obtained by requiring the CG model to reproduce 
a set of full-monomer (i.e., computed in the microscopic model) target 
distribution functions determined in the limit of zero 
density. As one needs to simulate only a small number of polymers to 
compute the target distributions, one can consider large molecules, thereby 
working in the large degree-of-polymerization, universal regime. 

The simplest CG approach consists in replacing the whole polymer with a 
monoatomic molecule, obtaining the so-called single-blob (SB) model. The
effective interactions for this simple representation have been 
either postulated by using phenomenological 
arguments\cite{LLWAKAR-98,JDLvFL-01} 
or obtained by accurate numerical simulations.\cite{HG-04} 
These models provide the correct qualitative picture,
\cite{LLWAKAR-98,WLL-98,WLL-99,MP-13} but are not
meant to provide quantitatively accurate predictions unless the system is in
the dilute regime, i.e., only for $\Phi_p \ll 1$, where $\Phi_p = {4\over 3}
\pi \rho_p \hat{R}_g^3$ ($\rho_p = N_p/V$, $N_p$ is the number of star
polymers in a box of size $V$, and 
$\hat{R}_g$ is the average zero-density radius of gyration) 
is the polymer volume
fraction. In this work we study the simplest generalization of the 
SB model that captures the star topology. Each star polymer
is replaced by a CG molecule with $(f+1)$ blobs, see Fig.~\ref{Mbpicturestar}:
one of them represents the
center of the star, while the other $f$ blobs are associated with the
arms.
It is important to derive the limits of validity of this multiblob (MB) model. 
In Ref.~\onlinecite{PCH-07} it was shown 
that a multiblob model for linear polymers with $n$ effective sites is 
accurate up to $\Phi_p\lesssim n^{3\nu-1}$ ($\nu\approx 0.588$ is the 
usual Flory exponent), a result that was confirmed by the 
simulations of Refs.~\onlinecite{DPP-12-Soft,DPP-12-JCP}. 
This argument should be applied with care to our star-polymer system.
Although we take $(f+1)$ sites to describe each star polymer, 
due to its geometrical construction, the level of resolution of our model is 
not higher than that of a trimer representation, thus we should take $n=3$
in the general expression. Therefore, we expect accurate 
results only up to $\Phi_p\lesssim 3^{3\nu-1}\approx2.4$. For this reason, we 
investigate the thermodynamic behavior up to $\Phi_p \approx 2$.
This is not a serious limitation, as most of the interesting 
phenomena---for instance, the fluid-solid transition for large 
functionalities---occur for much lower densities, i.e., for
$\Phi_p\lesssim 1.0$. 

In this work we consider
$f=6$, $f=12$, and $f=40$. For the two lowest values of $f$ the behavior 
should be qualitatively similar to that observed for linear chains, while 
for $f=40$ we expect a significantly different behavior, with the presence 
of ordering fluid-solid transitions.

The paper is organized as follows. In Sec.~\ref{sec2} we discuss the 
CG representation. In Sec.~\ref{sec2.1} we give the definitions
and present the general strategy, which follows closely what has been done 
for linear polymers in Ref.~\onlinecite{DPP-12-Soft}. Then, we 
present our results for the target distribution functions 
in Sec.~\ref{sec2.2}. In Sec.~\ref{sec3} we compute the effective interactions
for the CG model. In Sec.~\ref{sec4} we use the model to investigate the 
thermodynamic behavior. In Sec.~\ref{sec5} we focus on the location
of the fluid-solid transitions (more details are reported in the 
supplementary material) and give preliminary results for the 
stable crystal structures. In Sec.~\ref{sec6} we discuss how 
the polymer size changes at finite density. Finally, in Sec.~\ref{sec7} we draw
our conclusions. In the Appendix we give some details on the 
inversion precedure used to determine the intermolecular potentials.

\section{The coarse-grained representation} \label{sec2}

\subsection{Definitions and general strategy} \label{sec2.1}

\begin{figure}[t]
\begin{center}
\centering
\includegraphics[scale=0.5]{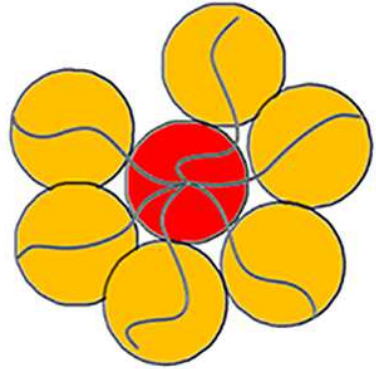}
\caption{Schematic representation of a star polymer with $f$ arms:
the polymer is divided into $(f+1)$ subunits (blobs), one for the center 
and one for each arm.}
\label{Mbpicturestar}
\end{center}
\end{figure}

In this section we define the CG model, following the same 
strategy used for linear polymers in Ref.~\onlinecite{DPP-12-Soft} and 
recently reviewed in Ref.~\onlinecite{DMPP-15-SpecTopics}. The simplest
model that captures the star topology is obtained by replacing the polymer
with a CG molecule made of $(f+1)$ atoms (we call them blobs): 
one of them represents the 
center of the star, while the other $f$ blobs are associated with the 
arms, see Fig.~\ref{Mbpicturestar}. 

To be specific we consider a star polymer with $f$ arms and $L$ monomers 
per arm.
We label each arm with a Greek index $\alpha$, $\alpha = 1,\ldots, f$, and the
monomer positions with ${\bm r}_{\alpha,i}$, $i=1,\ldots,L-1$. The 
central monomer position is ${\bm r}_0$. To each star, we associate a 
CG representation $\{{\bm R}_0,{\bm R}_1,\ldots,{\bm R}_f\}$, where 
\begin{eqnarray}
{\bm R}_0 &=& {1\over mf}\left( f {\bm r}_0 + \sum_{\alpha=1}^f 
      \sum_{i=1}^{m-1} {\bm r}_{\alpha,i} \right) ,
\nonumber \\
{\bm R}_\alpha &=& {1\over mf} \sum_{i=m}^{L-1} {\bm r}_{\alpha,i} ,
\end{eqnarray}
are the centers of mass of the central blob and of the blobs associated with 
the arms, respectively, and
$m = L/(f+1)$ [$L$ is assumed to be a multiple of $(f+1)$]. Note that,
for convenience, we have given a weight $f$ to the core monomer. 
As discussed in Ref.~\onlinecite{HNG-04}, this is irrelevant in the large-$L$
limit. For this representation, we define a set of 
intramolecular and intermolecular distribution functions that are used 
as targets for the CG model. First, we consider the adimensional 
intramolecular distribution functions
\begin{eqnarray}
P_{\alpha\beta}(b) &=& \hat{R}_g 
      \langle \delta (|{\bm R}_\alpha - {\bm R}_\beta| - r) \rangle,
\label{Pintra-FM}
\end{eqnarray}
where $\hat{R}_g$ is the zero-density radius of gyration 
(here and in the following we will use a hat to indicate that 
a given quantity is computed at zero density) and the average is 
over all conformations of a single isolated star. In the large-$L$ limit 
these quantities are functions of the ratio $b = r/\hat{R}_g$ and
are universal, in the sense that they do not depend on the 
microscopic polymer model. We 
also consider the distribution of the angle between two arms 
($1\le \alpha,\beta \le f$)
\begin{equation}
P_{\rm th,\alpha\beta}(\cos \theta) = 
      \left\langle \delta \left(  
      {({\bm R}_\alpha - {\bm R}_0)\cdot ({\bm R}_\beta - {\bm R}_0) \over
       |{\bm R}_\alpha - {\bm R}_0||{\bm R}_\beta - {\bm R}_0|} - 
    \cos \theta \right) \right \rangle.
\label{Pth-FM}
\end{equation}
Finally, we consider the intermolecular blob-blob potentials of mean
force
\begin{equation}
\beta W_{\alpha\beta}(b) = 
   - \ln \langle e^{-\beta U_{\rm inter}} 
    \rangle_{{\bm R}_\alpha^{(1)} = 0,R_\beta^{(2)} = r}\, ,
\label{PMF-FM}
\end{equation}
where $U_{\rm inter}$ is the intermolecular potential 
energy, and the average is over all configurations of 
pairs of noninteracting star polymers such that blob $\alpha$ of the 
first polymer is in the origin (${\bm R}_\alpha^{(1)}=0$), and 
blob $\beta$ of the second polymer is located in any
point at a distance $r$ from the origin. 

\begin{figure}[t]
\begin{center}
\centering
\includegraphics[scale=0.45]{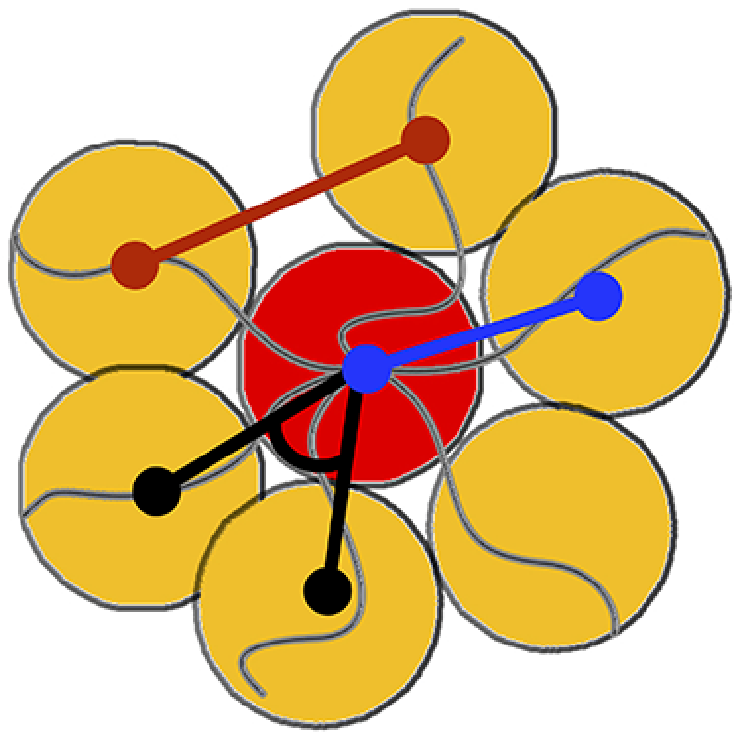}
\includegraphics[scale=0.55]{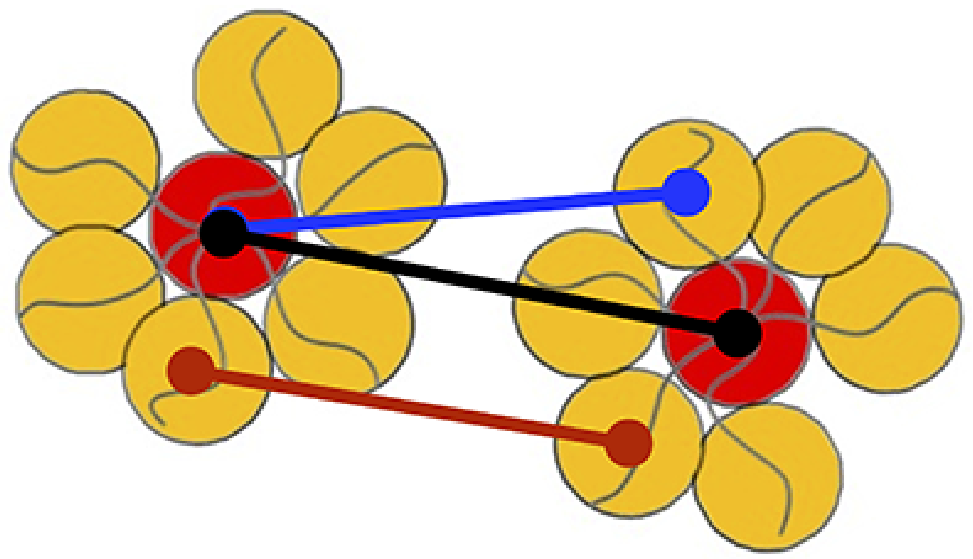}
\caption{Top: Intramolecular interactions; we consider potentials 
depending on the center-arm (blue line) and arm-arm distance (red line), 
and on the arm-arm angle (black line). Bottom: Intermolecular interactions:
we consider pair potentials depending on the 
center-arm (blue line), arm-arm (red line), and center-center (black line)
distance. 
}
\label{Starintr}
\end{center}
\end{figure}

In the CG model the basic object is a polyatomic molecule
with $(f+1)$ atoms located in ${\bf R}_0,\ldots,{\bf R}_f$.
All length scales are expressed in terms of $\hat{R}_g$,
hence potentials and distribution functions depend on the
adimensional combination ${\bm b} = {\bm R}/\hat{R}_g$.
In an exact CG procedure the intramolecular potential is 
a function of the coordinates of all atoms. 
As in our previous work
on linear polymers,\cite{DPP-12-Soft} we perform a drastic simplification. 
We only consider potentials that depend on a single scalar variable and,
moreover, we neglect all interactions that involve more than three blobs. 
In spite of these simplifications, we expect this parametrization of the 
intramolecular interactions to provide reasonably accurate results. 
In the case of linear polymers, we observed the appearance of a well-defined
hierarchy among the $n$-body interactions: 
$V_{\rm 2-body} \gg V_{\rm 3-body} \gg V_{\rm 4-body} \ldots$ We expect (and 
we will provide evidence below) the 
same here, so that the neglect of all $n$-body interactions with  $n\ge 4$
should have a limited impact on the accuracy of the model. 
Guided by the parametrization of the intramolecular interactions
used for linear polymers,\cite{DPP-12-Soft}
we write the intramolecular potential energy as,
see Fig.~\ref{Starintr}, 
\begin{eqnarray}
U^{\rm intra} &= &
   \sum_{1 \le \alpha < \beta \le f} V_{aa}(b_{\alpha\beta}) + 
   \sum_{1 \le \alpha \le f} V_{ca}(b_{0\alpha}) \nonumber \\
&& + 
   \sum_{1 \le \alpha < \beta \le f} V_{\rm th}(\cos \theta_{\alpha\beta}),
\end{eqnarray}
where $b_{\alpha\beta} = |{\bm R}_\alpha - {\bm R}_\beta|/\hat{R}_g$, and 
$\theta_{\alpha\beta}$ is the angle between ${\bm R}_\alpha - {\bm R}_0$ and 
${\bm R}_\beta - {\bm R}_0$.
The three independent (arm-arm, center-arm, and angular) 
intramolecular potentials are determined by requiring 
the adimensional distributions 
$P_{\alpha\beta}$ and $P_{\rm th,\alpha\beta}$
to be identical in the polymer microscopic model and in the CG model.

In the case of the intermolecular interactions, we only consider 
blob-blob pair potentials. Again this choice is motivated by feasibility 
and justified by our previous work on linear polymers and polymer-colloid
mixtures.\cite{DMPP-15-SpecTopics}  In all cases, this simple parametrization
provided accurate results. For the CG model of star polymers, we must
introduce three different pair interactions, as we must distinguish 
the central blob from those associated with the arms. 
Therefore, the intermolecular potential energy for two CG star polymers is 
written as, see Fig.~\ref{Starintr}:
\begin{eqnarray}
U^{\rm inter} &= &
   \sum_{\alpha,\beta=1}^f {\widetilde V}_{aa}(b_{\alpha\beta}) + 
\nonumber \\
&&
   \sum_{\alpha=1}^f [{\widetilde V}_{ca}(b_{0\alpha}) + 
    {\widetilde V}_{ca}(b_{\alpha0})] +  
    {\widetilde V}_{cc}(b_{00}),
\end{eqnarray}
where $b_{\alpha\beta} = 
|{\bm R}_{\alpha}^{(1)} - {\bm R}_{\beta}^{(2)}|/\hat{R}_g$, and
$\{{\bm R}_{\alpha}^{(1)}\}$, $\{{\bm R}_{\alpha}^{(2)}\}$ are the 
coordinates of the blobs belonging to the two polymers.
The three independent (arm-arm, center-arm, and center-center) 
potentials are determined so as to reproduce 
the potentials of mean force (\ref{PMF-FM}) in the full-monomer model.

\subsection{Determination of the full-monomer distributions} \label{sec2.2}

\begin{figure}[!ht]
\begin{center}
%%\advance\leftskip-2cm
\begin{tabular}{c}
\includegraphics[angle=-90,scale=0.35]{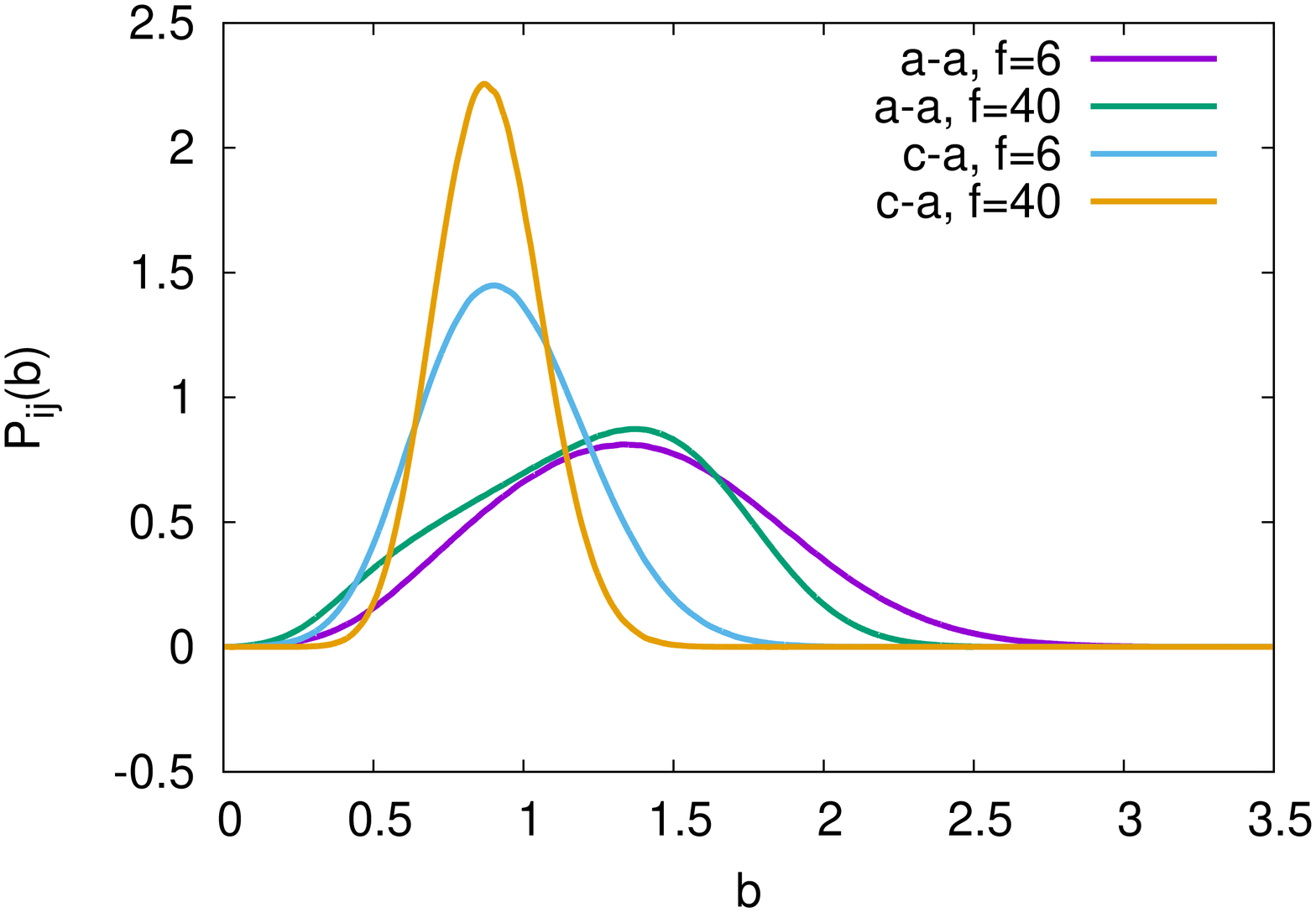} \\
 \includegraphics[angle=-90,scale=0.35]{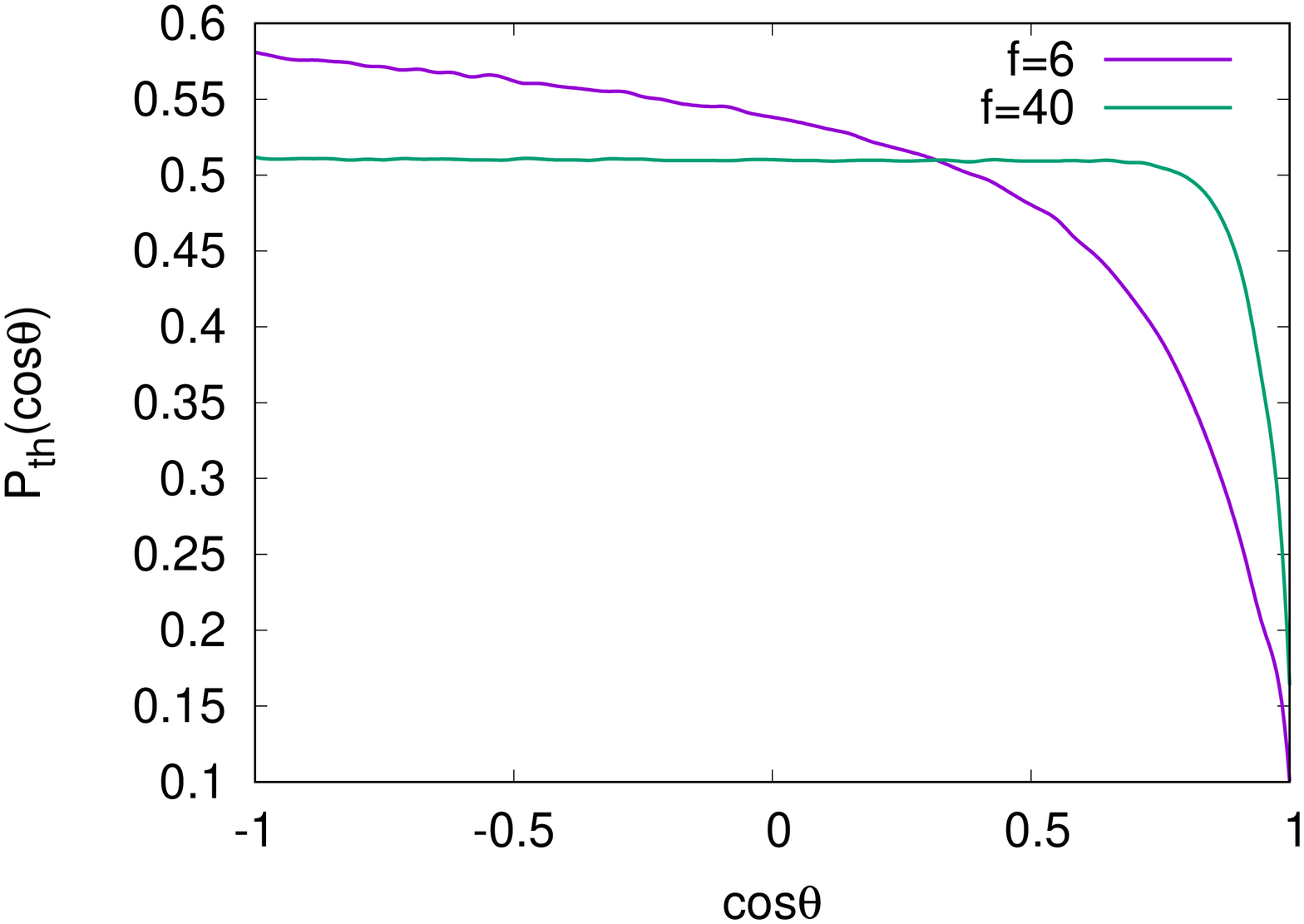}
\end{tabular}
\caption{Intramolecular distribution functions for $f = 6$ and 40.
Top: arm-arm (a-a) and center-arm (c-a) distance distributions 
as a function of $b=r/\hat{R}_g$. 
Bottom: angle distributions as a function of $\cos\theta$.}
\label{Intradistrib}
\end{center}
\end{figure}

To compute the distributions (\ref{Pintra-FM}) and 
(\ref{Pth-FM}), and the potentials of mean force (\ref{PMF-FM}),
we perform Monte Carlo simulations of 
the lattice Domb-Joyce model \cite{DJ-72} at a particular value of the
repulsion parameter
that guarantees the absence of the leading large-$L$
corrections\cite{CMP-06,DPP-16} (see the supplementary material 
for details).
Such a model, that we already used in Ref.~\onlinecite{RP-13}, 
is particularly
convenient, as it allows us to simulate very large systems and to obtain 
asymptotic large-$L$ results with good accuracy. 
Although in this 
paper we focus on $f= 6, 12$, and 40, we 
also determine the intramolecular distributions and the 
potentials of mean force for $f=18,24$, and $30$.
We perform simulations for several values of 
$L$, in the range $100\lesssim L \lesssim 1000$, 
without observing a significant $L$ dependence within the statistical accuracy.
This is clearly due to the optimality of the model that minimizes 
the $L$-dependent corrections.\cite{footnote:optimality}
The results we use in the CG procedure are those corresponding 
to star polymers with $L\approx 1000$. 

In Fig.~\ref{Intradistrib} we report the center-arm distribution 
$P_{ca}(b)$, defined as the average of $P_{0\alpha}(b)$ over all 
arms ($\alpha \ge 1$), and the arm-arm distribution 
$P_{aa}(b)$, which is the average of $P_{\alpha\beta}(b)$ over all
arm pairs ($\alpha,\beta \ge 1$). The distribution $P_{ca}(b)$ becomes 
more peaked as $f$ increases, indicating that the radial fluctuations of the 
arm centers of mass are suppressed for $f\to \infty$. Instead, the position 
of the maximum changes only slightly. If we compute the average 
squared distance $R^2_{ca}$ 
between the center of mass
of the central blob (for large $f$ it coincides with the center of the star) 
and that of a polymer arm, we obtain 
$R_{ca}/\hat{R}_g = 0.98, 0.93, 0.91, 0.90$ for $f = 6,12,30$, 40,
respectively.
We can compare these results with the predictions of the Daoud-Cotton model.
\cite{DC-82} 
Let us assume that each arm is confined in a cone of solid angle $4\pi/f$
with the vertex in the center of the star and that the arm monomers
are distributed with density $\rho(r)$, that only depends on the distance $r$.
For large $f$ we can neglect the central blob, so that 
\begin{equation}
R^2_{ca} = {1\over L} \int r^2 dr \int_0^{\theta_0} \sin\theta 
     d\theta\int_0^{2\pi} d\phi 
      \rho(r) (r\cos\theta)^2,
\end{equation}
with $1 - \cos\theta_0 = 2/f$. For large $f$, this expression becomes 
\begin{equation}
R^2_{ca} = {4\pi\over Lf} \int dr r^4 \rho(r) = 
     {1\over Lf} \int d^3 r\, r^2 \rho(r) = \hat{R}_g^2.
\end{equation}
This simple argument predicts $R_{ca}/\hat{R}_g = 1$, which 
slightly overestimates the correct result. Clearly, arms are not confined 
in a single cone for large $f$, 
but are significantly intertwined, especially close to the 
center of the star. As a result, the arm center of mass is closer to the 
center than the Daoud-Cotton picture predicts.

The distribution $P_{aa}(b)$ is not very sensitive to $f$. It 
moves slightly towards smaller values of $b$
as $f$ increases, probably as a consequence of the fact that the 
arm centers of mass are closer to the star center. 
If we compute the average arm-arm 
square distance $R_{aa}^2$, we obtain $R_{aa}/\hat{R}_g = 1.45,1,35,1.30,
1.29$ for $f=6,12,30,40$. This is
consistent with $R_{aa} = \sqrt{2} R_{ca}$,
which is obtained by assuming that the arm centers of mass are randomly 
distributed on a sphere of radius $R_{ca}$. 

In Fig.~\ref{Intradistrib} we also report the angle distribution
$P_{\rm th}(\cos\theta)$. For $f = 40$ it is approximately equal to 1/2 
except for $\theta \lesssim 15^\circ$, 
which confirms that the arm centers of mass are 
distributed randomly around the center for large $f$.
For $\theta \approx 0$ the distribution shows a dip. 
Note, however, that $P_{\rm th}(\cos\theta=1)$ increases 
as $f$ increases: we have $P_{\rm th}(1) \approx 0.1$, 0.15 for $f=6,40$,
respectively. This suggests that, for large values of $f$, also the dip for 
$\theta \approx 0$ disappears and therefore the distribution becomes flat, 
implying the absence of correlations among the angular positions of the
blobs. This result indicates that our parametrization of the intramolecular 
interactions should become increasingly accurate as $f$ increases: the 
neglected $n$-body correlations become irrelevant for stars of high
functionality.

\begin{figure}[t]
\begin{center}
\centering
\includegraphics[angle=-90,scale=0.35]{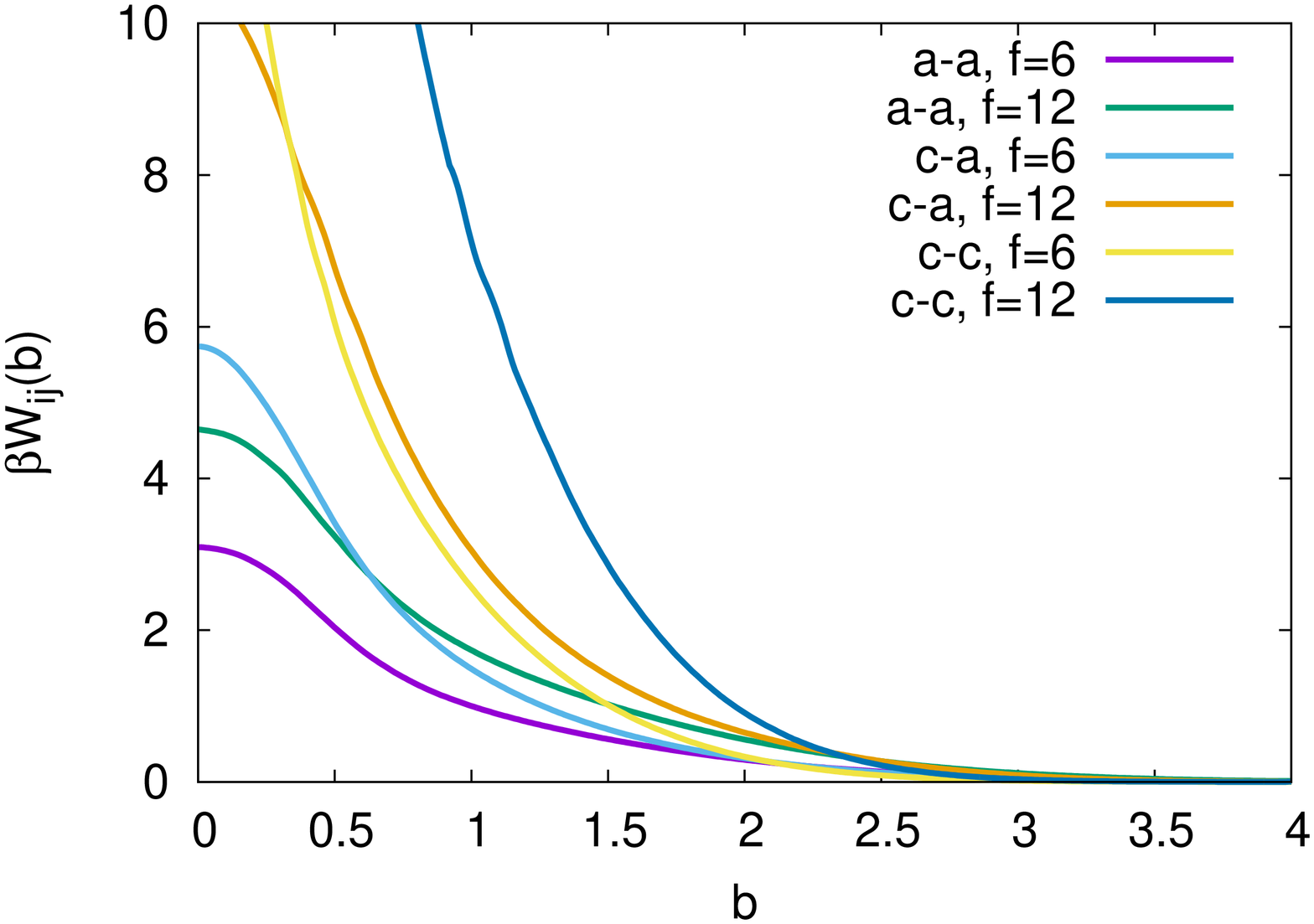}
\includegraphics[angle=-90,scale=0.35]{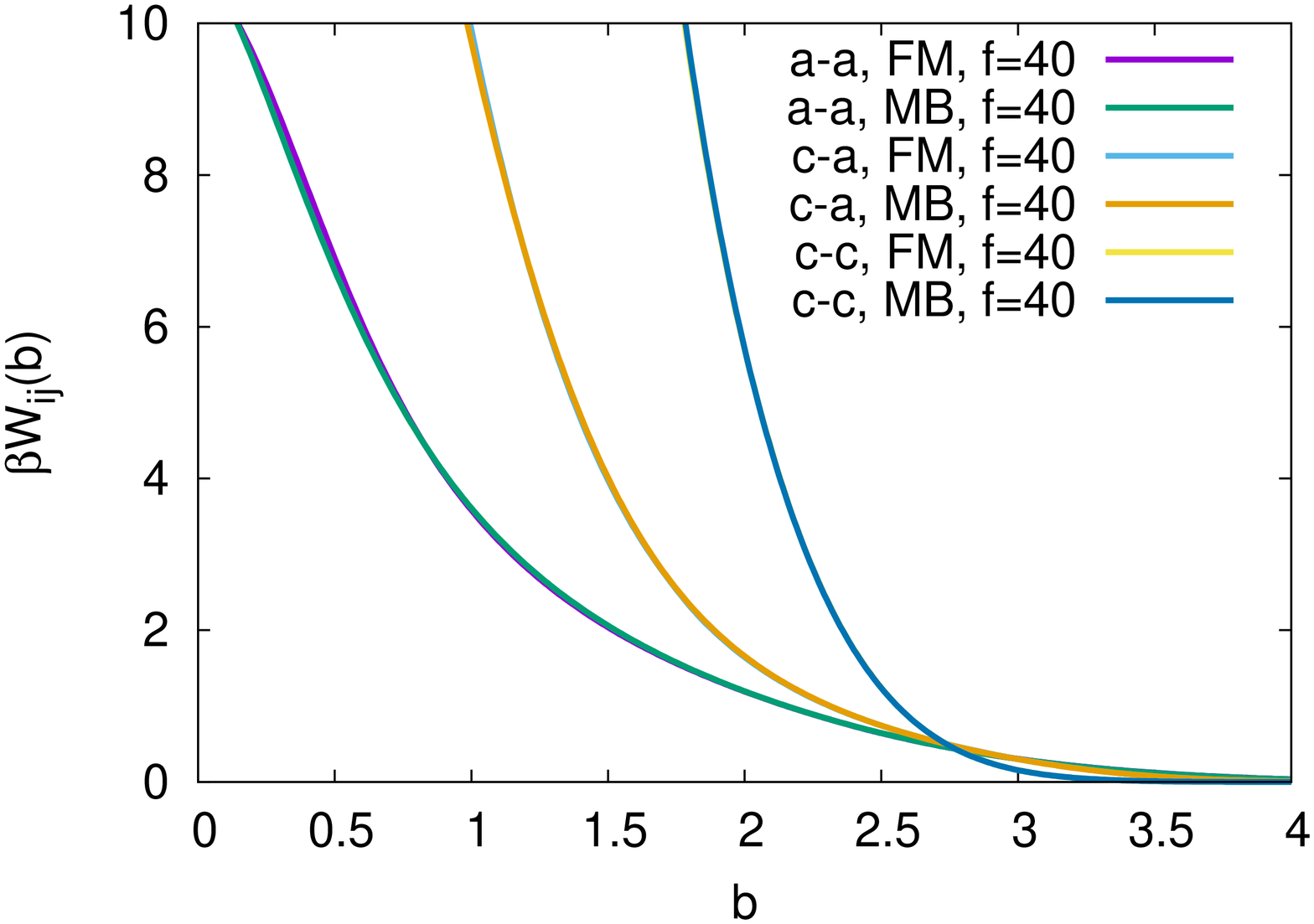}  \\
\caption{Potentials of mean force $\beta W_{{aa}}$ (arm-arm), 
$\beta W_{{ca}}$ (center-arm), and $\beta W_{{cc}}$ (center-center)
 as a function of $b=r/\hat{R}_g$, for star polymers with $f=6$ and 12 
(top), and with $f = 40$ (bottom, label FM). For $f=40$ we also 
report the 
potentials of mean force computed in the multiblob (MB) model, see 
Sec.~\ref{sec3.2}. 
They are essentially indistinguishable from the corresponding full-monomer 
quantities, confirming the accuracy of the inversion procedure.
}
\label{fig:VMF}
\end{center}
\end{figure}

The potentials of mean force (again we average over all equivalent arms) 
are reported in Fig.~\ref{fig:VMF}. They depend on $f$
and show that stars become significantly less penetrable as 
$f$ increases. For instance, for $f=6$ and 12 we have 
$\beta W_{aa}(b=0) = 3.1, 4.7$, while $\beta W_{aa}(b) \gtrsim 10$
for $f = 40$. As expected, for large values of $f$, blobs of two different 
stars cannot fully overlap: blobs are densely packed, and therefore, 
when a blob of one polymer moves closer to a blob of  a second polymer, 
it feels the strong repulsion of the close neighboring blobs, 
which forbids a significant overlap. This is also confirmed by the fact that,
for large $f$, $\beta W_{aa}(b)$ decays slowly with $b$. For $f=40$,
if $\hat{r}_g$ is the average radius of gyration of the blob 
($\hat{r}_g \approx 0.45 \hat{R}_g$), we have 
$\beta W_{aa}(2 \hat{r}_g/\hat{R}_g) \approx 4.0$: blobs of different
polymers rarely overlap. This is good news for our CG procedure, which is 
reliable only if there is little overlap among blobs of different chains.
The range of the potentials of mean force is approximately
$3 \hat{R}_g$, in agreement with the analysis of the monomer distribution 
of Ref.~\onlinecite{RP-13}. However, note the presence of a longer tail 
in $W_{aa}(b)$ which decreases slowly with $b$. Such tail has a simple 
geometrical interpretation. Since the center-arm distance is of order 
of $R_{ca}$, we expect contributions to $W_{aa}(b)$ at least up to 
$4 R_{ca} \approx 
4 \hat{R}_g$ (two arms at the opposite sites of the two interacting stars).

\section{Determination of the coarse-grained potentials } \label{sec3}

\subsection{Intramolecular potentials} \label{sec3.1}

\begin{figure}[t]
\begin{center}
%%\advance\leftskip-2cm
\begin{tabular}{cc}
\includegraphics[angle=-90,scale=0.30]{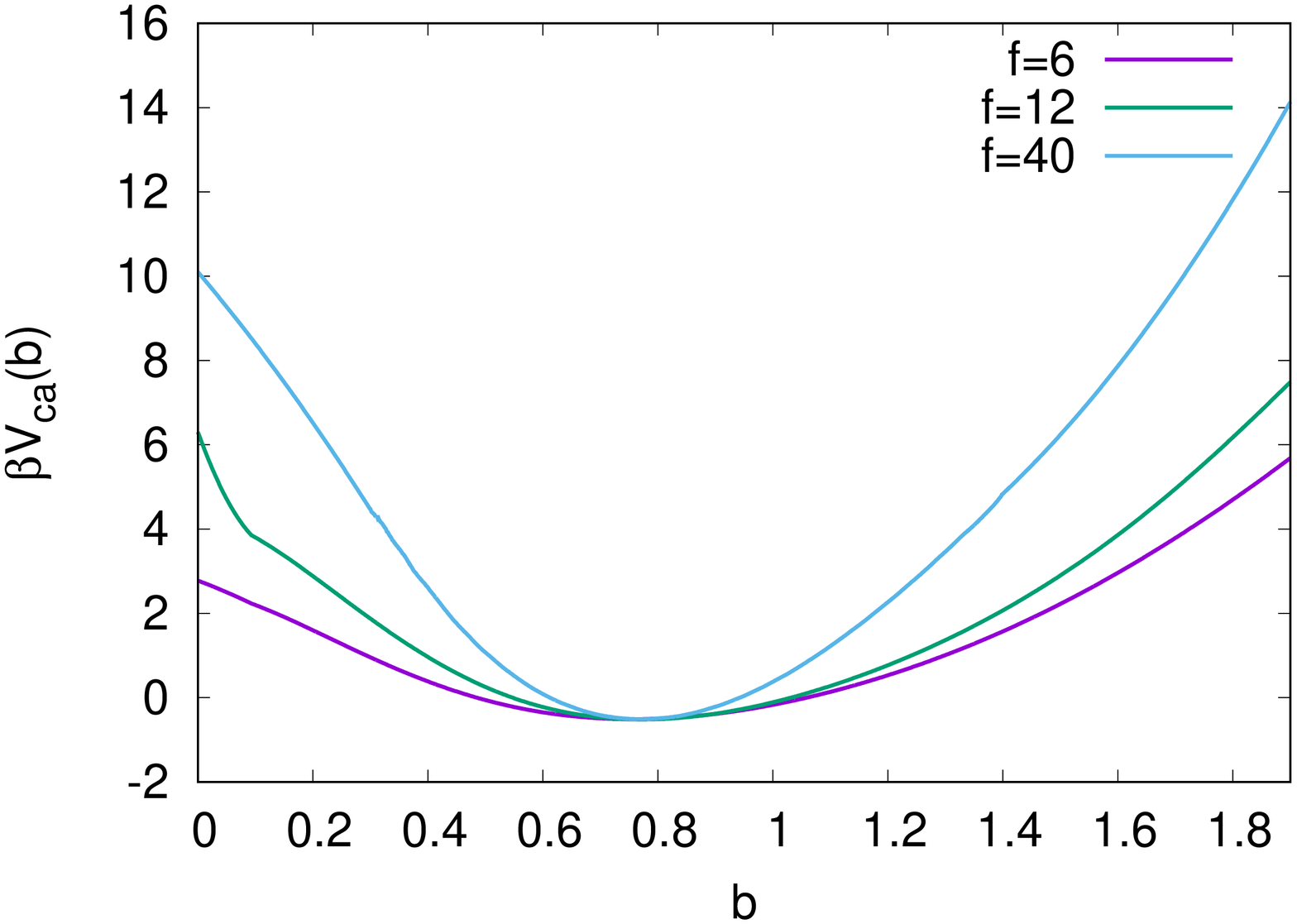} \\
\includegraphics[angle=-90,scale=0.30]{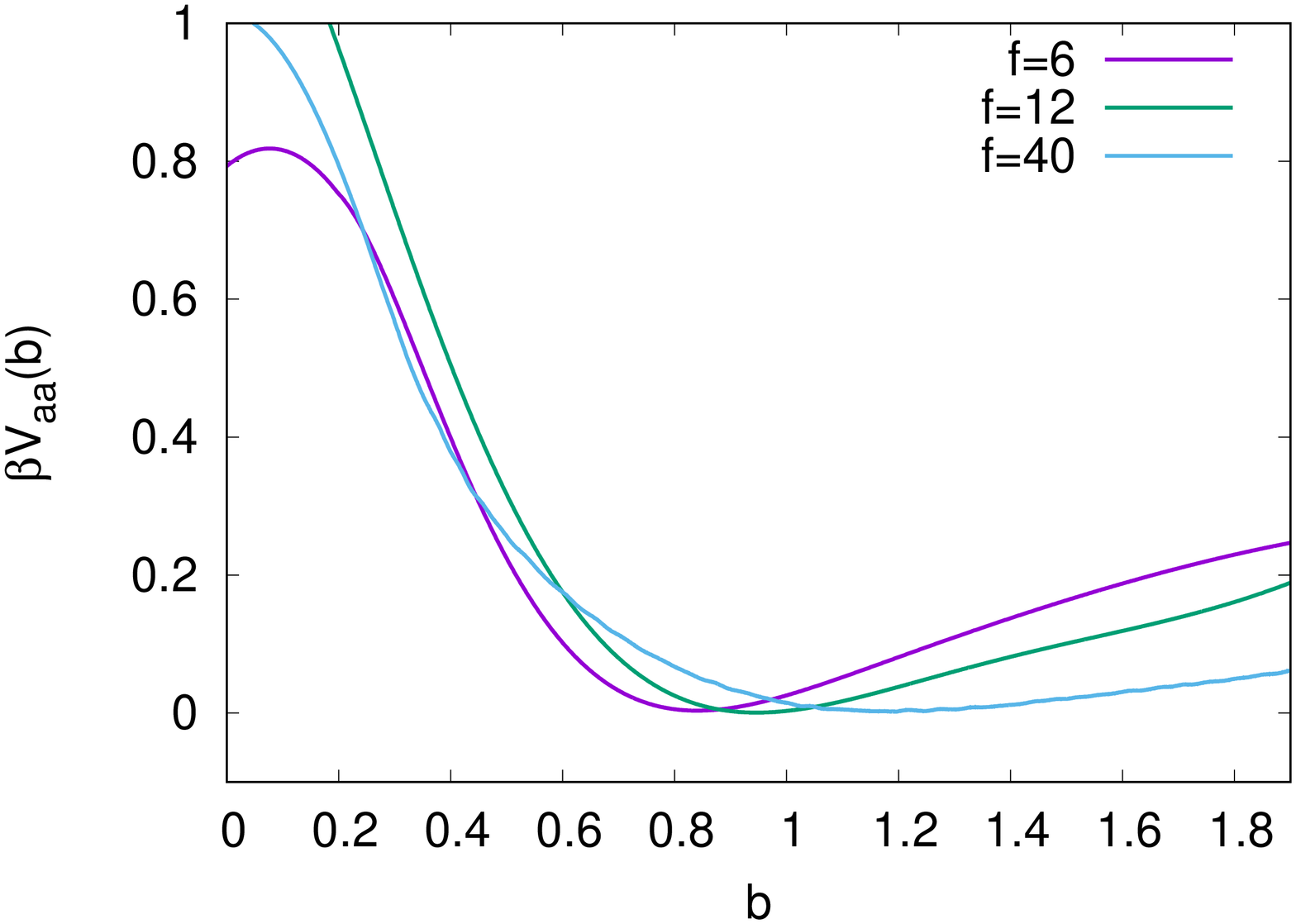} \\
\includegraphics[angle=-90,scale=0.30]{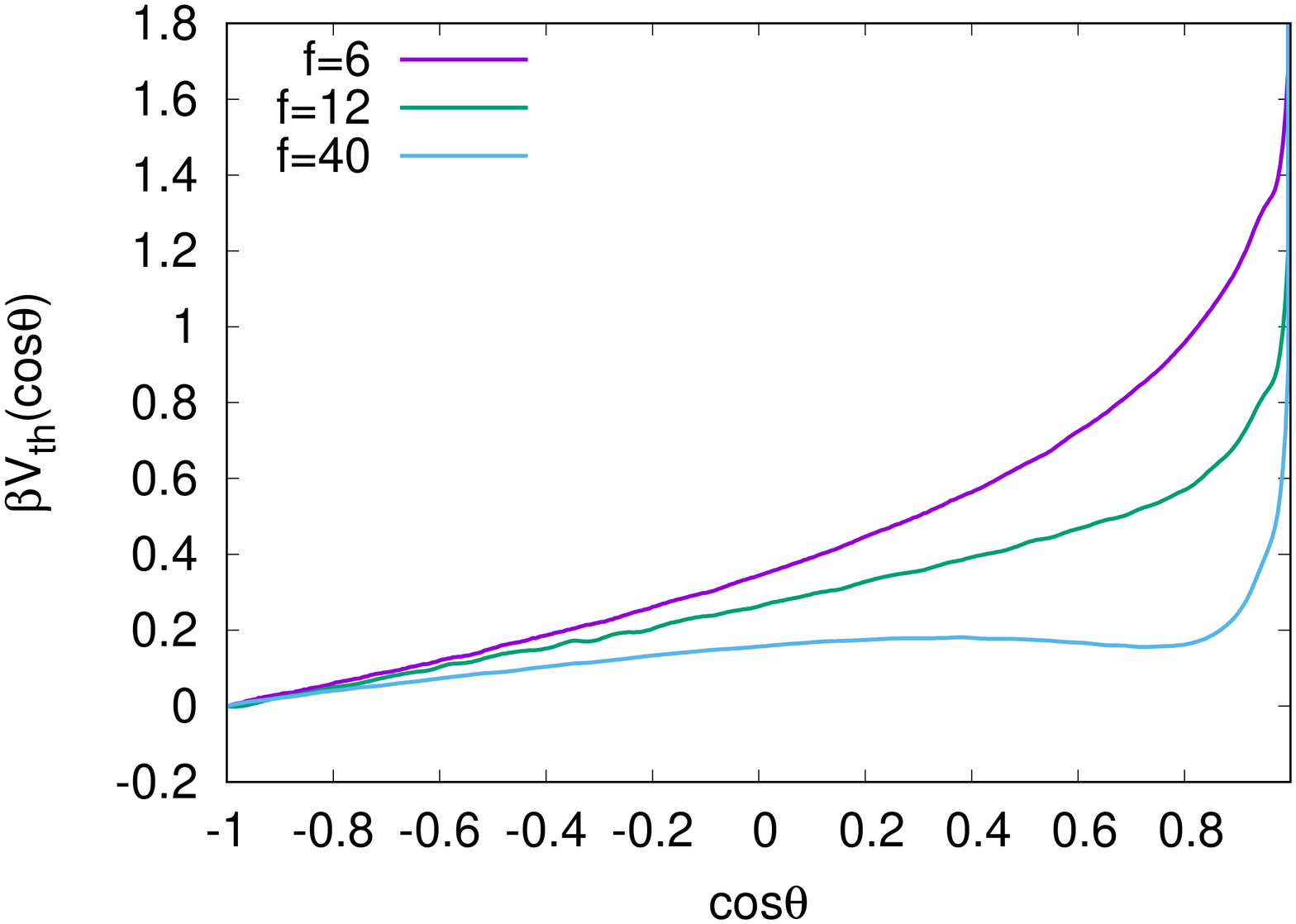} 
\end{tabular}
\caption{Intramolecular potentials for $f=6$, 12 and 40. 
Top: center-arm potential $\beta V_{ca}(b)$;
middle: arm-arm potential $\beta V_{aa}(b)$;
bottom: angular potential $\beta V_{th}(\cos\theta)$.
All potentials are normalized so that their minimum value is zero.
}
\label{IntrapotMBs}
\end{center}
\end{figure}

The CG multiblob potentials 
$\beta V_{ca}(b)$, $\beta V_{aa}(b)$, and $\beta V_{\rm th}(\cos\theta)$ 
are determined by enforcing the equality of the intramolecular distribution
functions
\begin{equation}
P^{FM}_i(x_i)=P^{CG}_i(x_i),
\end{equation}
where the superscripts FM and CG refer to the full-monomer atomistic
model and to the
coarse-grained model, respectively, and the distributions 
$P_i$, $i=1,2,3$, correspond to the three functions $P_{aa}$, $P_{ca}$,
 and $P_{\rm th}$ defined in the previous Section. For this purpose, we use
the iterative Boltzmann inversion (IBI) 
method.\cite{Schommers-83,MP-02,RPMP-03} We set 
$\beta V_{(0,i)}^{CG}(x_i) = - \ln P^{FM}_i(x_i)$ and then we use an iterative 
procedure that generates a sequence of approximations $\beta V_{(n,i)}^{CG}(x_i)$,
$n=1,2,\ldots$,
\begin{equation}
\beta V_{(n+1,i)}^{CG}(x_i)= 
\beta V_{(n,i)}^{CG}(x_i)
-a\ln{\left(\frac{P_i^{FM}(x_i)}{P_{(n,i)}^{CG}(x_i)}\right)},
\label{IBI-eq}
\end{equation}
where $P_{(n,i)}^{CG}(x_i)$ is the distribution function computed in the 
CG model with potentials $V_{(n,i)}^{CG}(x_i)$. Here $a$ is a mixing parameter,
which is tuned to optimize the convergence of the iterations.
For $f=40$, we start with $a=0.2$ and decrease it systematically till 
$a=0.1$. Convergence is achieved after 100 iterations.

The potentials are reported in Fig.~\ref{IntrapotMBs}  
for $f=6$, 12, and 40.  They are normalized so that their minimal value is 
zero. We are not able to determine them accurately 
for small $b$ ($b\lesssim 0.2$, say),
as the corresponding distribution functions are very small, and therefore 
not accurate, in this region.  However, for the same reason, the probability
that two blobs are so close is tiny. Hence, this lack of accuracy is not 
relevant in practice.
The center-arm potential $V_{ca}(b)$ is repulsive at short distances
where blobs overlap, takes its minimum for $b \approx 0.8$, 
and has a strong attractive tail at large distances, 
which is necessary to bind each arm to the center of the star. 
It is soft for $f=6$, with $V_{ca}(b=0)\approx 2.8k_BT$ at 
full overlap, but it becomes harder and narrower for $f=40$. 
For this value of $f$, we have
$V_{ca}(b=0)\gtrsim 10k_BT$, confirming the  increasing conformational 
rigidity of the star polymer as $f$ increases. 
The arm-arm potential $V_{aa}(b)$ has a soft, repulsive core, which 
increases slightly with $f$, and an attractive tail,
whose strength decreases as $f$ increases. 
The angular potential $V_{\rm th}(\cos\theta)$ 
is repulsive and takes its maximum for $\theta = 0$,
as a consequence of the repulsive interaction among different arms. 
As $f$ increases, $V_{\rm th}(\cos\theta=1)$ 
decreases and we expect the potential 
to vanish everywhere for large values of $f$.

\begin{table}[t]
\begin{center}
\begin{tabular}{ccc}
\hline\hline
 $f$ & FM & CG \\
 \hline
 6 & 0.759 & 0.760\\
12 & 0.772 & 0.771\\
40 & 0.793 & 0.795\\
\hline\hline
\end{tabular}
\vspace{0.7cm}
\caption{Ratio $\hat{R}^2_{g,b}/\hat{R}^2_g$ for $f=6$, 12, and 40, computed in
the full-monomer (FM) and (CG) models, for an isolated
star polymer (zero-density limit). 
Here $\hat{R}_{g,b}$ is the zero-density blob radius of gyration,
see Eq.~(\ref{radiusgyrMB}) for the definition.
}
\label{Rgbtable}
\end{center}
\end{table}

In order to assess the accuracy of the inversion procedure, we consider the
average radius of gyration $R_{g,b}$ of the CG molecule, which should 
coincide with  
\begin{equation}
R^2_{g,b}=
\frac{1}{2(f+1)^2}
\left\langle \sum_{i,j=0}^{f}({\bm R}_{i}-{\bm R}_{j})^2\right\rangle
\label{radiusgyrMB}
\end{equation}
in the CG representation of the polymer.  Note that 
it does not coincide with the 
radius of gyration ${R}_g$ of the original molecule.
The two quantities satisfy the exact relation\cite{DPP-12-Soft}
\begin{equation}
{R}^2_{g}={R}^2_{g,b}+{r}^2_{g,b},
\label{rgb}
\end{equation}
which also holds for each polymer chain.
Here ${r}^2_{g,b}$ is the average blob radius of gyration
\begin{align}
{r}^2_{g,b}&=\frac{1}{(f+1)}\sum_{\alpha=0}^f {r}^2_{g,b,\alpha},
\nonumber\\
{r}^2_{g,b,\alpha}&=\frac{1}{2mf}
    \left<\sum_{j,k=1}^{mf}( {\bm r}_j^\alpha-{\bm r}_k^{\alpha})^2\right>,
\end{align}
where ${\bm r}_k^\alpha$ are the positions of the monomers belonging 
to blob $\alpha$ and $m=L/(f+1)$.
In Table~\ref{Rgbtable}, we report the ratio $\hat{R}^2_{g,b}/\hat{R}^2_g$
for an isolated polymer (we remind the reader that a hat indicates that the 
corresponding quantity has been determined in the zero-density limit), 
comparing full-monomer and multiblob results. Differences are small, 
confirming the accuracy of the inversion procedure. 
From the data reported in Table~\ref{Rgbtable}, we can also estimate the average
radius of gyration of the blobs in the zero-density limit.
Using Eq.~(\ref{rgb}), we obtain 
$\hat{r}_g/\hat{R}_g = 0.49,0.48,0.45$ for $f=6,12,40$.  
Note that, in spite of the fact that the star polymer is significantly more 
packed and dense as $f$ increases, $\hat{r}_g$ changes only slightly 
with the number of arms, another indication that different blobs
strongly overlap.

\subsection{Intermolecular interactions} \label{sec3.2}

We now compute the three intermolecular potentials,
$\beta \widetilde{V}_{ca}(b)$, $\beta \widetilde{V}_{aa}(b)$, and 
$\beta \widetilde{V}_{cc}(b)$, 
requiring the CG model to reproduce the 
full-monomer mean-force potentials, i.e., enforcing
\begin{equation}
\beta W^{CG}_{ij}(b)=\beta W^{FM}_{ij}(b),
\end{equation}
where $ij = cc$, $aa$, and $ca$. The determination of the potentials 
satisfying this condition is not trivial, and we have used a 
different strategy for the low-functionality stars ($f=6$ and $12$) and 
for $f=40$. In the first case we have essentially used the IBI method, 
\cite{Schommers-83,MP-02,RPMP-03} while for large functionalities a 
different procedure has to be employed. Details are given in the Appendix.

In all cases we obtain relatively short-range 
potentials, which are able to 
reproduce well the long tails that are present in the potentials 
of mean force. This is very relevant for the thermodynamic consistency 
of the CG approach, as tails give an important 
contribution to the pressure. Therefore, matching the tails is essential
to obtain the correct thermodynamics. 

\begin{figure}[t]
\begin{center}
\centering
\includegraphics[angle=-90,scale=0.35]{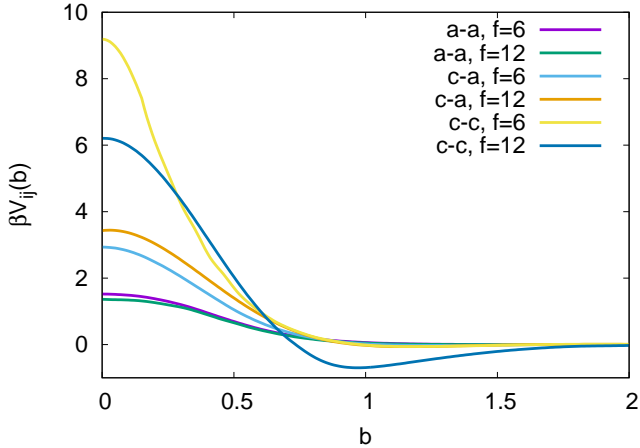}
\caption{Intermolecular potentials for $f=6$ and 12. 
We report the arm-arm potential $\beta\widetilde{V}_{aa}$, 
the center-arm potential 
$\beta\widetilde{V}_{ca}$, and the center-center potential 
$\beta\widetilde{V}_{cc}$. The latter is only meaningful for 
$b\gtrsim0.2$ and 0.8 for $f=6$ and 12, respectively.}
\label{InterpotMBs}
\end{center}
\end{figure}

The results for $f=6$ and 12 are shown in Fig.~\ref{InterpotMBs}.
It is important to stress that we are able to compute each potential only 
for those values of $b$ for which $W^{FM}_{ij}(b)$ is not too large
(in practice we only consider the regions in which 
$\beta W^{FM}_{ij}(b)\lesssim 10$).
In particular, the center-center potential is determined only for 
$b\gtrsim0.2$, 0.8 for $f=6$, 12, respectively. This is not a limitation,
however, since the probability that the centers of the two 
stars are so close is always extremely low, see Fig.~\ref{fig:VMF}.
For both values of $f$,
the arm-arm interaction is purely repulsive and 
soft at contact: $\widetilde{V}_{aa}(b=0)\approx 1.5k_BT$. 
It increases only slightly from $f=6$ to $f=12$. 
The center-arm interaction is more repulsive than the arm-arm interaction, 
$\widetilde{V}_{ca}(b=0)\approx 3k_BT$, $3.5k_BT$ for $f=6$ and 12, 
respectively, 
but the two potentials have approximately the same range. 
The center-center potential is the most repulsive interaction. Note that the 
potentials we have determined show the unexpected feature that 
$\widetilde{V}_{cc}(b)$ decreases for small values of $b$ as $f$ increases. 
This behavior, however, occurs in the region in which the $f=12$ potential
cannot be trusted, and it is most likely incorrect.
More interestingly, the center-center potential has 
an attractive tail for $r/\hat{R}_g\gtrsim 1$, 
whose depth increases slightly from $f=6$ to $f=12$.

\begin{figure}[t]
\begin{center}
\advance\leftskip-1.7cm
\begin{tabular}{c }
\includegraphics[angle=-90,scale=0.3]{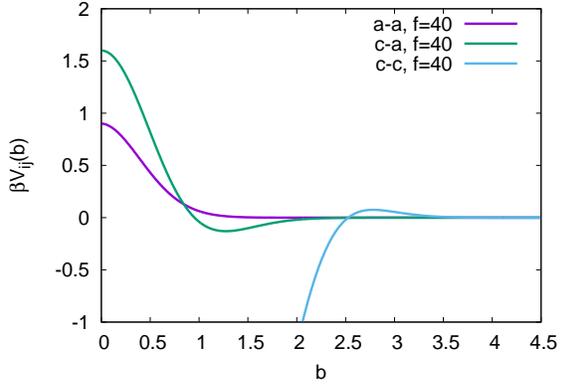} \\
\end{tabular}
\caption{Intermolecular potentials of the multiblob model for $f=40$. 
We show
the arm-arm potential $\beta\widetilde{V}_{aa}$, the center-arm potential
$\beta\widetilde{V}_{ca}$ (valid for $b\gtrsim 1.0$), 
and the center-center potential $\beta\widetilde{V}_{cc}$
(valid for $b\gtrsim 1.9$).
}
\label{Interpot40}
\end{center}
\end{figure}

The potentials $\widetilde{V}_{ij}(b)$ for $f=40$ are reported in 
Fig.~\ref{Interpot40}. They are only relevant where 
$\beta W_{ij}(b)\lesssim 10$. Therefore, the
center-arm and the center-center potentials cannot be trusted for distances 
$b\lesssim1$ and $b\lesssim1.9$, respectively. The potentials reproduce
quite accurately the target distributions. A detailed analysis of the errors,
see Appendix, indicates that $\widetilde{V}_{aa}(b)$ and
$\widetilde{V}_{ca}(b)$ have an accuracy of 1\%. The center-center potential
is less accurate, with an error of at most 5\%.

The arm-arm potential is 
soft, $\beta \widetilde{V}_{aa}(b=0)\approx 0.9$, and decreases rapidly.
For $b=1$ we have $\beta \widetilde{V}_{aa}(b)\approx 0.062$, 
which is very close to the value it assumes for $f=6$ and 12. 
This is consistent with the idea that the range of the potential is of 
the order of $2\hat{r}_g \approx \hat{R}_g$. 
The center-arm potential 
is small and attractive in the relevant region $b\gtrsim 1$. For $b=1$,
$\beta \widetilde{V}_{ca} \approx -0.04$, while at the minimum (located 
at $b\approx 1.28$) we have $\beta \widetilde{V}_{ca} \approx -0.13$. 
Finally, the center-center potential is attractive for $b\lesssim 2.5$---this 
also occurs for $f=6$ and 12---and then it has a small repulsive tail.

\section {Thermodynamic behavior} \label{sec4}

In this section, we analyze the thermodynamic and structural properties 
of star polymer solutions under good-solvent conditions, 
both at zero and finite density. 
The results we obtain using the CG multiblob (MB) model will then be compared 
with those obtained using the SB model, in which 
stars are represented by monoatomic molecules interacting by means of 
pair potentials. We use two different SB representations. First, we 
consider the midpoint (MP) representation, in which we take the 
center of the star as interaction site (this is the one mostly considered 
in SB studies, see, e.g., Refs.~\onlinecite{LLWAKAR-98,Likos-01}). 
For linear polymers (they can be viewed as stars with $f=2$ arms), 
it corresponds to taking the central monomer as interaction site.
A second possibility 
consists in taking the polymer center of mass (CM) as interaction 
site. For linear polymers (therefore, for $f=2$), 
the CM representation is more accurate than the 
MP one. \cite{DMPP-15-SpecTopics} As $f$ increases, the CM of the star 
converges to the polymer center and therefore the two representations 
become equivalent for large $f$. Therefore, we present results for 
the two different choices only for $f=6$ and $f=12$. For $f=40$ there
is little difference between them. For the MP case, we use the 
pair potentials reported in Ref.~\onlinecite{HG-04}.
For the CM representation we have determined the 
potentials numerically using the optimal Domb-Joyce lattice model.\cite{DJ-72}

\subsection{Zero density} \label{sec4.1}
\label{ZDstar}

\begin{table*}[t]
\begin{center}
\begin{tabular}{ c c c c c c c}
\hline\hline
& \multicolumn{2}{ c }{$f=6$} & \multicolumn{2}{ c }{$f=12$} & \multicolumn{2}{ c }{$f=40$} \\
\cline{2-3}\cline{4-5}\cline{6-7}
& $A_2$ & $A_3$ & $A_2$ & $A_3$ & $A_2$ & $A_3$\\
SB-CM & 14.15 & 88.43 & 23.70 & 288.96 & 42.39 & 1040 \\
SB-MP & 14.66 & 85.6 & 23.97 & 288 &  \\
MB & 14.231(4) & 88.6(6) & 23.78(5) & 289(2) & 41.91(2) & 1028(6)  \\
FM & 14.202(12) & 90.3(4) & 23.54(3) & 290(2)  & 41.90(11) & 1031(13)\\
\hline
\end{tabular}
\vspace{0.7cm}
\caption{Universal adimensional combinations $A_2$ and $A_3$ for $f=6$, 12. 
We give results obtained by using single-blob models, in the 
center-of-mass (SB-CM) and center (SB-MP) 
representations, and the multiblob model (MB). We also report 
the full-monomer (FM) results of Ref.~\onlinecite{RP-13}.}
\label{Virialcoeffcomp}
\end{center}
\end{table*}

At zero density the accuracy of the CG model can be tested by comparing CG and
full-monomer estimates of the universal adimensional coefficients 
$A_n=B_n/\hat{R}^{3(n-1)}_g$, where $B_n$ is the 
$n$-th virial coefficient. As discussed in Ref.~\onlinecite{DPP-12-Soft},
the comparison of $A_2$ provides a test of the inversion procedure.
Results (virial coefficients are computed as discussed in 
Refs.~\onlinecite{CMP-06,CMP-08-virial})
are reported in Table~\ref{Virialcoeffcomp}.
The results obtained by using the SB model in the CM representation and the 
MB model are very close to the FM ones of Ref.~\onlinecite{RP-13}. 
Differences are less than 1\%. Slightly larger differences (approximately 
3\%) are obtained in the MP representation. They are due 
to the slight inaccuracy of the
interpolating formula of Ref.~\onlinecite{HG-04}.

Let us now compare the third virial combination $A_3$, which gives information
on the relevance of the three-body interactions.\cite{DMPP-15-SpecTopics}
As it can be seen from 
Table~\ref{Virialcoeffcomp}, all CG results are close to the FM ones. 
Clearly, at least in the small-density region, the thermodynamics is 
controlled by pair interactions, at variance with what happens with 
linear polymers (in that case $A_3$ is understimated by 21\% in the CM
representation and by 50\% in the MP representation
\cite{DMPP-15-SpecTopics}). 
This is probably due to the fact that stars are more compact
objects, so multiple overlaps are rare. These results confirm the assumption
that the description in terms of pair interactions becomes more accurate as 
$f$ increases in the very dilute regime.

\subsection{Finite-density results: $f=6$ and $f=12$} \label{sec4.2} 

\begin{figure}[!t]
\begin{center}
\centering
\includegraphics[angle=0,scale=0.43]{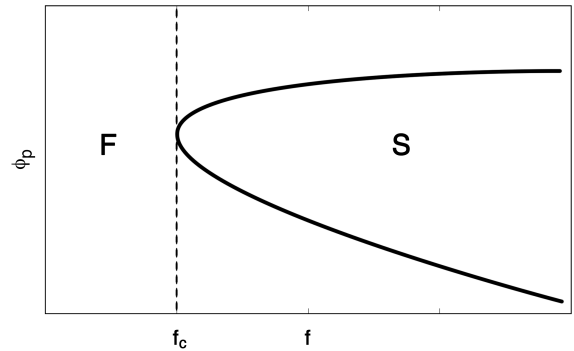}
\caption{Expected phase diagram for star polymers. Here $f$ is
the functionality of the polymer and $\Phi_p$ the polymer volume
fraction. ``S" corresponds to a solid intermediate phase, ``F" to a fluid phase.
}
\label{phase-diag}
\end{center}
\end{figure}

\begin{figure}[!t]
\begin{center}
\centering
\includegraphics[angle=-90,scale=0.35]{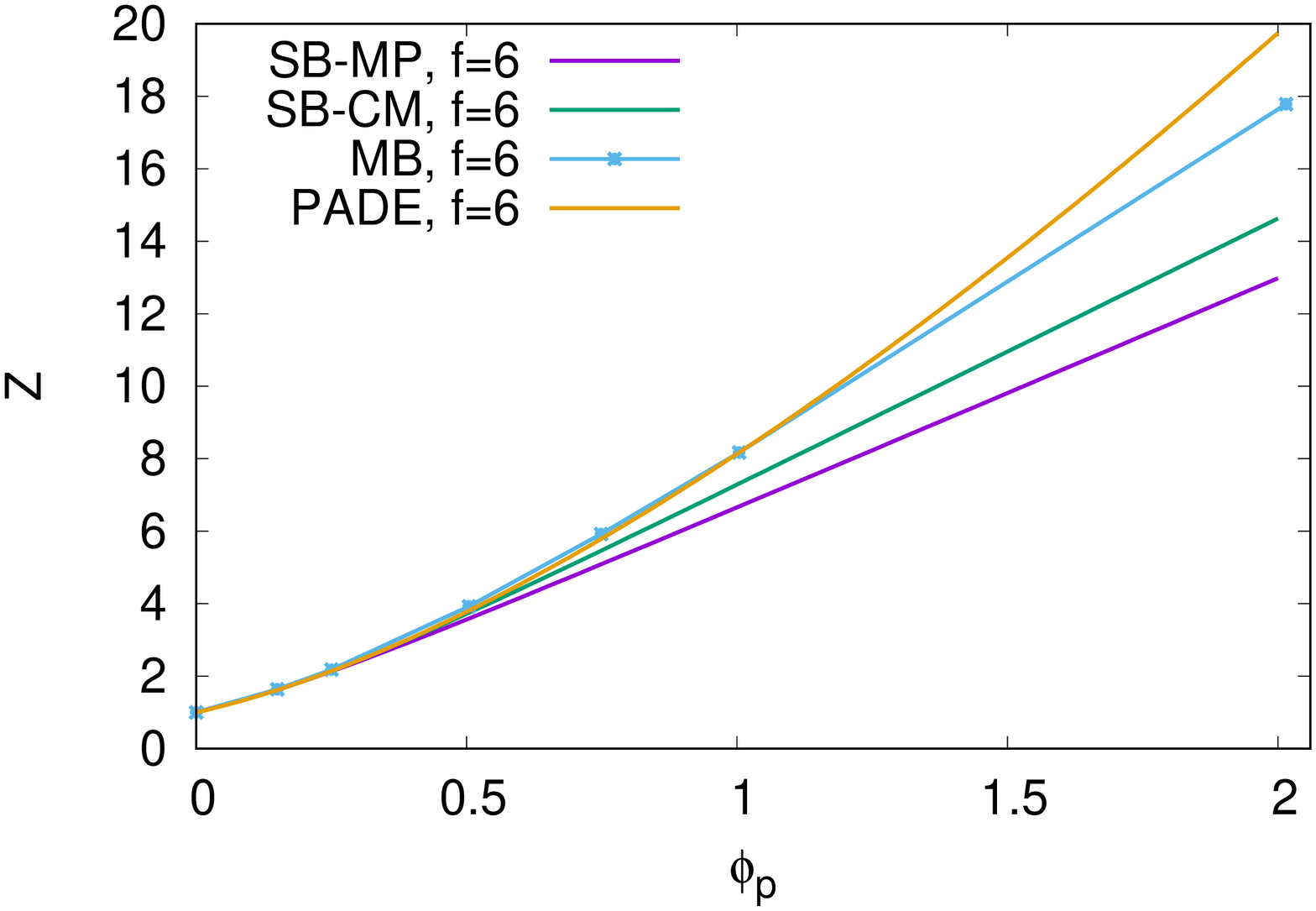}
\includegraphics[angle=-90,scale=0.35]{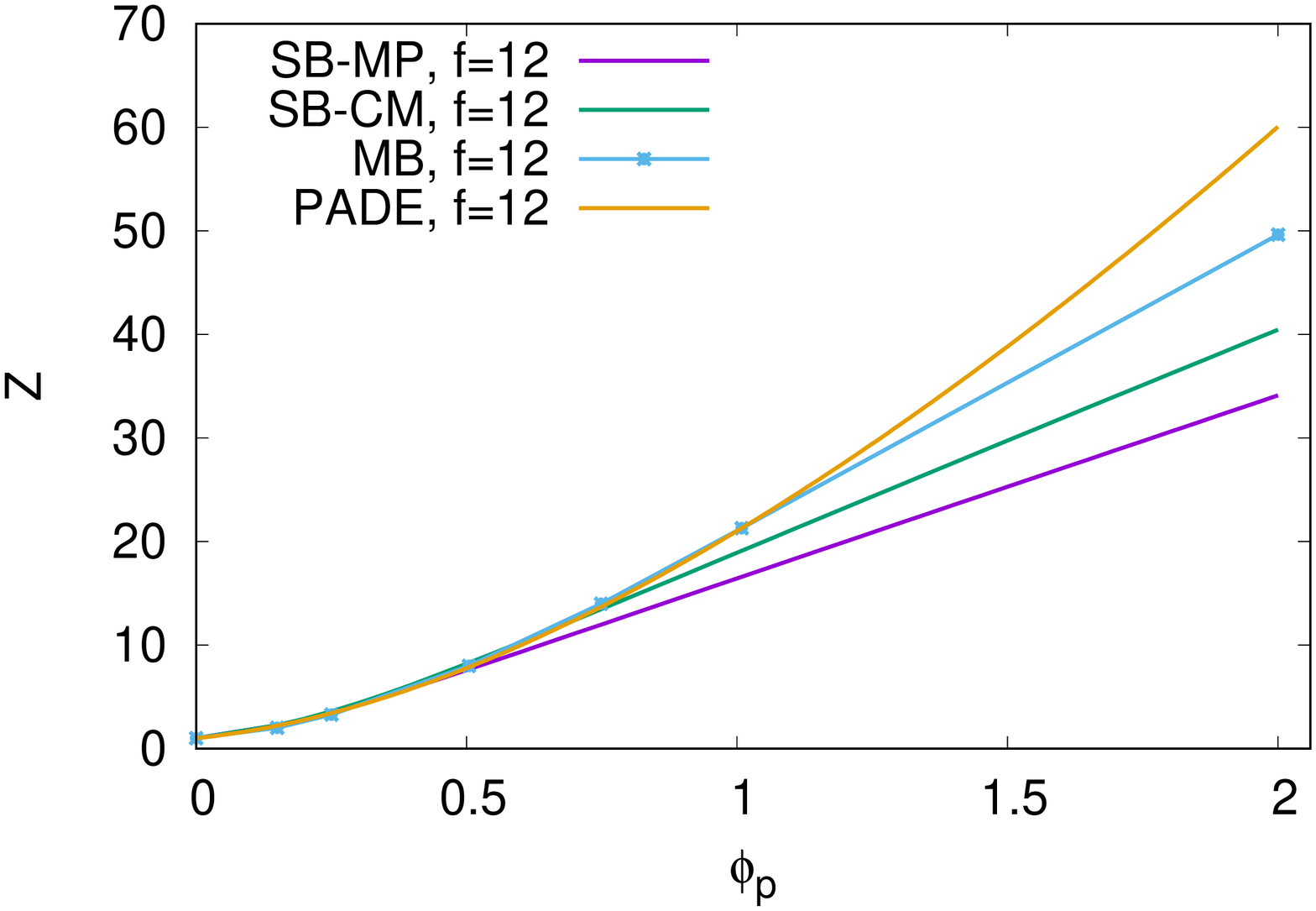}
\caption{Osmotic coefficient $Z=\beta P/\rho_p$ as a function of 
the volume fraction $\Phi_p$, for $f=6$ (top) and $f=12$ (bottom). 
We report multiblob (MB) estimates, single-blob results in the 
center of mass (SB-CM) and center (SB-MP) representations, 
and the extrapolation (\ref{Pade}) (PADE).
}
\label{osmo}
\end{center}
\end{figure}

At finite density, star polymers are supposed to have two distinct behaviors
depending on the functionality $f$. For small values of $f$, the density 
can be increased at will and polymers smoothly go from the dilute
regime to the semidilute one, and finally, to the melt. 
On the other hand, for large $f$, a fluid-solid transition occurs in the 
dilute regime, with the appearance of a solid intermediate phase. 
\cite{WPC-86,WP-86,WLL-98}
The transition is expected to occur for $f > f_c$, see Fig.~\ref{phase-diag},
where $f_c$ 
is predicted by SB models to be in the range\cite{WLL-98,MP-13}
$30\lesssim f_c \lesssim 40$.
For this reason, we consider here only the results for 
$f=6$ and 12.  Results for $f=40$ are discussed in the next section.

We determined the osmotic coefficient 
\begin{equation}
Z = {\beta P\over \rho_p},
\end{equation} 
which is a universal function of $\Phi_p$ in the large-$L$ limit.
For SB models, $Z$ was computed by using integral equation methods
\cite{hansen:liquids} 
and the Rogers-Young closure.\cite{RY-84} 
For a few selected values of $\Phi_p$, we also 
computed $Z$ by Monte Carlo simulations, 
obtaining results in perfect agreement with the 
Rogers-Young estimates, see supplementary material.
In the case of the MB model, 
for $f=6,12$ 
we performed Monte Carlo canonical simulations in a finite
cubic box of size $L\approx 30 \hat{R}_g$ and periodic boundary 
conditions. The pressure was determined by using the 
molecular virial. \cite{Ciccotti-virial,AC-2004}

We report the estimates of $Z$ in Fig.~\ref{osmo}.
The three CG models are fully consistent for small volume fractions, 
confirming the results of Sec.~\ref{ZDstar} on their
thermodynamic consistency in the limit of zero density.
Differences begin to appear for $\Phi_p \approx 0.5$. This is not unexpected,
since the SB representation is only valid as long as there are no overlaps 
among the stars, i.e., for $\Phi_p\lesssim 1$. The inaccuracy of the 
approximation can be understood by considering the SB models in the 
two different representations: deviations from the results obtained
using the MB model occur exactly where the 
two SB models give significantly different predictions.
As it happens for linear chains, 
the largest deviation is observed for the SB model in the MP
representation. At $\Phi_p=1$, we have 
$Z_{\text{MP}}/Z_{\text{MB}}- 1\approx20\%$ and
$Z_{\text{CM}}/Z_{\text{MB}}-1\approx12\%$, for both $f=6$ and $f=12$. 
At $\Phi_p=2$, we have $Z_{\text{MP}}/Z_{\text{MB}}-1\approx30\%$
and $Z_{\text{CM}}/Z_{\text{MB}}-1\approx20\%$ ,
again for both $f=6$ and 12.
It is interesting to note that deviations are essentially independent of $f$,
so that the SB approximation does not become more accurate as $f$ increases,
at variance with what occurs for very small densities.

\begin{table}[t]
\begin{center}
\begin{tabular}{ c | c c c c c}
\hline\hline
$f$ & $a_1$ & $a_2$ & $a_3$ & $a_4$ & $a_5$\\
\hline
6 & 3.70768 & 5.78599 & 1.1215 & --- & ---\\
12 & 5.26919 & 13.9592 & 0.982019 & --- & --- \\
40 & 11.2498 & -7.07114 & 829.674 & -1095.29 & 436.441\\
\hline
\end{tabular}
\vspace{0.7cm}
\caption{Coefficients of the Pad\'e approximant \eqref{Pade} to the 
compressibility factor $Z(\Phi_p)$, for $f=6$ and 12. 
For $f=40$, we report the coefficients of the fifth-order polynomial 
expansion of the compressibility factor $Z(\Phi_p)$, as explained in the 
text.}
\label{Compparam}
\end{center}
\end{table}

Ref.~\onlinecite{CMP-06} extrapolated the virial expansion for 
linear polymers to the semidilute  
regime, by using a simple Pad\'e approximant and enforcing
\cite{deGennes-79,Freed-87,dCJ-book} 
the large-$\Phi_p$ behavior $Z(\Phi_p)\sim \Phi_p^{1.311}$. 
Comparison \cite{Pelissetto-08}
with full-monomer results in the semidilute limit later 
showed that the parametrization was reasonably accurate with
an error of 2.5\% and 5\% for $\Phi_p =5$ and 10, respectively. 
We perform here the 
same extrapolation, using also the MB finite-density results 
to improve the approximation. The compressibility factor is parametrized as
\begin{equation}
\label{Pade}
Z(\Phi_p)=\left(\frac{1+a_1\Phi_p+a_2\Phi_p^2}{1+a_3\Phi_p}\right)^{1.311}.
\end{equation}
The coefficients $a_1$, $a_2$, and $a_3$ 
are fixed by requiring  $Z$ to reproduce the full-monomer universal 
combinations $A_2$ and $A_3$, and the MB compressibility factor $Z$ 
for $\Phi_p=1$. The coefficients are reported in 
Table~\ref{Compparam} for $f=6$ and 12. The extrapolated curves 
are reported in Fig.~\ref{osmo}. The Pad\'e extrapolation formula
predicts larger pressures than the MB 
model, although differences are only relevant for 
$\Phi_p\gg 1$ (see the supplementary material for the numerical results). 
For $\Phi_p = 2$ we find 
$Z_{\rm Pade}/Z_{MB} - 1 = 0.11$ and 0.21 for $f=6$ and 12, respectively.
If we take these differences seriously, they indicate that the accuracy 
of the MB model decreases with increasing values of $f$, a result that 
we will better explain in Sec.~\ref{sec6}.

It is interesting to compare our results with the renormalization-group
perturbative predictions obtained in the dimensional Wilson-Fisher expansion.
\cite{CBMF-86,DRF-90,MBLFG-93} As discussed in 
Refs.~\onlinecite{DRF-90,MBLFG-93}, this approach is only reliable for 
small values of $f$. This is fully confirmed by the comparison of 
the renormalization-group predictions with our MB data, see supplementary
material. For $f=6$, they differ by less than 6\% in the dilute regime 
$\Phi_p\le 1$. On the other hand, 50\% discrepancies are observed for $f=12$.

\subsection{Finite-density results: the fluid phase for $f=40$} \label{sec4.3}

For $f=40$ SB models predict a transition 
\cite{WLL-99,MP-13} 
between the  fluid phase
and a bcc solid phase: The fluid phase 
extends only up to a packing fraction $\bar{\Phi}_{p,s}$ of the order of 1.
More precisely, one obtains\cite{WLL-99} $\bar{\Phi}_{p,s}\approx 1$ 
by using the phenomenological potentials of Ref.~\onlinecite{LLWAKAR-98} and 
$\bar{\Phi}_{p,s}\approx 1.5$,\cite{MP-13} by using an extrapolation of the 
potentials of Ref.~\onlinecite{HG-04}. 
For this reason, we have performed MB canonical simulations only up 
to $\Phi_p = 0.9$. We consider fluid systems---we start the simulation from 
disordered configurations---of size
$L\approx 20\hat{R}_g$ with periodic boundary conditions.

\begin{figure}[!tb]
\begin{center}
\centering
\includegraphics[angle=-90,scale=0.35]{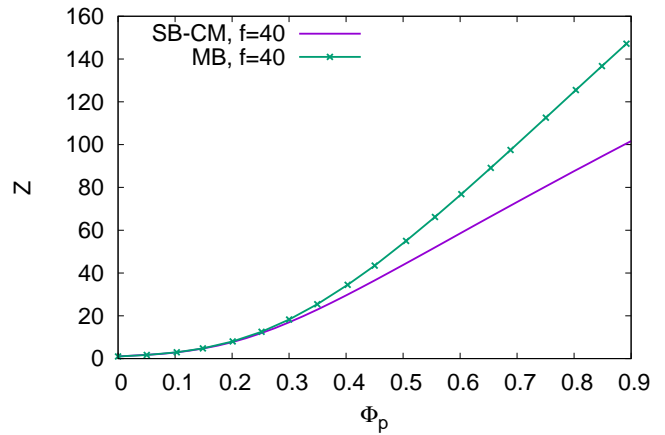}
\caption{Osmotic coefficient $Z=\beta P/\rho_p$ as a function 
of the volume fraction $\Phi_p$, for $f=40$. 
We report multiblob (MB) results and single-blob (SB-CM) results in the 
center of mass  representation.
All results are obtained in the fluid phase, which is metastable 
for $\Phi_p\gtrsim 0.4$ (MB) and for $\Phi_p\gtrsim 0.6$ (SB-CM).
The MB results (points) have been obtained by Monte Carlo simulations;
we also report (line) the corresponding polynomial interpolation 
(see Table~\ref{Compparam} for the coefficients). The SB results have been
obtained by integral-equation methods.
}
\label{osmo40}
\end{center}
\end{figure}

In Fig.~\ref{osmo40} we report the osmotic coefficient
$Z(\Phi_p)$ for the SB (CM representation) and MB model. 
While the results obtained in the two CG models are 
fully consistent in the zero-density limit (where $Z$ can be reasonably 
represented by means of the virial expansion, see Sec.~\ref{ZDstar}), 
significant deviations are observed for $\Phi_p\gtrsim 0.3$. 
For $\Phi_p\approx0.5$, 
$Z_{\text{CM}}/Z_{\text{MB}}-1\approx 20\%$. Note that deviations are larger 
for $f=40$ than for $f=6,12$, indicating again that it is not true that 
SB models become more accurate 
as $f$ increases. This is only true for very small densities, where overlaps 
are rare. As $\Phi_p$ increases, the opposite occurs. 
This is due to the fact that, as $f$ increases, 
the internal structure of the polymer, 
which is not taken into account by the SB model, 
plays an increasingly important role. We shall come back to this point in 
Sec.~\ref{sec6}.

As before, we also determine an interpolation formula for $Z$. In this case, 
the expression should only apply up to the fluid-solid transition. 
Therefore, we simply
parametrize the results by means of a 
fifth-order polynomial $Z(\Phi_p)=1+\sum_{k=1}^5 a_k\Phi^k$. 
The coefficients $a_k$ are again reported in Table~\ref{Compparam}.
This interpolation holds up to $\Phi_p = 0.8$, well 
within the solid phase (see Sec.~\ref{sec5}).

It is also interesting to verify whether the polymer equation of state can be 
reasonably approximated by the hard-sphere one, provided one chooses an 
appropriate effective hard-sphere radius $R_c$. As discussed in 
Ref.~\onlinecite{RP-13}, this approximation works nicely for large $f$ in the 
very dilute limit $\Phi_p\to 0$, as the first two virial coefficients $B_2$ and 
$B_3$ satisfy the 
relation $B_3/B_2^2 \approx 5/8$, appropriate for hard spheres. 
The comparison of the finite-density results is performed in the 
supplementary material. It shows that the identification of large-$f$ star
polymers with hard spheres holds only for very dilute solutions. For $f=40$,
at 
$\Phi_p=0.15$ the hard-sphere approximation overestimates $Z$ already
by 25\%.

\section{Fluid-solid transition for $f=40$} \label{sec5}

\subsection{The solid phase: qualitative predictions} \label{sec5.1}

For $f=40$, the analysis of SB models, both using 
phenomenological  \cite{WLL-99} and numerically-determined\cite{MP-13} 
potentials, predicts a solid density window.
As a first step, we investigate
whether such solid phase also occurs when one uses the MB model.
For this purpose we perform canonical simulations, using ordered starting
configurations. We generate 
configurations such that the centers of each star polymer are located on 
the site of a bcc, fcc, and diamond lattice (the lattice structures that were
found to be stable for different values of $f$ in Ref.~\onlinecite{WLL-99}),
while the arms are randomly located around the centers.
Then, the system evolves under a standard Metropolis dynamics 
(see the supplementary material for details).
The volume fractions investigated are $\Phi_p=0.8$, $1$, $1.6$, and 2, 
on cubic simulation boxes with $n=7,8$ elementary cells per side
for all three different types of lattice
(correspondingly, the number of polymers is $2n^3$, $4n^3$, and $8n^3$,
for the bcc, fcc, and diamond case).
For all densities, the fcc and diamond lattices are unstable. 
After $N\approx2\cdot 10^3$ iterations,
there is no presence of the original solid structure. On the other 
hand, the solid bcc structure appears to be stable for $\Phi_p=0.8$ and $1.0$. 
After $10^5$ iterations, the star polymers are still approximately located 
on the sites of a bcc lattice, the molecules only oscillating around the 
lattice sites. 
This confirms the (meta)stability of the bcc structure for these volume 
fractions, so that the fluid-to-solid transition, if present, 
takes place in the dilute regime, hence at a significantly lower density 
than that predicted by SB models.

SB models, moreover, predict reentrant melting. \cite{WLL-99} 
The same phenomenon occurs by
using the MB model. For $\Phi_p=1.6$ and 2
the lattice bcc structure is not stable.
After approximately $10^3$ iterations, the solid is completely melted.
Therefore, we confirm the qualitative 
picture obtained by using SB models. Quantitative differences are 
however observed: as we now discuss, 
the solid phase occurs at lower densities than 
those predicted by SB models.

\subsection{Location of the fluid-solid transition} \label{sec5.2}

In order to determine the location of the transition lines,
we perform isobaric simulations, using the strategy discussed 
in Refs.~\onlinecite{LW-77-78,CW-77,NVM-08,ESVV-13}. 
We prepare a starting configuration in a box of 
size $L_1\times L_2\times L_3$ with periodic boundary conditions, 
such that half of the system is in the 
solid phase while the second half is in the fluid phase. Then, we let the 
system evolve at constant pressure. In most of the simulations we keep
$L_1 = L_2$ and let $L_1$ and $L_3$ evolve independently. However, to identify
possible orthorombic stable crystal structures we also perform 
isobaric simulations in which all box dimensions vary independently.
Details on the simulations and on the analyses performed are reported in 
the supplementary material.

We first analyze the SB model in the CM representation.
We have performed simulations for $\widetilde{P} = \beta P \hat{R}_g^3 = 
8.5, 10, 12.6$, and 17.4, which, according to the equation of state obtained
by performing simulations in the fluid phase (disordered starting
configuration), should correspond to $\Phi_p \approx 0.6, 0.64, 0.7$, and 0.8,
respectively. 
For $\widetilde{P} = 8.5$ the solid part of the system melts, while in 
all other cases the system freezes. The density stays essentially constant in 
the simulation: the packing fractions corresponding to the solid and to the 
metastable fluid phase differ by less than 1\%. These results allow us to 
conclude that 
the low-density fluid-solid transition occurs at (we report the error on the 
result in parentheses)
\begin{equation}
\widetilde{P}_{fs} = 9.2(8).
\end{equation}
The corresponding fluid-solid coexistence density 
interval $[\Phi_{p,f},\Phi_{p,s}]$ satisfies
\begin{equation}
   0.60 \lesssim \Phi_{p,f} < \Phi_{p,s} \lesssim 0.64.
\end{equation}
At the transition, the solid phase corresponds to a bcc lattice. However, as 
the pressure is increased we identify other (meta)stable lattice structures.
For $\widetilde{P} = 12.6$, corresponding to $\Phi_p \approx 0.7$, beside
the bcc lattice, we also identify two different (meta)stable monoclinic 
body-centered structures. If we write the vectors identifying the unit cell as 
\begin{eqnarray}
{\bm v}_1 &=& (a,0,0), \nonumber \\
{\bm v}_2 &=& (b\tan\phi,b,0), \nonumber \\
{\bm v}_3 &=& (0,0,c), 
\end{eqnarray}
the two stable structures have $b/a \approx 0.97$, $c/a \approx 1.59$,
$|\phi| \approx 14^\circ$, and $b/a \approx 0.93$, $c/a \approx 1.66$, 
$|\phi| \approx 22^\circ$. We have not determined which is the most stable 
structure at this value of the pressure. 
We have however indications that asymmetric lattice structures 
become the stable ones as $\widetilde{P}$ increases. For $\widetilde{P} \approx
24$ (corresponding to $\Phi_p \approx 0.9$) we find that the bcc lattice 
structure is unstable. Indeed, if we start the simulation 
from a system which contains a bcc
solid, we end up with a final (body)-centered tetragonal structure,
corresponding to $b/a\approx 1$, $c/a\approx 1.4$, $\phi\approx 0$.
This 
tetragonal structure is stable with respect to orthorombic deformations, 
at variance with what happens 
for $\widetilde{P} = 12.6$. At the latter value of the pressure, 
tetragonal structures are 
unstable.  Therefore, we predict an additional solid-solid structural 
transition at $\widetilde{P}_{ss}$ with 
$12.6 < \widetilde{P}_{ss}\lesssim 24$. 

We have also estimated the location of the solid-fluid transition that 
separates the solid from the high-density fluid phase. Isobaric 
simulations at $\widetilde{P} = 70$, 100, 110, and 120, allow us to 
estimate a transition for $\widetilde{P}_{sf} = 110 (10)$. 
The corresponding coexistence interval $[\Phi_{ps},\Phi_{pf}]$ satisfies 
$1.27\lesssim \Phi_{ps} < \Phi_{pf} \lesssim 1.66$. 

The same analysis has been performed for the MB model, see the supplementary
material for details. We observe freezing for 
$\widetilde{P} = 10.9$, 6.6, and 5.0. The packing fraction of the corresponding 
solid phase is $\Phi_p \approx 0.60, 0.50, 0.46$, respectively. On the other
hand, for $\widetilde{P} = 3.38$ the system melts: the density of the 
corresponding fluid is $\Phi_p \approx 0.40$. Therefore, in the MB case, 
the fluid-(bcc) solid transition occurs for 
\begin{equation}
\widetilde{P}_{fs} = 4.2(8),
\end{equation}
and the fluid-solid coexistence density interval satisfies
$
   0.40 \lesssim \Phi_{p,f} < \Phi_{p,s} \lesssim 0.46;
$
correspondingly we have $Z_{fs} = 41(8)$.
Note the significant difference between the SB and MB results, both for the 
pressure and for the densities at coexistence. 

In the MB case, we have not performed a detailed analysis of other possible
metastable asymmetric lattice structures. 
We have, however, identified 
another, at least metastable, lattice structure for 
$\widetilde{P} = 6.6$. The structure is monoclinic
with $\phi\approx 12$-$13^\circ$, as in the SB case.
Therefore,
we conclude that the existence of these monoclinic 
(meta)stable structures is not due
the coarse-graining procedure, but that it is a property of the
underlying star-polymer system.

Also in the MB case, we have estimated
 the location of the solid-fluid transition that 
separates the solid from the high-density fluid phase. We perform 
isobaric simulations for $\widetilde{P} = 55$, 70, 85. For the two largest 
values of $\widetilde{P}$ the fluid phase is the most stable one, 
while the bcc crystal is apparently stable for $\widetilde{P} = 55$. 
These results 
allow us to locate 
the transition at 
\begin{equation} 
\widetilde{P}_{sf} = 62(8). 
\end{equation}
The corresponding coexistence interval $[\Phi_{ps},\Phi_{pf}]$ satisfies 
$1.13\lesssim \Phi_{ps} < \Phi_{pf} \lesssim 1.26$.  
Note again that 
the MB results for both the pressure and the coexisting densities
are significantly different from those obtained by using the SB model.

\section{Polymer size} \label{sec6}

The residual flexibility of the MB model 
allows us to investigate the density dependence of the structural 
intramolecular properties of a single star polymer as the
functionality $f$ grows. 

\begin{figure}[!t]
\begin{center}
\centering
\includegraphics[angle=-90,scale=0.35]{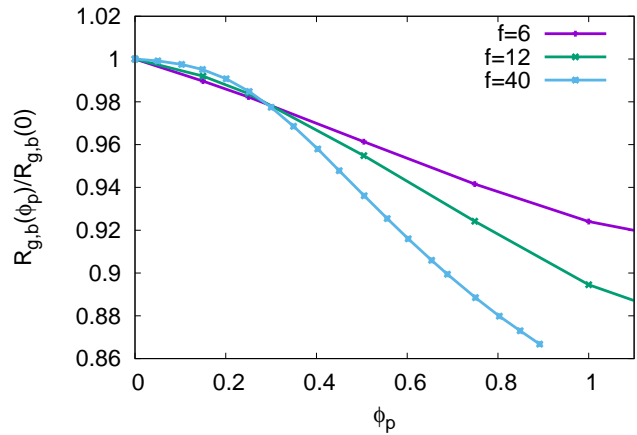}
\caption{Adimensional combination ${R}_{g,b}/\hat{R}_{g,b}$ 
in the multiblob (MB) model as a function of $\Phi_p$, for $f=6$, 12, and 40.
The results for $f=40$ are obtained in the fluid phase, which becomes 
metastable for $\Phi_p\gtrsim 0.4$.
}
\label{Rgdensity}
\end{center}
\end{figure}

The size reduction of the molecules as the density increases can be 
quantified by 
means of the average blob radius of gyration $R_{g,b}(\Phi_p)$ 
defined in Eq.~\eqref{radiusgyrMB}. This quantity differs from the radius of 
gyration of the microscopic model. Nonetheless, 
it also provides a measure of the 
size of the polymer and, therefore, an indication of how stiff the 
polymer is for different values of $f$.
In Fig.~\ref{Rgdensity} we report the adimensional 
universal combination ${R}_{g,b}(\Phi_p)/\hat{R}_{g,b}$ 
($\hat{R}_{g,b}$ is the zero-density quantity) for the 
MB model with $f=6$, 12, and 40. 
The results have been obtained in canonical 
simulations starting from a disordered configuration. Therefore, they always 
refer to polymers in the fluid phase. For $f=40$ this phase is unstable for 
$\Phi_p\gtrsim 0.4$. For $f = 40$ we have also computed this ratio in the 
solid bcc phase, without observing any significant difference. 
For $\Phi_p = 0.5$ we find
$(R_{g,b,\rm fluid}/R_{g,b,\rm solid} - 1) \lesssim 0.1\%$. 

The results for $R_{g,b}$ show the presence of two distinct regimes.
For $\Phi_p \lesssim 0.3$, we have the {\em stiff} regime. Molecules are 
stiff  and they become more rigid as $f$ increases. For these values of 
the density, star polymers can be considered as compact hard objects, 
whose size does not vary with density. This is the regime in which we expect
star polymers to be 
well represented by SB models, which should become more 
accurate as $f$ increases. However, when the packing fraction
exceeds 0.3, one enters a new regime, that we may call 
{\em compressed} regime. This regime is absent for linear polymers: 
as the density increases, linear polymers simply begin to overlap. 
Star polymers instead are very compact objects and therefore it is 
very difficult for them to overlap, especially for large functionalities.
Therefore, they start to shrink, reducing the volume effectively
occupied by each molecule and the overlap between two different 
polymers. The compression increases as a function of $f$, 
as a consequence of the 
increasing difficulty in overlapping.
In this compressed regime, SB models should not be
accurate, and indeed, it is exactly for these values of $\Phi_p$
that we observe significant differences in the predictions of the 
osmotic factor $Z$ between the SB and the MB model. MB models are somehow 
able to describe this compressed phase. However, this is not enough 
to guarantee that the MB model is accurate, since the strong compression of 
the molecule might also change significantly the size and shape of the 
blobs, which are instead assumed to be essentially independent of the density 
in the 
multiblob picture. Therefore, even if we are in the dilute regime, it
is possible that
our MB model loses its accuracy as $f$ increases. 

\section{Conclusions} \label{sec7}

In this work we study the thermodynamics and phase behavior of star
polymers of different functionalities $f$. We consider polymers with 
$f=6$ and 12 arms---in
this case the behavior resembles that of linear polymers---and stars with 
$f=40$ arms, which show a more complex behavior, with a solid phase for 
intermediate values of the density. We use a CG model that captures the star
topology, with $f$ blobs representing the polymer arms and one blob
corresponding to the star center. We use structurally consistent effective 
pairwise interactions, 
which are defined so that the CG model reproduces a set of 
zero-density distribution functions computed in the microscopic model. 
Since the determination of the target distributions only requires simulations
of two polymers, we are able to consider long chains. Thanks also to the 
optimality of the chosen microscopic model---we use the 
Domb-Joyce model at a particular value of the repulsion parameter such that 
\cite{CMP-06,DPP-16} 
the leading finite-length corrections are negligible---we are able 
to obtain accurate asymptotic estimates of the distribution functions with 
relatively short polymers (approximately 1000 monomers per arm). 
As a consequence, our model provides
an accurate CG description of star polymers in the universal, large
degree-of-polymerization limit. It is important to stress that, as 
we only consider the universal behavior, we implicitly 
assume that $L$ is so large that 
the average monomer density is tiny.
In particular, the size of the core 
should be small compared to $\hat{R}_g$. In the Daoud-Cotton model, 
this requires $L\gg f^{1/2}$: as $f$ increases, larger polymers are 
needed to observe the universal behavior. This is consistent with the 
numerical results of Ref.~\onlinecite{RP-13}, that observed an 
increase of the finite-size corrections with $f$.  

The multiblob (MB) model is used to study the thermodynamics of a solution
of star polymers in the dilute and semidilute regime, for $\Phi_p\lesssim 2$.
We find that the MB estimates of the pressure coincide with those 
obtained by using single-blob (SB) models for volume fractions
$\Phi_p\lesssim \Phi_{p,SB}$, where $\Phi_{p,SB} \approx 0.5$ for $f=6,12$ 
and $\Phi_{p,SB} \approx 0.3$ for $f=40$. Thus, the density interval in which 
SB models are quantitatively accurate decreases as $f$ increases: At finite
density, the SB model becomes less accurate as $f$ increases. This is at 
variance with what happens in the limit of zero density.
For instance, the SB estimates of the virial combination $A_3$
become more accurate as $f$ increases.\cite{MP-13,RP-13}

The MB model allows us to study the density dependence of the polymer size.
In the very dilute limit polymers are stiff and 
become more rigid as $f$ increases. In this regime, for large values of
$f$ polymers are well represented by hard spheres, i.e., by rigid molecules 
whose size does not depend on density. This is in agreement with the results 
of Ref.~\onlinecite{RP-13} for the ratio $A_3/A_2^2$ that was found consistent
with the hard-sphere value 5/8 for $f\to\infty$. The density range in which 
this occurs is, however, quite small and moreover it decreases as $f$
increases. As $\Phi_p$ increases, the soft nature of the polymers becomes 
more evident, and for $\Phi_p\gtrsim 0.3$, we enter a different, 
{\em compressed} regime, in which the ratio $R_{g,b}(\Phi_p)/\hat{R}_g$ 
decreases significantly with increasing $f$: in this regime, 
large-functionality star polymers shrink 
more than linear polymers at the same polymer volume fraction.
The origin of this compressed phase 
can be understood quite easily.  As the density increases, 
the available free space decreases. As 
it is difficult for polymers to overlap, it is more convenient for them
to shrink, an effect that becomes more 
significant as $f$ increases. In this regime 
star polymers can no longer be modelled
as hard spheres. This significant size change explains why SB models are 
not accurate for $\Phi_p \gtrsim 0.3$: they are simply not able to 
take this size reduction into account, as they have no internal structure. 

We also study the phase diagram for $f=40$. We confirm the presence of an
intermediate solid phase, predicted using SB models.\cite{WLL-99} 
However, the fluid-solid transitions occur at lower densities than
those predicted by SB models. In particular, the transition separating 
the solid phase from the low-density fluid phase occurs at $\Phi_p\approx 0.4$,
while the one between the solid and the high-density fluid phase occurs
at $\Phi_p \approx 1.2$. The solid phase is therefore in the 
dilute regime in which star polymers can be viewed as nonoverlapping
compressible spheres. In the semidilute regime in which polymers must
necessarily overlap, the solid phase is unstable and one observes a dense
polymeric fluid.
We have also performed a preliminary investigation of the nature of the 
solid phase. We find several, at least metastable, crystal structures both in
the SB and MB model. In the SB case, we also observe 
a structural transition, separating a low-density 
bcc phase, and a high density solid phase, in which 
the stable structure is apparently a centered tetragonal lattice. 
Further work is needed to completely characterize the nature of the solid
phase. In particular, it would be interesting to determine the 
relative stability of the different crystal structures by a direct 
comparison of their free energies or enthalpies. 
This analysis should be feasible
using thermodynamic integration methods. 

The MB model we have considered here can also be used to determine the 
phase behavior of 
mixtures of star polymers with other soft particles, 
improving the results obtained by using SB representations.
\cite{ALE-02,DLL-02,CL-10,LCSLZWLR-11,LCL-16} It would also
be interesting to improve the MB model by increasing the number of interaction
sites. This, however, is probably a very hard problem, which requires 
more sophisticated inversion methods. Finally, it would be interesting to 
investigate SB compressible models,\cite{VBK-10,DPP-12-compressible}
in which the interaction potentials depend on the size of the polymer.
In principle, these models are able to describe the compression of the 
polymer and therefore they might be reasonably accurate also for 
$\Phi_p \gtrsim 0.3$. Work in this direction is in progress.

\section*{Supplementary material}

In the supplementary material we report numerical data and details on the 
computations we performed. In particular: i) we compare integral-equation 
and Monte Carlo results for SB models; ii) we compare the MB results 
for the pressure with theoretical predictions;
iii) we define the Domb-Joyce model and its asymptotic behavior;
iv) we give some algorithmic details on the simulations of the
multiblob model;
v) we provide a detailed description 
of our results for the determination of the transitions separating the 
fluid phases from the intermediate solid phase for star polymer
systems with 40 arms. 

\section*{Acknowledgements}

We thank Hsiao-Ping Hsu, Raffaello Potestio, and Tristan Bereau 
for a careful reading of the manuscript and useful comments.

\appendix

\section{Inversion procedure for the intermolecular potentials.}

The determination of the intermolecular potentials becomes 
increasingly difficult as $f$ increases. 
For $f=6$ and 12 the standard IBI method \cite{Schommers-83,MP-02,RPMP-03} 
or a small variant thereof works reasonably.
For $f=40$ instead we have not able to use the IBI method and we have 
used an {\em ad hoc} procedure, close
in spirit to the approach of Ref.~\onlinecite{AB-01}. 
Below we give a few details of the approach.

\subsection{Low-functionality stars: $f=6$ and $f = 12$}

For $f=6$ the IBI method \cite{Schommers-83,MP-02,RPMP-03} 
used for the intramolecular interactions still works reasonably,
provided one uses an appropriate mixing parameter $a$ (we use 
$a \approx 0.1$-0.2) as in 
Eq.~(\ref{IBI-eq}).  We first perform a set of IBI iterations in which we fix 
$\widetilde{V}_{ca}(b) = \widetilde{V}_{cc}(b) = \widetilde{V}_{aa}(b)$ 
and determine $\beta \widetilde{V}_{aa}(b)$ by requiring the CG model to reproduce 
$\beta W_{aa}^{FM}(b)$. Once this is done, we perform a second set of 
IBI iterations in which we vary independently
$\widetilde{V}_{aa}(b)$ and $\widetilde{V}_{ca}(b)$, 
setting at each step $\widetilde{V}_{cc}(b) = \widetilde{V}_{ca}(b)$. 
We stop when the CG model reproduces the full-monomer
arm-arm and center-arm potentials of mean force. Finally, we perform IBI 
iterations, varying independently the three potentials, until we 
reproduce all potentials of mean force. In total, we perform 300 iterations.

For $f=12$ the IBI method \cite{Schommers-83,MP-02,RPMP-03} works, although
not straightforwardly. 
The main reason for the failure of the standard IBI method 
is the presence of long tails, which extend up to 
$4\hat{R}_g$ (see the discussion at the end of Sec.~\ref{sec2.2}), 
in the mean-force potentials.
Indeed, if we use Eq.~(\ref{IBI-eq}) with the distributions $P$ 
replaced by $e^{-\beta W}$, in a few iterations also 
the intermolecular potentials $\widetilde{V}$ develop a similar tail, which 
increases with the number of iterations. Such a tail is clearly unphysical.
We expect the potentials to have a range of the order of a few blob radii
of gyration. As $\hat{r}_g \approx 0.5 \hat{R}_g$, we expect the potentials 
to be small beyond, say, $2 \hat{R}_g$. 

To avoid the appearance of these tails, we have modified the IBI procedure,
introducing an appropriate mixing function $a_{ij}(b)$. We therefore perform a 
set of iterative steps in which 
the potentials are updated using 
\begin{equation}
\label{NewIBIinter}
\widetilde{V}_{(n+1),ij}(b) = \widetilde{V}_{n,ij}(b) 
   -a_{ij}(b)[W_{n,ij}^{CG}(b)-W^{FM}_{ij}(b)].
\end{equation}
We choose Gaussian functions
\begin{equation}
a_{ij}(b)=A e^{-b^2/\sigma^2}, 
\end{equation}
with the same amplitudes $A$ and widths $\sigma$ for all potentials.
We use $A = 0.2$ and $\sigma = 1.8$.
In practice, we proceed as follows. 
We start the IBI procedure from the $f=6$ potentials and 
perform several iterations (\ref{NewIBIinter}), 
updating all potentials at each step. Once the
potentials of mean force are approximately reproduced, we perform
some standard IBI iterations with a constant mixing parameter $a$. 
We start with $a = 0.1$ and then we reduce it, taking $a=0.05$ in the last set
of iterations. The total number of IBI iterations is 250.

\subsection{High-functionality stars: $f=40$}

\begin{figure}[t]
\begin{center}
\includegraphics[scale=0.3]{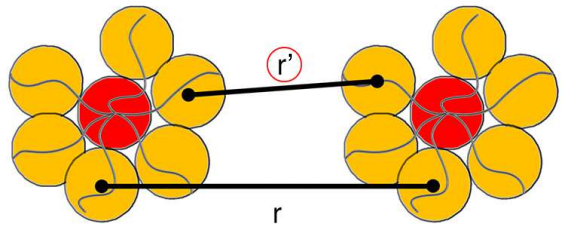}
\caption{A typical configuration for the determination of the arm-arm 
potential of mean force. We fix the distance $r$ between two arbitrary blobs 
and compute the total intermolecular potential energy. The summation 
contains contributions $\widetilde{V}_{aa}(r')$, 
in which $r'$ (the distance between 
different pairs of blobs) varies significantly (significant 
contributions are obtained up to $r'$ of the order of $4\hat{R}_g$).
}
\label{Interprob}
\end{center}
\end{figure}

The determination of the intermolecular potentials  
$\beta \widetilde{V}_{ij}$ for $f=40$ has been more 
difficult than for $f=6,12$, and deserves a separate discussion. 
In this case the IBI method does not work, even in the modified form 
(\ref{NewIBIinter}). To understand the origin of the problems, we 
should first realize that, in the computation of the intermolecular 
potential energy,
there are $f^2$ contributions depending on $\widetilde{V}_{aa}$, $2 f$
contributions depending on $\widetilde{V}_{ca}$, and only one term depending 
on $\widetilde{V}_{cc}$. The presence of this hierarchy implies that 
even a small change of $\widetilde{V}_{aa}$ gives rise to a significant change 
of all $\beta W_{ij}$, so that such potential should be only
slightly changed at each step. The second problem stems from 
an intrinsic property of the IBI method. At each step, the potential
at distance $b$ is corrected by using 
the difference  $W_{(n,ij)}^{CG}(b)-W^{FM}_{ij}(b)$ at the same 
value of $b$, see Eq.~(\ref{NewIBIinter}).
Implicitly, this assumes that, if we only change the 
potential in a tiny region around $b = b_0$, the potentials of mean
force change significantly only in a small region close to $b = b_0$. 
This assumption does not hold in our case.
Indeed, the mean force potentials are computed from 
Eq.~(\ref{PMF-FM}), that is fixing the distance $r$ between 
two blobs and then averaging $e^{-\beta U_{\rm inter}}$ 
over the positions of all
other blobs. The distance among all other blobs varies among all 
possible values in the range of the potentials, see Fig.~\ref{Interprob}. 
This means that 
$W_{aa}(b)$ for $b=b_0$ depends on the value of $\widetilde{V}_{aa}(b)$ 
for any value $b$,
as it gets contributions from $f^2$ pair of blobs whose distance 
is quite different from that of the two blobs we have kept fixed.
The problem is less severe, but still present, for the center-arm 
potential that gives $2 f$ contributions to the calculation of 
$W_{ca}(b)$, while
it is not present for $\widetilde{V}_{cc}(b)$.

As the IBI method cannot be used, we adopt a different strategy, close
in spirit to the approach of Ref.~\onlinecite{AB-01}.
We choose appropriate parametrizations of the potentials and optimize
the coefficients, so as to obtain the best possible agreement between
the potentials of mean force computed in the 
full-monomer and CG models.
We start from a set of Gaussian potentials
$\beta \widetilde{V}_{ij}^{(0)}(b)=A_{ij}\exp[-(b/\sigma_{ij})^2]$,
fixing $A_{ij}$ and $\sigma_{ij}$ to some reasonable values --- we are
guided by the results for $f=6$ and 12 --- 
so that every potential of mean force $\beta W_{ij}(b)$ computed in the 
CG model is not very different from its full-monomer counterpart
in the relevant region, i.e., for the  values of $b$ 
where $\beta W_{ij}(b)\lesssim10$. Then, keeping 
the center-arm and center-center potentials fixed, we perform systematic 
changes of $A_{aa}$ and $\sigma_{aa}$ in order to improve the 
agreement between the CG and FM estimates of $\beta W_{aa}(b)$.
We stop the procedure 
when $\beta W_{aa}^{FM}(b)$ is roughly reproduced in the CG model.
After this optimization, we compute the other potentials of mean force, 
finding that both $\beta W_{ca}^{CG}(b)$ and $\beta W_{cc}^{CG}(b)$
significantly overestimate the corresponding full-monomer quantity.
At this point, we optimize the center-arm potential. We change the 
functional form of the center-arm potential and parametrize it
as $\beta \widetilde{V}_{ca}^{(0)}(b)=A_{ca}\exp[-(b/\sigma_{ca})^2](1-k_{ca}b^2)$.
The introduction of the term proportional to $k_{ca}$ makes it possible 
for the potential to have an attractive tail.
The parameters 
$A_{ca}$, $\sigma_{ca}$, and $k_{ca}$ are then optimized in such a way to 
roughly reproduce $\beta W_{ca}^{FM}(b)$. The arm-arm potential of mean force 
$\beta W_{aa}^{CG}(b)$
changes: it slightly underestimates $\beta W_{aa}^{FM}(b)$ at the end 
of the procedure. In particular, tails appear to be understimated. 
At this point we optimize again the arm-arm potential,
choosing a slightly different parametrization:
$\beta \widetilde{V}_{aa}(b)=A_{aa}\exp[-(b/\sigma_{aa})^t]$. The exponent
$t$ is introduced to obtain a better agreement for $b\gtrsim 2$.
The parameters $t$, $A_{aa}$, and $\sigma_{aa}$ are again optimized.
At the end of the procedure, we obtain $t = 1.86$---the potential 
decays slightly slower than a Gaussian---and the two potentials of mean
force, $\beta W_{aa}(b)$ and $\beta W_{ca}(b)$, are accurately reproduced in 
the relevant region. Only $\beta W_{cc}^{CG}(b)$ differs from the full-monomer
target quantity.  At this point we change $\widetilde{V}_{cc}(b)$
using Eq.~\eqref{NewIBIinter} and setting $a_{cc} = 1$.
Since the contribution of this potential to the total intermolecular 
potential energy is of order 1, 
within our target precision (1\%) 
this change does not modify the arm-arm and
the center-arm mean-force potentials. 

The potentials of mean force for the full-monomer
and CG models with $f=40$ are compared in Fig.~\ref{fig:VMF}. 
As it can be seen, 
the CG model with the obtained potentials reproduces quite accurately
the target distributions.

It is interesting to discuss the accuracy of the results. If we vary 
$\widetilde{V}_{aa}(b)$ by 1\%, i.e., we consider either
$1.01\widetilde{V}_{aa}(b)$ or $0.99\widetilde{V}_{aa}(b)$, keeping
$\widetilde{V}_{ca}(b)$ and $\widetilde{V}_{cc}(b)$ fixed, 
the potentials of mean force $W_{aa}^{CG}(b)$ and $W_{ca}^{CG}(b)$
change at most by 1\%, which is the accuracy with which we reproduce 
the full-monomer potentials of mean force. Analogously, if we change 
$\widetilde{V}_{ca}(b)$ by 1\%, $W_{aa}^{CG}(b)$ and $W_{ca}^{CG}(b)$ 
change by less than 0.6\%. Therefore, we conclude that 
$\widetilde{V}_{aa}(b)$ and $\widetilde{V}_{ca}(b)$ are determined with 
an error of approximately 1\%. As for the potential $\widetilde{V}_{cc}(b)$, 
its value is strictly correlated to that of 
$\widetilde{V}_{aa}(b)$ and $\widetilde{V}_{ca}(b)$. 
Indeed, when we change one of these two potentials by 1\%, we should change 
$\widetilde{V}_{cc}(b)$ by 3-4\% to guarantee that 
$W_{cc}^{FM}(b)$ is reproduced. Therefore, we estimate an error of at most
5\% on $\widetilde{V}_{cc}(b)$.

\clearpage

\section{Supplementary material}

\subsection{Summary}
In this supplementary material we report numerical data and details on the
computations we performed. In particular: i) we compare integral-equation
and Monte Carlo results for SB models; ii) we compare the MB results
for the pressure with theoretical predictions;
iii) we define the Domb-Joyce model and its asymptotic behavior;
iv) we give some algorithmic details on the simulations of the 
multiblob model;
v)  we provide a detailed description
of our results for the determination of the transitions separating the
fluid phases from the intermediate solid phase for star polymer
systems with 40 arms.

\subsection{Integral equations}

\begin{table}[t]
\begin{center}
\begin{tabular}{ c | c c c c c c c c}
\hline\hline
& \multicolumn{2}{ c }{$\Phi_p=0.25$} & \multicolumn{2}{ c }{$\Phi_p=0.80$} & \multicolumn{2}{ c }{$\Phi_p=2.0$}\\
\hline
$f$ & RY & MC & RY & MC & RY & MC\\
6 & 2.146 & 2.143 & 5.832 & 5.846 & 14.626 & 14.617\\
12 & 3.601 & 3.621 & 14.573 & 14.707 & 40.450 & 40.630 \\
40 & 11.781 & 11.850 & 87.595 & 86.930 & --- & ---\\
\hline
\end{tabular}
\vspace{0.7cm}
\caption{Compressibility factor $Z(\Phi_p)$ for the SB model 
in the CM representation, for $f=6$, 12 and 40. We compare RY and MC results.
The results for $\Phi_p = 0.80$ and $f= 40$ are obtained in the 
metastable fluid phase.
}
\label{compare-RY-suppl}
\end{center}
\end{table}

In the integral-equation approach one determines the 
pair correlation function
$h(r)$  and the 
direct correlation function $c(r)$, by requiring the validity of 
the Ornstein-Zernike relation \cite{HMD-suppl}
\begin{equation}
\hat{h} (k) = \hat{c}(k) +
   \rho \hat{c}(k) \hat{h}(k)
\end{equation}
(here $\hat{f}(k)$ is the (three-dimensional) Fourier transform
of $f(r)$) and of a closure relation. We use the Rogers-Young closure
defined by \cite{RY-84-suppl}
\begin{equation}
g(r) = e^{-\beta V(r)} \left[
 1 + {\exp[(h(r) - c(r)) f(r)] - 1
      \over f(r) } \right],
\end{equation}
where the function $f(r)$ is given by
\begin{equation}
   f = 1 - e^{-\chi r}.
\end{equation}
The closure relation depends on the parameter $\chi$ which is fixed by
requiring thermodynamic consistency for the pressure, i.e., the 
equivalence of the virial and of compressibility route.\cite{HMD-suppl}

This closure has been extensively used in the analysis of single-blob (SB) 
models.
\cite{LLWAJAR-98-suppl,WLL-99-suppl,MP-13-suppl}
To verify its accuracy we have compared its predictions for the 
compressibility factor $Z$ with 
Monte Carlo results for different values of $\Phi_p$. The results 
are reported in Table~\ref{compare-RY-suppl}. In all cases, differences 
are less than 1\%.

\subsection{Comparison of the thermodynamic results with other 
predictions} 

\subsubsection{Renormalization-group predictions}

Here we wish to compare our results for the osmotic factor $Z$ with the 
renormalization-group predictions of Ref.~\onlinecite{CBMF-86-suppl} (they 
are reviewed and compared with experiments in
Refs.~\onlinecite{DRF-90-suppl,MBLFG-93-suppl}). These predictions are obtained 
using the standard dimensional expansion in powers of $\epsilon = 4-d$, 
where $d$ is the space dimension. The result is expressed in terms of 
the adimensional concentration
\begin{equation}
\overline{c} = B_2 {N_p\over V} = {3\over 4\pi} A_2 \Phi_p,
\end{equation}
where $B_2$ is the second virial coefficient and $N_p$ is the number of star polymers
in the box of size $V$. At first order in $\epsilon$, the compressibility 
factor $Z$ is 
\begin{equation}
Z = 1 + \overline{c} \left( 1 - {\epsilon \over 16 \overline{c}^2} 
    I(\overline{c}) \right),
\label{Z_RG-suppl}
\end{equation}
where 
\begin{eqnarray}
I(\overline{c}) &=& \int_0^\infty y^3 dy\left\{
   \vphantom{1\over2}
   \ln[1 + 4 \overline{c} g(y)]  \right.   \nonumber \\
&& \quad \left. - {4 \overline{c} g(y) \over 1 + 4 \overline{c} g(y)} - 
     8 \overline{c}^2 g(y)^2 \right\}, \\
g(y) &=&   {1\over y^2} + {h(y)\over y^4} \left[1 + {1\over 2} (f-1) h(y)
     \right], \\
h(y) &=& e^{-y^2/f} - 1.
\end{eqnarray}
This expression is supposed to be accurate only for small values of $f$, 
say $f\lesssim 6$.
For larger values, the coefficients of the expansion increase significantly,
so that one-loop results are not
predictive.\cite{DRF-90-suppl,MBLFG-93-suppl} 
If $Z_{MB}$ is the multiblob (MB) Monte Carlo result
and $Z_{RG}$ is the expression reported above with $\epsilon = 1$, we obtain 
for $f=6$:
\begin{center}
\begin{tabular}{lll}
$\Phi_p = 0.15$:  & $\qquad Z_{MB} = 1.64$ & $\qquad Z_{RG} = 1.73$, \\
$\Phi_p = 0.25$:  & $\qquad Z_{MB} = 2.18$ & $\qquad Z_{RG} = 2.36$, \\
$\Phi_p = 0.50$:  & $\qquad Z_{MB} = 3.93$ & $\qquad Z_{RG} = 4.16$, \\
$\Phi_p = 0.75$:  & $\qquad Z_{MB} = 5.92$ & $\qquad Z_{RG} = 6.14$, \\
$\Phi_p = 1.00$:  & $\qquad Z_{MB} = 8.13$ & $\qquad Z_{RG} = 8.23$. \\
\end{tabular}
\end{center}
The agreeement is reasonable, with differences that are at most of 6\%. 
We also compared the multiblob results with Eq.~(\ref{Z_RG-suppl}) 
for $f=12$: in this case discrepancies are very large, of the 
order of 50\%, confirming the unreliability of the $\epsilon$-expansion 
for large functionalities.

For larger values of $\Phi_p$ the renormalization-group expression 
(\ref{Z_RG-suppl}) is 
not predictive. Indeed, for large $\overline{c}$ we obtain
\begin{equation}
Z = 1 + \overline{c} \left\{1 + {\epsilon\over 4}  
  \left[ \log (4 \overline{c}) + a(f) \right] \right\},
\end{equation}
($a(f) \approx 2.7$ for $f=6$),
which differs from the expected $Z \sim \overline{c}^{1.311}$. The correct
behavior can be obtained by exponentiating the logarithmic term and 
noting that $(2 - d\nu)/(d\nu-1) \approx \epsilon/4 + O(\epsilon^2)$. 
However, there is no unambiguous way to obtain the prefactor. 

\subsubsection{Star polymers as hard spheres}

It is interesting to verify whether there exists a range of densities
in which star polymers can be modelled as hard spheres with an 
effective radius $R_c$.
We fix $R_c$ by requiring 
$Z$ to be exactly reproduced for $\rho_p\to 0$ at order $\rho_p$, 
where $\rho_p$ is the 
polymer number density. Therefore, we require 
\begin{equation}
Z_{\rm pol} \approx 1 + A_2 \hat{R}_g^3 \rho_p =
    1 + {16\pi\over 3} R_c^3 \rho_p,
\end{equation} 
which implies 
\begin{equation}
{R_c\over \hat{R}_g} = \left({16\pi\over 3 A_2}\right)^{1/3}.
\end{equation}
We then compare the hard-sphere equation of state (we use the 
Carnahan-Starling\cite{HMD-suppl} expression) with the MB results.
For $\Phi_p = 0.15$ we obtain:
\begin{center}
\begin{tabular}{ccc} 
$f= 6$: & $\qquad Z_{HS} = 1.72$ & $\qquad Z_{MB} = 1.64$; \\
$f= 12$:& $\qquad Z_{HS} = 2.53$ & $\qquad Z_{MB} = 2.11$; \\
$f= 40$:& $\qquad Z_{HS} = 6.00$ & $\qquad Z_{MB} = 4.79$. \\
\end{tabular}
\end{center}
Even for such a small density, the hard-sphere equation of state 
is not accurate. Even worse, the accuracy decreases as $f$ increases:
the density interval in which star polymers can be considered as 
hard spheres decreases as $f$ increases.

\subsection{Domb-Joyce model}

The atomistic (full-monomer) results have been obtained by using the 
lattice Domb-Joyce model.\cite{DJ-72-suppl} A linear polymer chain of 
length $L$ is represented by a lattice random walk 
$\{ {\bm r}_1, \ldots, {\bm r}_L\}$ on a cubic lattice with 
$|{\bm r}_{i} - {\bm r}_{i+1}| = 1$. An effective local repulsion is introduced 
by penalizing self-intersections. Chains are indeed averaged with weight
$e^{-w E}$, where $w$ is a free parameter that plays the role of 
inverse temperature and the energy $E$ is the number of 
self-intersections, 
\begin{eqnarray}
E[\{{\bf r}_{i}\}] &=&
\sum_{1\le i < j\le L}
            \delta({\bm r}_{i},{\bm r}_{j}),
\end{eqnarray}
where $\delta({\bm r},{\bm s}) = 1$ if ${\bm r} = {\bm s}$ and
$\delta({\bm r},{\bm s}) = 0$ otherwise.
The athermal self-avoiding walk model
is obtained as the limit $w\to\infty$ of the Domb-Joyce model. 
Self-intersections are forbidden and only walks such that
${\bm r}_{i} \not= {\bm r}_{j}$ are sampled.
The model can be generalized to star polymers.\cite{HNG-04-suppl,RP-13-suppl}
A star polymer
with $f$ arms and degree of polymerization $Lf$ is represented by
$f$ random walks of length $L$ on a cubic lattice that have a common
origin, i.e. by the
set of lattice points $\{{\bf r}_{\alpha,i}\}$,
$\alpha:1,\ldots,f$, $i:1,\ldots,L$, with ${\bf r}_{1,1} = {\bf r}_{2,1} =
\ldots = {\bf r}_{f,1}$ and $|{\bf r}_{\alpha,i} - {\bf r}_{\alpha,i+1}| = 1$.
The corresponding energy is 
\begin{eqnarray}
E[\{{\bf r}_{\alpha,i}\}] &=&
   \sum_{1\le\alpha<\beta\le f}
   \sum_{i,j} \delta({\bf r}_{\alpha,i},{\bf r}_{\beta,j}) \nonumber \\
&& +
   \sum_\alpha \sum_{1\le i < j\le L}
            \delta({\bf r}_{\alpha,i},{\bf r}_{\alpha,j}).
\end{eqnarray}
We have considered here a single chain. The generalization to 
several interacting chains (this is needed for the computation of the 
potentials of mean force) is completely analogous.
For any $w>0$, the Domb-Joyce model describes polymers under good-solvent 
conditions for large enough values of $L$. 
However, the asymptotic limit is observed 
for values of $L$ that significantly depend on $w$. Therefore, $w$ 
represents a crucial parameter that should be optimized to obtain 
asymptotic results with the least computational effort. 

To make the discussion quantitative, consider a generic large-scale
adimensional quantity $A$, which depends on $L$, $f$, and $w$. 
For large $L$, renormalization group predicts 
\begin{equation}
    A(L,f,w) = A^*(f) + a_A(f,w)/L^\Delta + b_A(f,w)/L^{\Delta_2} + \ldots
\end{equation}
where $A^*(f)$ is universal, i.e. model-independent: for any $w > 0$
the limiting value depends only on the functionality $f$.
The exponent $\Delta$ is also universal 
[simulations of linear SAWs give
\cite{Clisby-10-suppl} $\Delta = 0.528(12)$]. 
The amplitudes $a_A(f,w)$ and $b_A(f,w)$ depend
instead on $w$. However, given two different observables
$A$ and $B$, the amplitude ratios $a_A(f_1,w)/a_B(f_2,w)$ are
model independent.\cite{AA-80-suppl} 

In order to obtain the universal leading quantity $A^*(f)$ with the least 
computational effort, it is convenient to choose $w$ so that $a_A(w,f) = 0$. 
For this particular value $w^*$, convergence is faster. Note also that, 
if $a_A(w^*,f_1) = 0$ for a specific quantity $A$ and functionality 
$f_1$, we have $a_B(w^*,f_2) = 0$ 
for any quantity $B$ and any $f_2$, 
because of the universality of the amplitude ratios. 
Therefore, for $w = w^*$ convergence is faster for any observable
and value of $f$ (this was explicitly checked in 
Ref.~\onlinecite{RP-13-suppl}). 
The optimal value $w^*$ has been determined in several papers 
using linear polymers. 
\cite{BN-97-suppl,CMP-06-suppl,DP-16-suppl,Clisby-17-suppl}
The simulations we present here were performed taking $w = 0.5058$, 
which is close to the most recent estimate $w^* = 0.4828(13)$ of 
Ref.~\onlinecite{Clisby-17-suppl}.

The Domb-Joyce model for linear polymers 
can be efficiently simulated using the pivot
algorithm.\cite{MS-88-suppl,Kennedy-02-suppl,CMP-06-suppl} 
For star polymers an efficient algorithm is discussed in 
Ref.~\onlinecite{RP-13-suppl}, which combines pivot moves with a new set of 
moves that speed up the simulation close to the center of the star.

\subsection{Simulations of the coarse-grained multiblob model} \label{sec4}

In all simulations of the coarse-grained model we use a simple Metropolis 
dynamics. In the multiblob case we consider three different types of 
displacement moves: 
\begin{itemize}
\item[i)] displacement of an arm blob;
\item[ii)] displacement of a center blob;
\item[iii)] rigid displacement of a single star.
\end{itemize}
In all cases, the new position is chosen uniformly in a cube centered in the 
old blob (or molecule) position. The linear size $(2\Delta)$ 
of the cube is chosen to guarantee 
an average acceptance rate of approximately 40-50\% and turns to be little 
dependent on the density of the system.
For $f=40$, in the analysis of the freezing transition, we use 
$\Delta = 0.4\hat{R}_g$, $0.05\hat{R}_g$, $0.1\hat{R}_g$ for the 
moves of type i), ii), and iii), respectively.
In the same simulations we also perform random rotations of the star 
around the central blob. Their average acceptance is approximately 40\%,
with a small density dependence. This relatively large acceptance is 
probably due to the fact that the stars do not overlap significantly 
at the densities we have considered.

In the following we report multiblob results for $f=40$ in terms of 
{\em iterations}. In one iteration we attempt
$Nf/2$ moves of type i), $N$ moves of type ii), $N$ moves of type iii),
and $N$ random rotations, where $N$ is the number of molecules in the 
system. In all cases the molecule (and also the arm blob for moves i) is 
randomly chosen at each step. 
In the fluid phases particle diffuse 
quite rapidly. For $f=40$ 
the diffusion coefficient for the centers of mass of the 
stars is $(0.047 \hat{R}_g)^2$ per iteration at $\widetilde{P} = 85$ 
(high-density fluid phase)
and $(0.063 \hat{R}_g)^2$ per iteration at $\widetilde{P} = 3.38$
(low-density fluid phase).  

In the isobaric simulations we also attempt to change the box dimensions
$L_i$ (we consider parallelepipedal boxes of 
size $L_1\times L_2\times L_3$). We perform Metropolis moves proposing
\begin{equation}
\ln L'_i = \ln L_i + \Delta_L (r - 0.5),
\end{equation}
where $r$ is a uniform random number in $[0,1]$ and $\Delta_L$ is chosen
to have an acceptance of 50\%. We perform 5 volume-change trials every 
iteration. 

Single-blob simulations are done in a similar fashion. We perform random 
displacements of the molecules, choosing the linear size $(2\Delta)$ of the 
displacement cube so that 
the average acceptance is approximately 40-50\%. For $\widetilde{P} = 8.5$ 
we take $\Delta = 0.12 \hat{R}_g$, while for $\widetilde{P} = 120$, 
we take $\Delta = 0.02 \hat{R}_g$. Note that here $\Delta$ significantly
depends on density.  

\subsection{Freezing transition}

\subsubsection{Technical details} \label{Freezing-technical-suppl}

We give here some technical details on the analysis of the stability of 
the solid structures.
To understand whether the fluid or the solid phase is the most stable one,
we consider starting configurations such that half of the system is in the
solid phase, while the second half is fluid. Then, we let the system evolve at 
constant pressure. In most of the simulations we consider 
boxes of size $L_1\times L_1 \times L_3$,
updating independently $L_1$ and $L_3$, which should also allow us to observe,
at least in principle, cubic-to-tetragonal transitions. If the stable phase 
is the fluid one, the solid melts. In the opposite case, at the end of the 
simulation all molecules belong to a crystal structure. 

Let us now define more precisely the starting configurations that have
been used in most of the simulations. We consider a box of size $L_1\times
L_1 \times L_3$ and 
a centered tetragonal lattice with primitive lattice vectors
\begin{eqnarray}
{\bm u}_1 &=& \left( {a\over2} ,{a\over 2}, {c\over 2} \right), \nonumber \\
{\bm u}_2 &=& \left( {a\over2} ,-{a\over 2}, {c\over 2} \right), \nonumber \\
{\bm u}_3 &=& \left( {a\over2} ,{a\over 2}, -{c\over 2} \right).
\end{eqnarray}
The bcc lattice is obtained for $c=a$. The solid part of the system 
is a lattice of size $L_1 = n a$, $L_3^{(s)} = m c$, with 
$L_3^{(s)} = L_3/2$. The total number of molecules belonging to the 
solid is therefore $N_{\rm sol} = 2 n^2 m$. To specify fully the structure, 
we must fix the aspect ratio $c/a$ and the lattice spacing $a$. 
In more detail, if ${\bm r}_0^{(\alpha)} =
(x^{(\alpha)}, y^{(\alpha)}, z^{(\alpha)} )$,
$\alpha = 1,\ldots, 4 n^2 m$, 
are the coordinates of the centers of the polymers,
we set at time $t=0$ (starting configuration)
\begin{equation}
  {\bm r}_0^{(\alpha)}(t=0) = (k_1 a, k_2 a, k_3 c), 
\end{equation}
for $\alpha = 1 + k_1 + k_2 n + k_3 n^2$, where $0\le k_1 < n$, $0\le k_2 < n$,
$0\le k_3 < m$, and 
\begin{equation}
  {\bm r}_0^{(\alpha)}(t=0) = (k_1 a, k_2 a, k_3 c) + (a/2,a/2,c/2),
\end{equation}
for $\alpha = 1 + k_1 + k_2 n + k_3 n^2 + N_{\rm sol}/2$, where 
$k_1,k_2,k_3$ vary as before. For the fluid ($N_{\rm sol}+1\le \alpha \le 
2 N_{\rm sol}$) instead, we choose $x^{(\alpha)}$ and $y^{(\alpha)}$ randomly
in $[0,L_1]$ and $z^{(\alpha)}$ in $[L_3/2,L_3]$.
Once the centers of the polymers are fixed, the arms are positioned randomly
around them. The same procedure is applied in the SB case: here 
the vectors ${\bm r}_0^{(\alpha)}$ are the positions of the CG molecules.

Before performing the isobaric simulations, we 
perform two short canonical thermalization runs. 
In the first one, the positions of the
centers are kept fixed, while the arms are randomly displaced (20-50 iterations;
in each iterations we only try moves of type i),
see Sec.~\ref{sec4}). In the second one, we still do
not move the centers of the molecules belonging to the solid; all other blobs 
are instead randomly displaced (20-50 iterations). 
The resulting configuration
is the starting one for the isobaric simulation. Of course, in the SB case
only the second run is performed. 

To verify whether the system freezes, i.e., to verify whether the solid phase
is the stable one, we proceed as follows. First, 
we introduce adimensional coordinates 
\begin{equation}
{\bm s}^{(\alpha)} = \left( {n x^{(\alpha)} \over L_1}, 
       {n y^{(\alpha)} \over L_1}, {2 m y^{(\alpha)} \over L_3} \right),
\end{equation}
($n$ and $m$ refer to the number of lattice cells in the $x$ and $z$ directions,
respectively, see the definition of the starting configuration), so that 
the $x$ and $y$ components of ${\bm s}^{(\alpha)}$ vary between 0 and $n$, 
while the $z$ component varies between 0 and $2m$. 
If the crystal has not melted in the simulation, 
the positions of the polymers with 
$\alpha \le N_{\rm sol}$ should be close to 
a lattice structure.  However, since we use periodic 
boundary conditions, a global translation is possible. 
Therefore, we introduce 
\begin{eqnarray}
{\bm a}_{\alpha} &=& (k_1,k_2,k_3) \quad \alpha = 1 + k_1 + k_2 n + k_3 n^2 
\nonumber \\
{\bm a}_{\alpha} &=& (k_1 + 1/2,k_2+1/2,k_3+1/2)  \\
  & & \hphantom{(k_1,k_2,k_3) \quad }
    \alpha = 1 + k_1 + k_2 n + k_3 n^2 + N_{\rm sol}/2  \nonumber 
\end{eqnarray}
($k_1$, $k_2$, and $k_3$ vary as before)
and then fix a translation vector 
${\bm c}$ by minimizing 
\begin{equation}
\sum_{\alpha} \left({\bm s}^{(\alpha)} - {\bm a}_\alpha - {\bm c}\right)^2,
\end{equation}
where we sum over all molecules in the solid portion of the system which
do not belong to the 
fluid-solid interface, i.e., we do not consider the molecules 
with $k_3 = 0$ or $k_3 = m - 1$. As usual, we take into account the periodic
boundary conditions by using the minimum image convention. 
Once ${\bm c}$ has been determined,
we define the lattice positions
\begin{eqnarray}
 {\bm b}_\alpha &=& {\bm a}_\alpha + {\bm c}  \qquad\qquad \qquad
    1 \le \alpha \le N_{\rm sol}  \\
 {\bm b}_\alpha &=& {\bm a}_{\beta} + {\bm c} + (0,0,m) \qquad
    N_{\rm sol} + 1 \le \alpha \le 2 N_{\rm sol},  \nonumber 
\end{eqnarray}
with $\beta = \alpha - N_{\rm sol}$.
For each of the sites of this lattice, 
we determine the distance from the closest polymer center:
\begin{eqnarray}
 d_\gamma &=& \min_{\alpha} |{\bm s}^{(\alpha)} - {\bm b}_\gamma| .
\end{eqnarray}
Finally, we define $N_m(\epsilon)$, which gives the number of sites for which 
such distance is less than $\epsilon$. We distinguish between the solid 
and fluid phase, defining
\begin{eqnarray}
&& N_{ms}(\epsilon) = {1\over N_p} \sum_{\gamma \le N_{\rm sol} }
     \theta(\epsilon - d_\gamma)  \nonumber \\
&& N_{mf}(\epsilon) = {1\over N_p} \sum_{\gamma > N_{\rm sol} }
     \theta(\epsilon - d_\gamma),
\end{eqnarray}
where $N_p$ is the total numer of molecules.
For a bcc or a centered tetragonal lattice, the nearest-neighbor distance is 
$\sqrt{3}/2 \approx 0.866$, hence $\epsilon$ should be taken at least less
that 0.4 to obtain meaningful results. In the following, we shall report
results using $\epsilon = 0.2$.

\subsubsection{The single-blob model} 

\begin{figure}[!t]
\begin{center}
\centering
\includegraphics[angle=-90,scale=0.33]{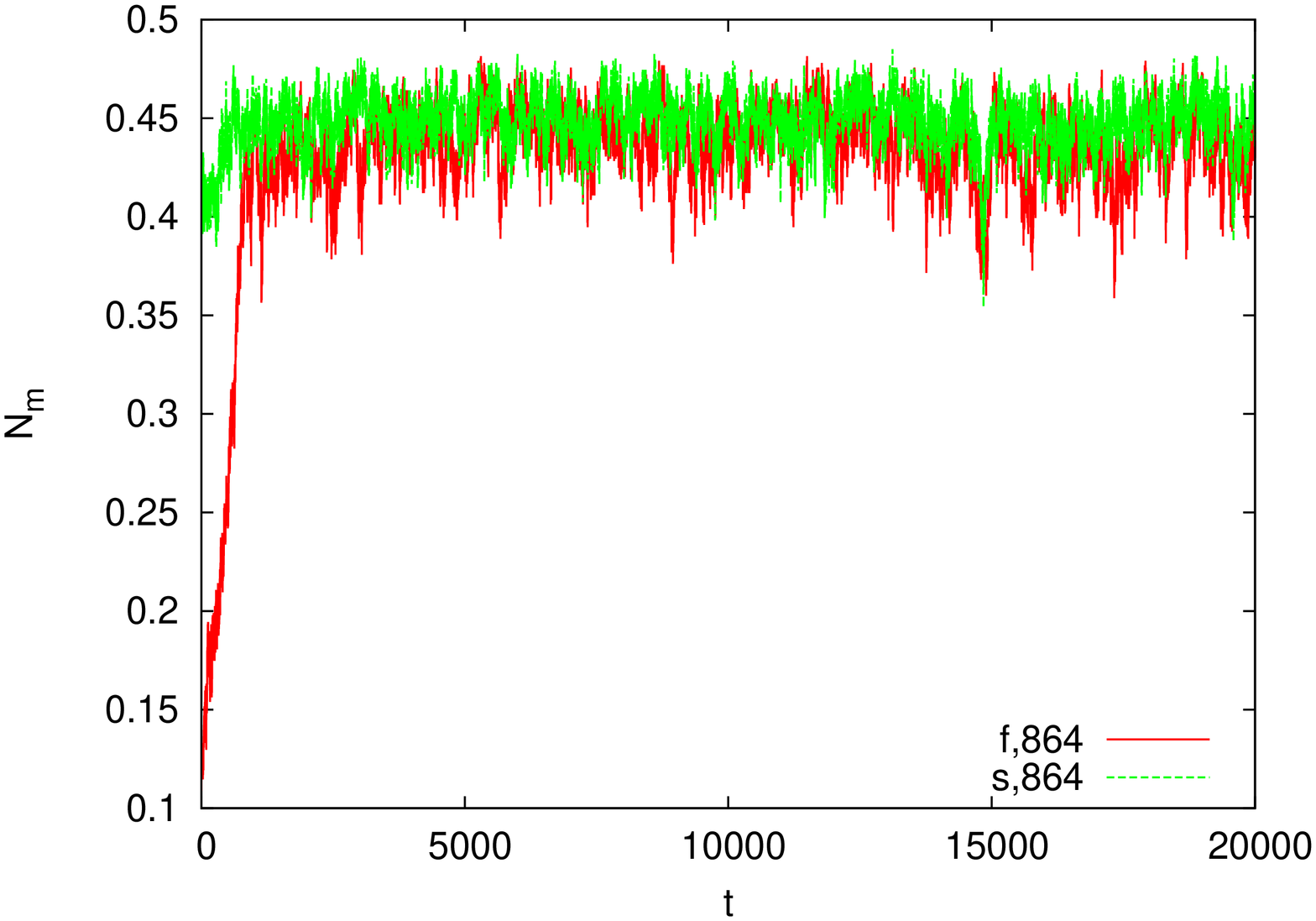} 
\includegraphics[angle=-90,scale=0.33]{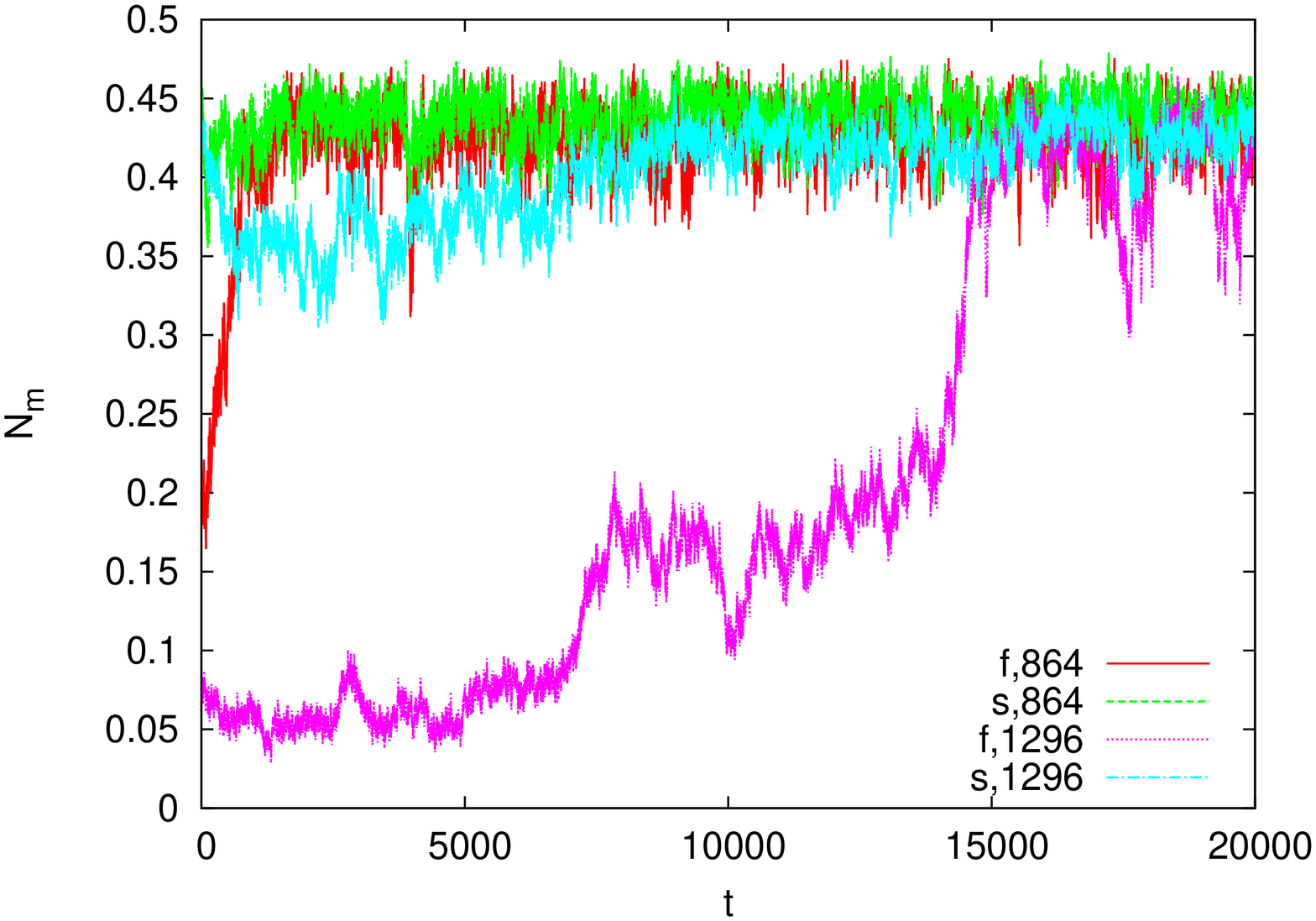} 
\includegraphics[angle=-90,scale=0.33]{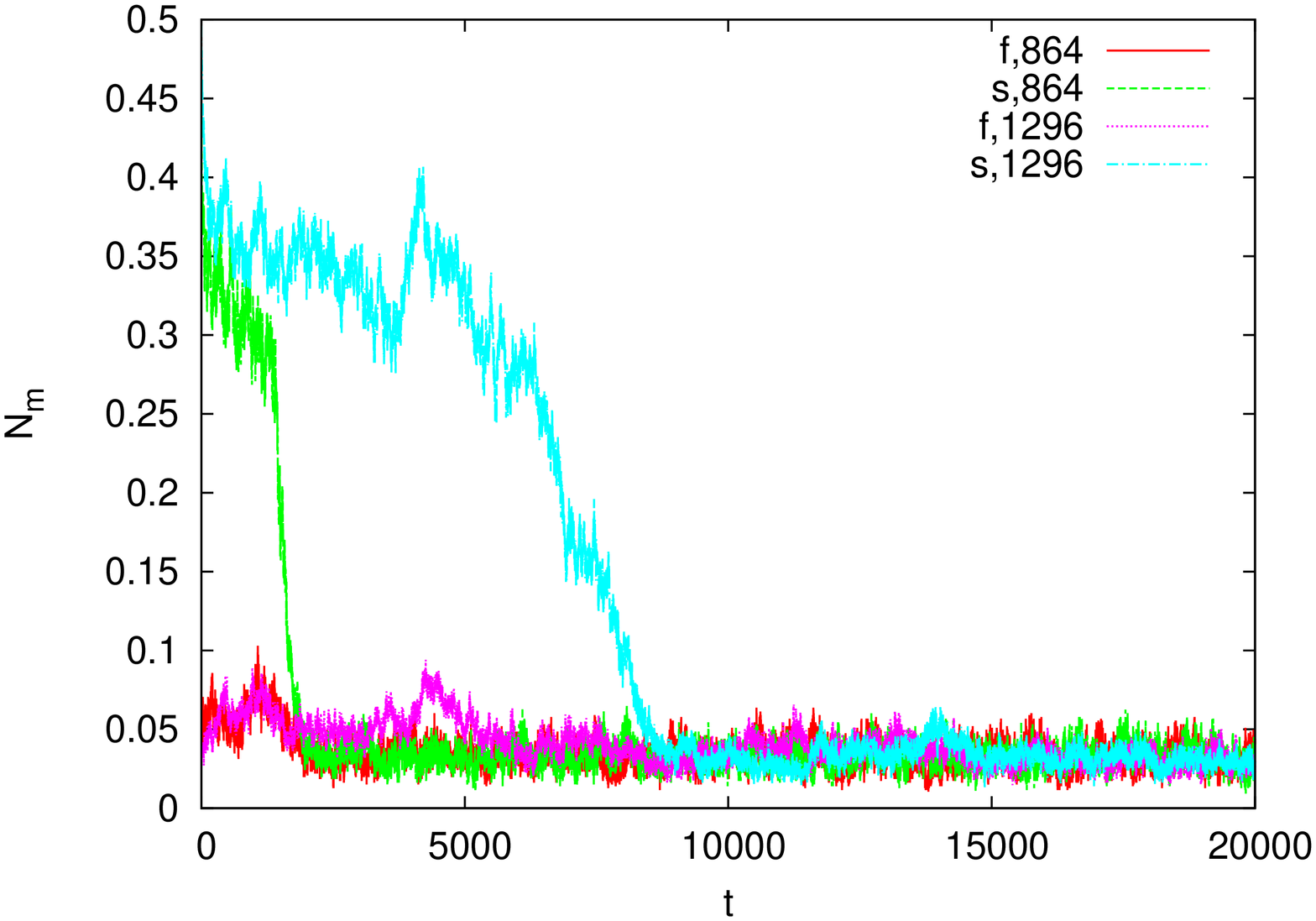} 
\caption{Plot of $N_{ms}(\epsilon)$ and of $N_{mf}(\epsilon)$ as 
a function of the number $t$ of iterations. Top: $\widetilde{P} = 12.6$; 
middle: $\widetilde{P} = 10.0$; bottom: $\widetilde{P} = 8.5$.
Here $\epsilon = 0.2$. Results for $864$ and 1296 molecules (for 
$\widetilde{P} = 10$ and 8.5 only).
For $\widetilde{P} = 12.6$ and 10, the solid is the stable phase;
for $\widetilde{P} = 8.5$ the stable phase is fluid.
}
\label{NSB_suppl}
\end{center}
\end{figure}

To determine the solid-fluid boundary we first perform preliminary 
canonical simulations for several values of $\Phi_p$, starting from a 
bcc crystal structure with 250 molecules. We find that the bcc structure
is apparently stable for $\Phi_p \gtrsim 0.7$: no evidence of melting 
is observed for runs consisting of $10^4$ iterations (in one iteration
we try to randomly move all molecules).
To identify the 
density range  in which the solid phase is stable, we perform isobaric 
runs, starting from mixed (solid-fluid) configurations, as described in 
Sec.~\ref{Freezing-technical-suppl}. 
We consider an $L_1\times L_1 \times L_3$ box and 
we generate three different  configurations with 
$c = a$ (bcc lattice), $n = m = 6$ (the total number of particles is 
864) and $\Phi_p = 0.6$, 0.7, 0.8. Since the equation of state
obtained using integral equations predicts
$\widetilde{P} \approx 8.5$, 12.6, and 17.4 
($\widetilde{P} = \beta P \hat{R}_g^3$) 
for these packing fractions,
the generated configurations are used to start
the runs at $\widetilde{P} = 8.5$, 12.6, and 17.4, respectively.
We find that the solid is the 
stable phase for the two largest values of $\widetilde{P}$, while 
the stable phase is fluid for $\widetilde{P} = 8.5$. This is evident from 
Fig.~\ref{NSB_suppl}, where we report $N_{mf}(\epsilon)$ and 
$N_{ms}(\epsilon)$ for $\widetilde{P} \approx 8.5$ and 12.6. For the smallest
value of the pressure the number of molecules that are close to the 
crystal structure decreases rapidly, while the opposite occurs for the largest
value of $\widetilde{P}$. To improve the estimate of the pressure at the 
transition, we perform a simulation at $\widetilde{P} = 10.0$, again with 
864 particles: the solid structure is stable, see Fig.~\ref{NSB_suppl}. 

We also study the size dependence of the results. For this 
purpose we repeat 
the runs for $\widetilde{P} = 10$ and 8.5 with a larger number of
particles. We start the simulation from a mixed configuration in 
which the solid is a bcc lattice with $n = 6$, $m =9$. The total 
number of particles is 1296.
The conclusions are identical, see Fig.~\ref{NSB_suppl}.

\begin{figure}[!t]
\begin{center}
\centering
\includegraphics[angle=-90,scale=0.33]{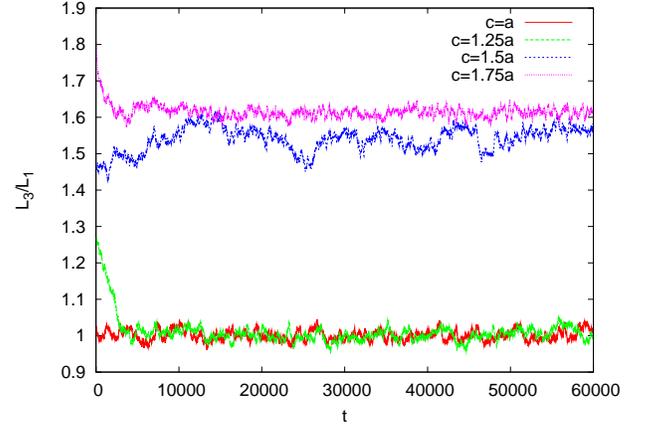}
\caption{Plot of $L_3/L_1$ as 
a function of the number $t$ of iterations (we compute this quantity every 10
iterations) for starting configurations
that differ in the value of $r_{\rm start}=c/a$. 
Results for systems of 1024 molecules ($n=m=8$) and $\widetilde{P} = 12.6$. }
\label{ratio12p6_suppl}
\end{center}
\end{figure}

\begin{figure}[!t]
\begin{center}
\centering
\includegraphics[angle=-90,scale=0.33]{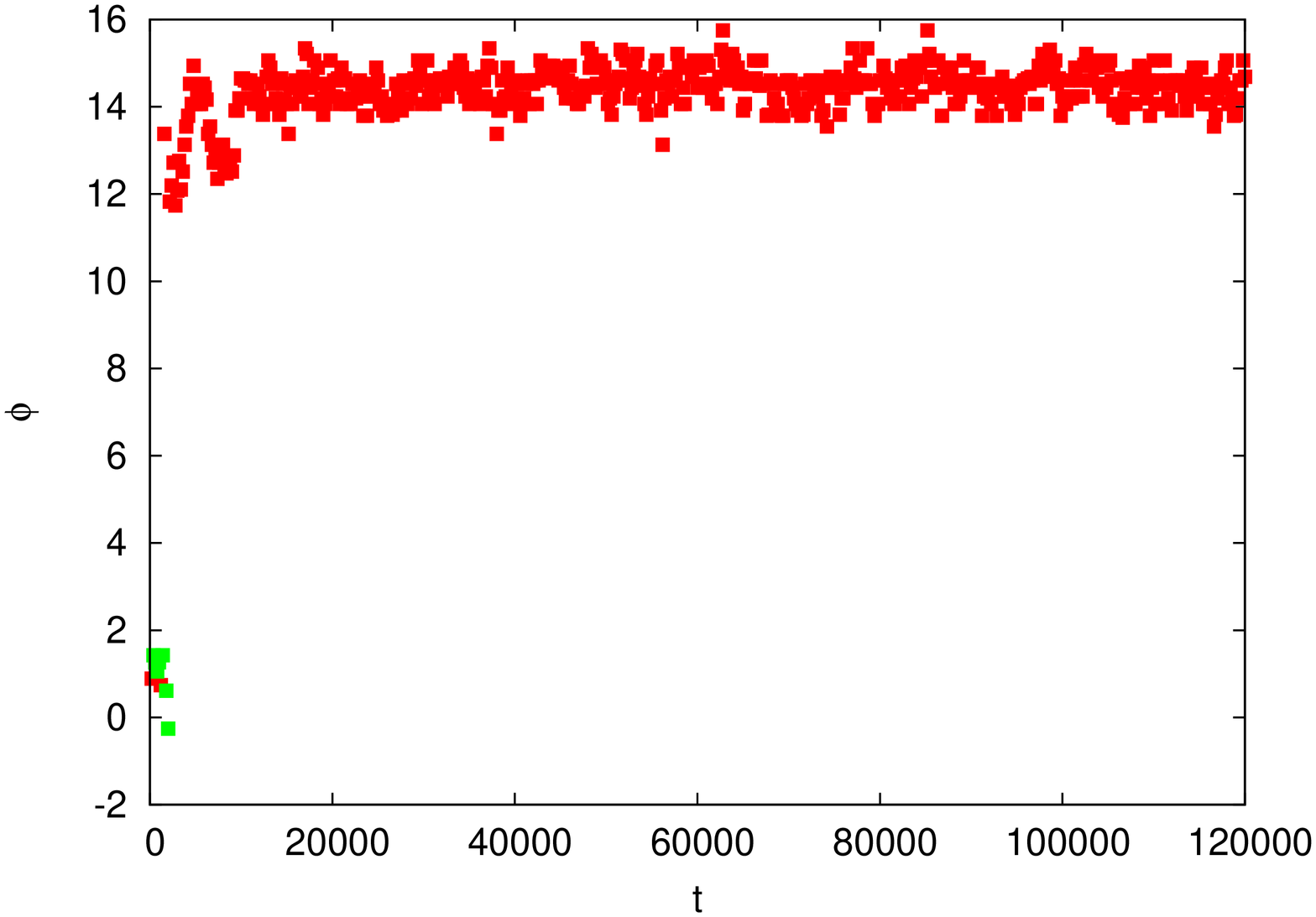}
\includegraphics[angle=-90,scale=0.33]{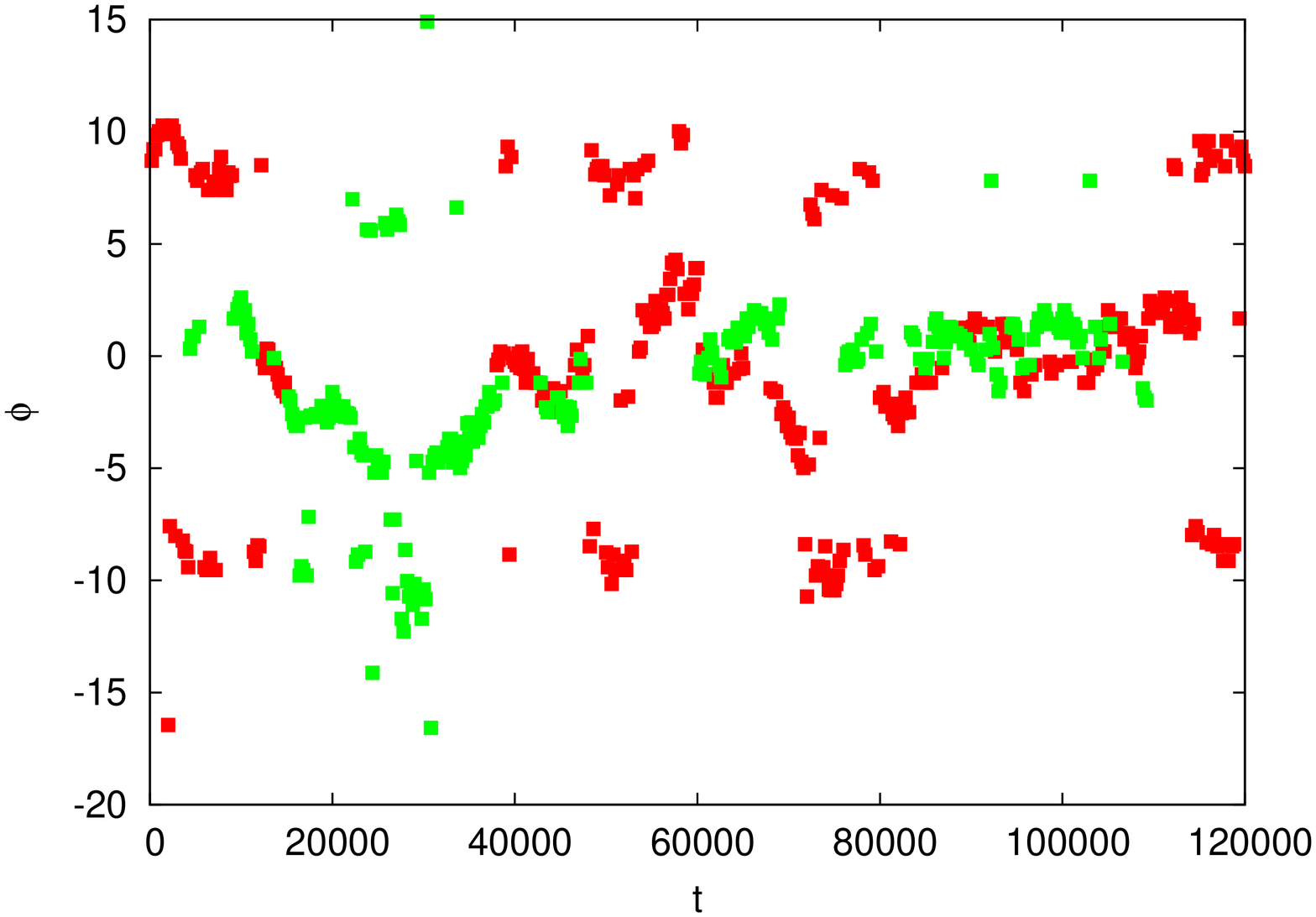}
\caption{Plots of $\phi$ (in degrees) as 
a function of the number $t$ of iterations (we estimate $\phi$ every 200 
iterations): red points correspond to 
systems with unit cell given in Eq.~(\ref{monoclinic-a}), green points
to systems with unit cell given in Eq.~(\ref{monoclinic-b}).
Results for systems of 1024 molecules ($n=m=8$)
and $\widetilde{P} = 12.6$. Top: run with $r_{\rm start} = 1.75$; 
bottom: run with $r_{\rm start} = 1.5$.}
\label{PhiSB_suppl}
\end{center}
\end{figure}

We also study if it is possible to have a tetragonal solid phase. 
For this purpose we perform five runs at $\widetilde{P} = 12.6$,
using different solid starting configurations. The starting configuration is 
a (body) centered tetragonal lattice with $n = m = 8$ (1024 molecules), 
with five different values of $r_{\rm start} = c/a$. 
The size $a$ is fixed by 
requiring $\Phi_p = 0.7$ (which is the equilibrium value of the 
packing fraction for such value of the pressure). Then, we monitor the 
ratio $L_3/L_1$, which can be identified with the ratio $c/a$.
For $r_{\rm start} = 0.75$, the crystal melts and we end up 
with a fluid system. 
For $r_{\rm start} = 1,1.25$ we observe that the system relaxes towards 
a bcc lattice, while for $r_{\rm start} = 1.5,1.75$, we observe the 
appearance of apparently stable lattice structures  with $c/a\approx 1.6$,
see Fig.~\ref{ratio12p6_suppl}.
The analysis of the configurations obtained in the two runs shows the 
presence of different lattice structures. Some configurations correspond to 
a centered tetragonal lattice, but we also observe monoclinic body-centered
structures. The elementary 
unit cell of these lattices can be parametrized by the vectors
\begin{eqnarray}
{\bm v}_1 &=& (a,0,0) \nonumber \\
{\bm v}_2 &=& (a \tan \phi,a,0) \nonumber \\
{\bm v}_3 &=& (0,0,c), 
\label{monoclinic-a}
\end{eqnarray}
or by 
\begin{eqnarray}
{\bm v}_1 &=& (a, a\tan\phi ,0) \nonumber \\
{\bm v}_2 &=& (0,a,0) \nonumber \\
{\bm v}_3 &=& (0,0,c). 
\label{monoclinic-b}
\end{eqnarray}
The tetragonal lattice correspond to $\phi = 0$. The ratio $c/a$ can be estimated 
by considering the ratio $L_3/L_1$ (note that in the simulated 
system the number of unit 
cells is the same in all directions), while $\phi$ is obtained by performing 
a minimization calculation. We minimize the deviations of the molecule positions
from the lattice structure with respect to $\phi$ and a global translation
vector ${\bm c}$. The results are shown in Fig.~\ref{PhiSB_suppl}. 
In the run with $r_{\rm start} = 1.75$ the lattice structure becomes monoclinic
after a few iterations, with $\phi \approx 14^\circ$. For $r_{\rm start} = 1.5$
the molecules oscillate instead between a tetragonal lattice and 
four equivalent monoclinic lattice structures with $|\phi| \approx 10^\circ$.
There are, therefore, at least three different metastable structures 
with $|\phi|\approx 0^\circ$ (tetragonal lattice), $10^\circ$, and $14^\circ$.

\begin{figure}[!t]
\begin{center}
\vspace{-1truecm}
\begin{tabular}{c}
\vspace{-2truecm}
\includegraphics[angle=-90,scale=0.33]{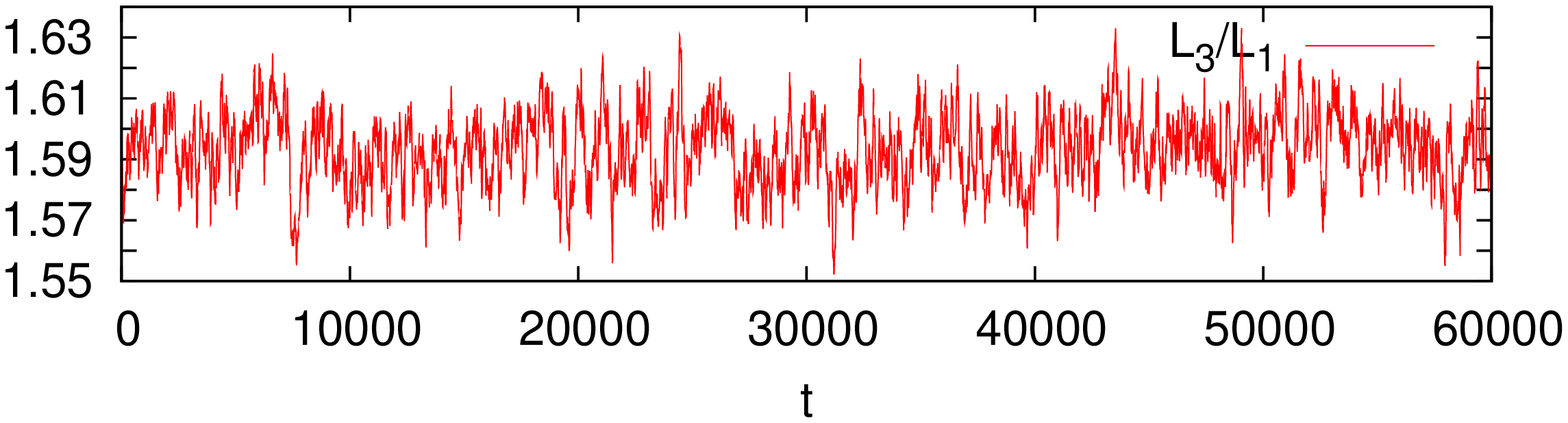} \\[-2truecm]
\includegraphics[angle=-90,scale=0.33]{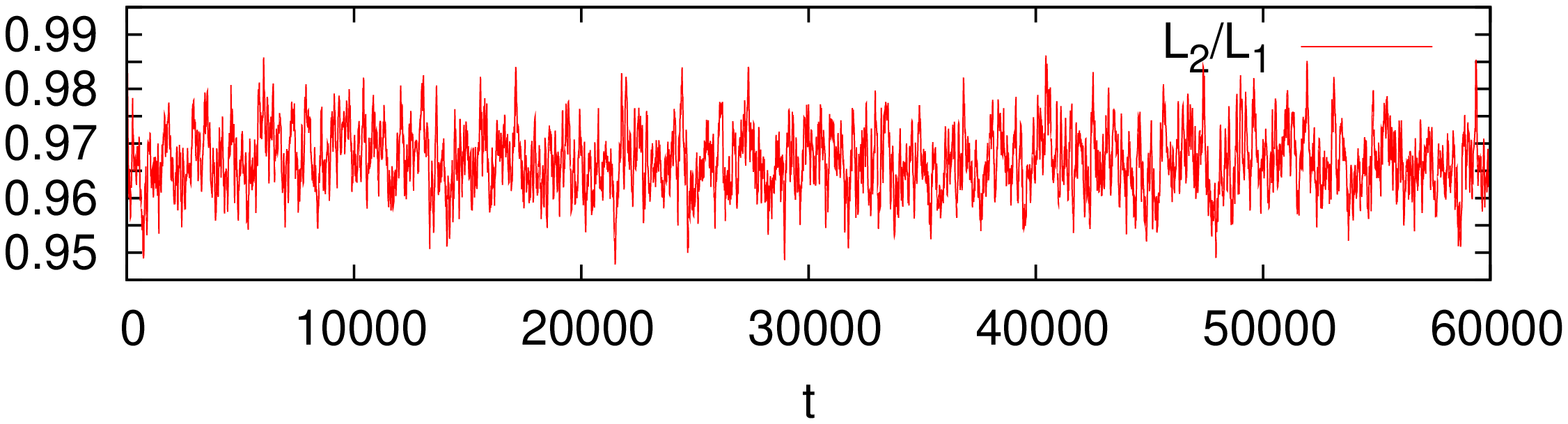} \\[-4truecm]
\includegraphics[angle=-90,scale=0.33]{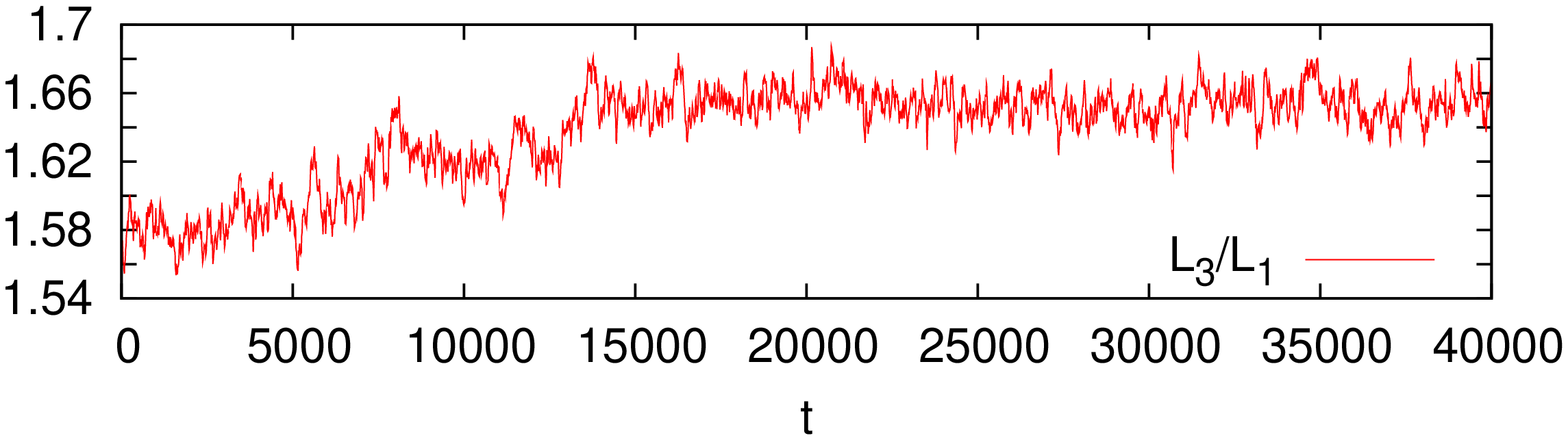} \\[-4truecm]
\includegraphics[angle=-90,scale=0.33]{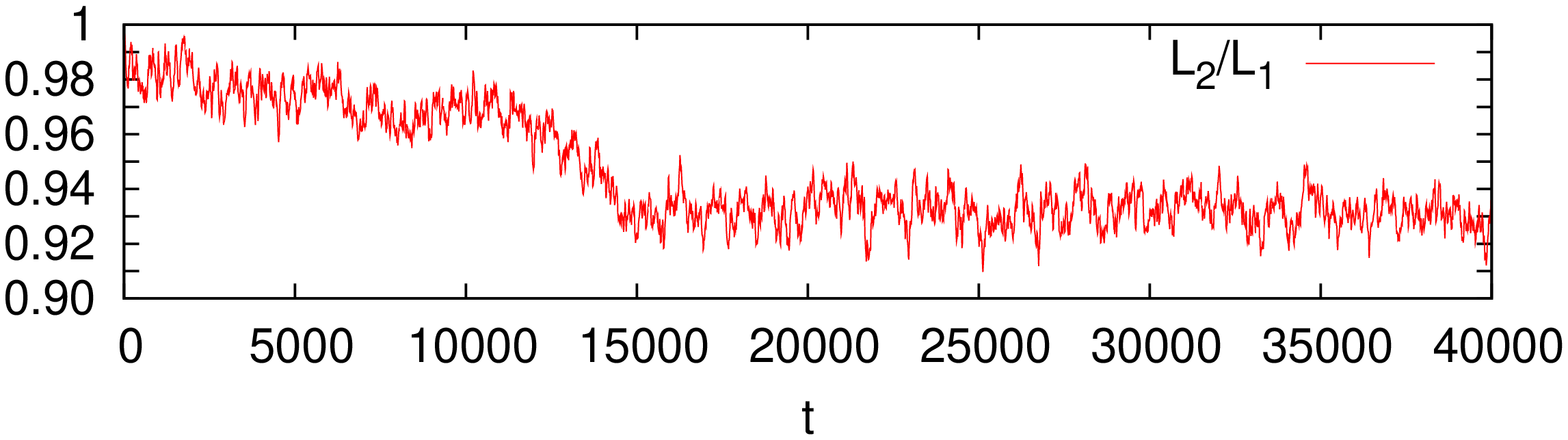} 
\end{tabular}
\vspace{-2truecm}
\caption{Ratios $L_3/L_1$ and $L_2/L_1$ as 
a function of the number $t$ of iterations (we compute the two ratios every 10
iterations). 
Results for runs of 1024 molecules ($n=m=8$)
and $\widetilde{P} = 12.6$. The starting configuration was obtained in runs
with $r_{\rm start} = 1.75$ (upper panels), and 
$r_{\rm start} = 1.5$ (lower panels). }
\label{bco_SB_suppl}
\end{center}
\end{figure}

\begin{figure}[!t]
\begin{center}
\centering
\includegraphics[angle=-90,scale=0.33]{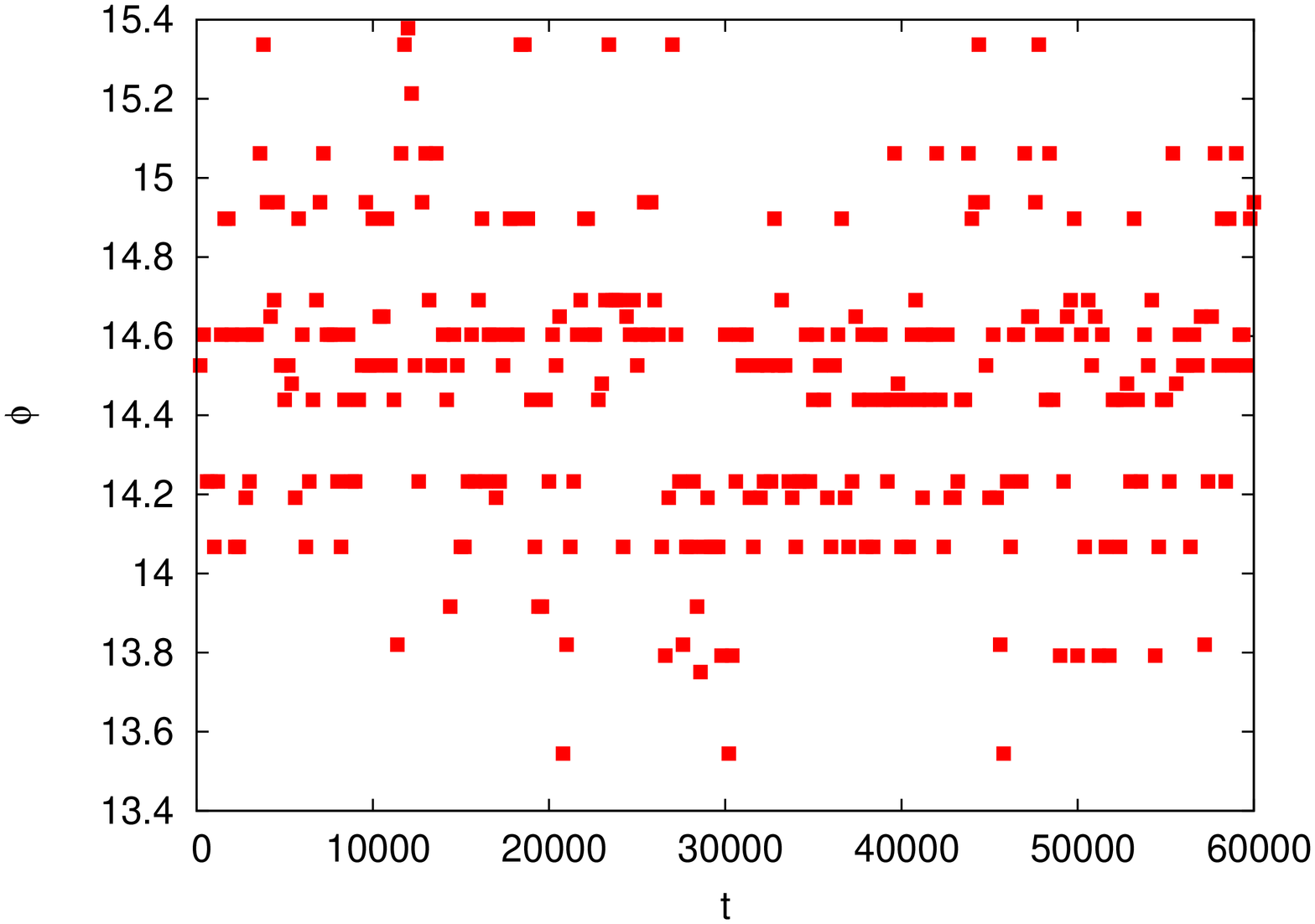}
\includegraphics[angle=-90,scale=0.33]{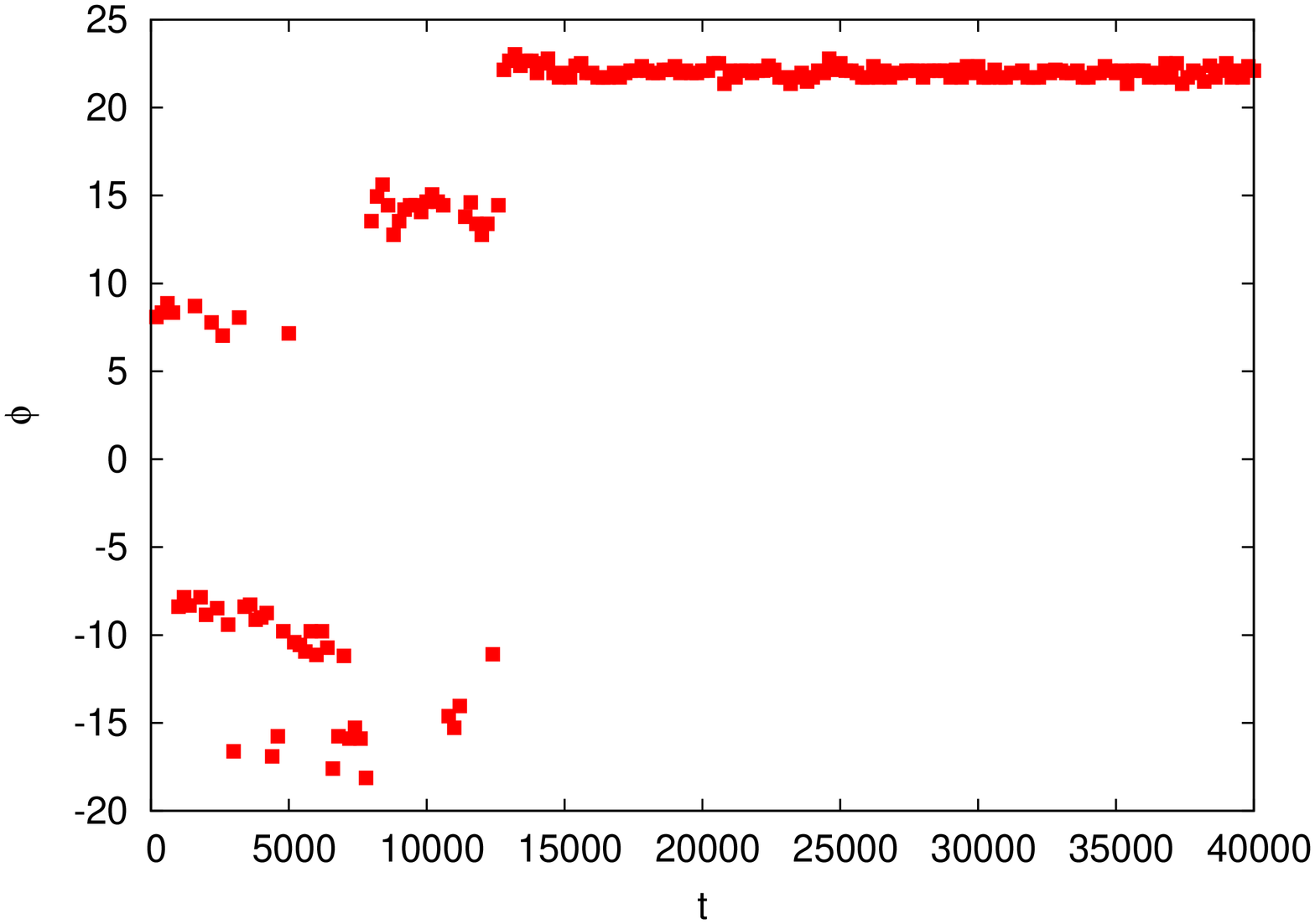}
\caption{Plots of $\phi$ as 
a function of the number $t$ of iterations (we estimate $\phi$ every 200 
iterations).  Results for systems of 1024 molecules
and $\widetilde{P} = 12.6$. Top: the simulation starts from a
configuration obtained in the run with $r_{\rm start} = 1.75$; 
bottom: the starting configuration was obtained in the 
run with $r_{\rm start} = 1.5$.}
\label{PhiSB_iso3_suppl}
\end{center}
\end{figure}

Since the new lattice structures break the symmetry between the $x$,$y$ 
directions and the
$z$ direction, it is natural to expect them to be unstable when we let the 
system change its size in the three directions independently. We thus perform
isobaric runs for systems of size $L_1\times L_2 \times L_3$, updating 
independently $L_1$, $L_2$, and $L_3$. We start the simulations from the final
configurations obtained in the two runs with $r_{\rm start} = 1.5$ and 1.75. 
In both cases, see Fig.~\ref{bco_SB_suppl}, after some iterations
the symmetry under the exchange of $L_1$ and $L_2$ is broken and we obtain 
a body-centered lattice with unit cell given by
\begin{eqnarray}
{\bm v}_1 &=& (a,0,0) \nonumber \\
{\bm v}_2 &=& (b \tan \phi,b,0) \nonumber \\
{\bm v}_3 &=& (0,0,c). 
\end{eqnarray}
The angle $\phi$ is again obtained by performing a minimization procedure.
The results are reported in Fig.~\ref{PhiSB_iso3_suppl}. For the run with 
$r_{\rm start} = 1.75$ we find $b/a \approx 0.97$, $c/a \approx 1.59$, and 
$\phi \approx 14^\circ$, while for the $r_{\rm start} = 1.5$ we obtain 
$b/a\approx 0.93$, $c/a \approx 1.66$, and $\phi\approx 22^\circ$.

\begin{figure}[!t]
\begin{center}
\centering
\includegraphics[angle=-90,scale=0.33]{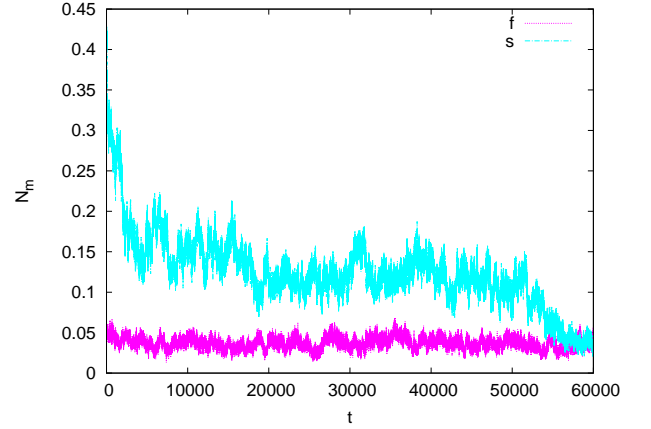} 
\caption{Plot of $N_{ms}(\epsilon)$ and of $N_{mf}(\epsilon)$,
$\epsilon = 0.2$, as 
a function of the number $t$ of iterations. 
Here $\widetilde{P} = 10.0$. The starting configuration is a mixed 
solid-fluid system (1296 molecules) in which the solid is 
a tetragonal structure with $n = 6$, $m = 9$, and $c/a = 1.5$.
}
\label{NSB2_suppl}
\end{center}
\end{figure}

\begin{figure}[!t]
\begin{center}
\vspace{-1truecm}
\begin{tabular}{c}
\vspace{-2truecm}
\includegraphics[angle=-90,scale=0.33]{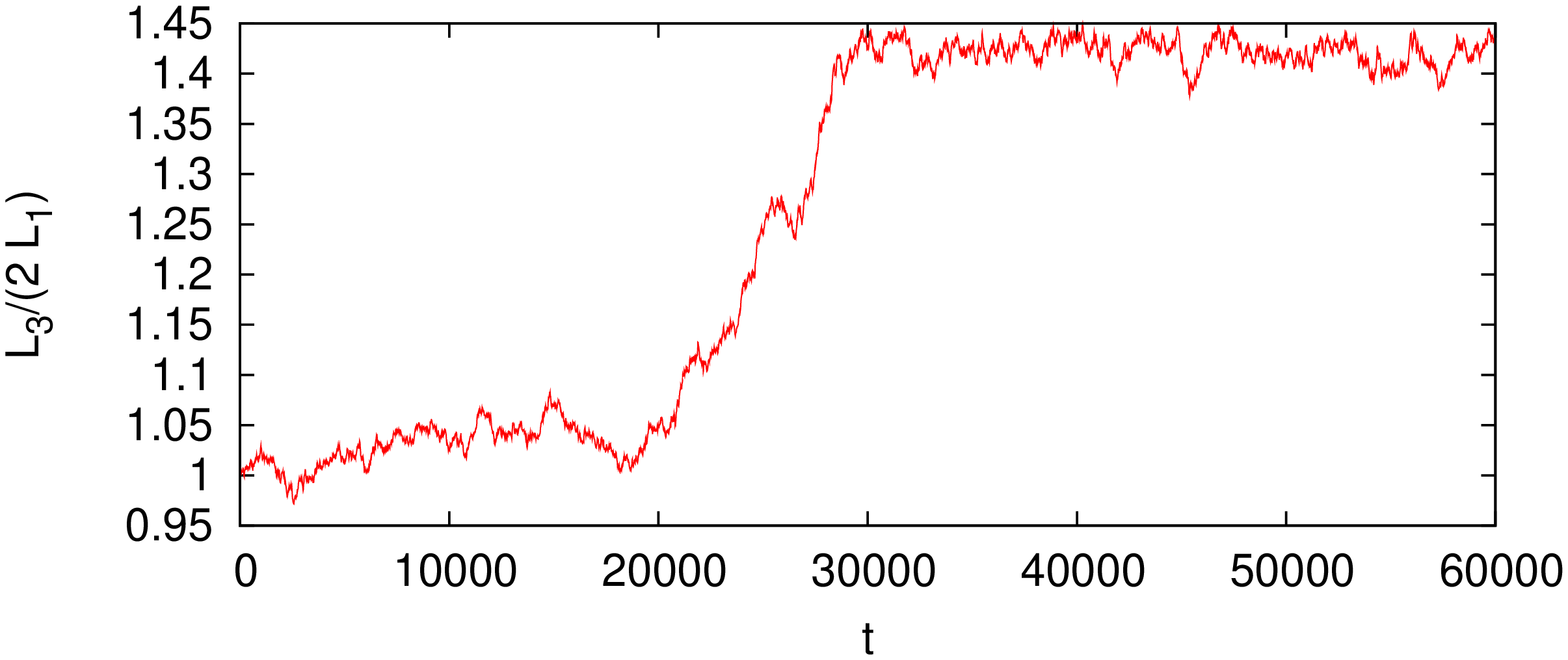} \\[-1.5truecm]
\includegraphics[angle=-90,scale=0.33]{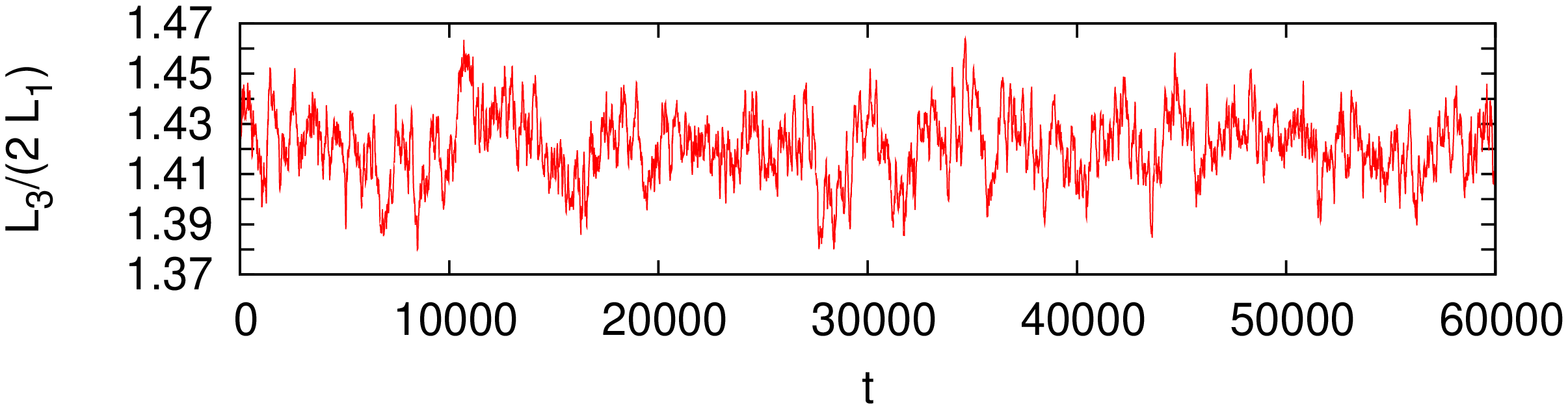} \\[-3.99truecm]
\includegraphics[angle=-90,scale=0.33]{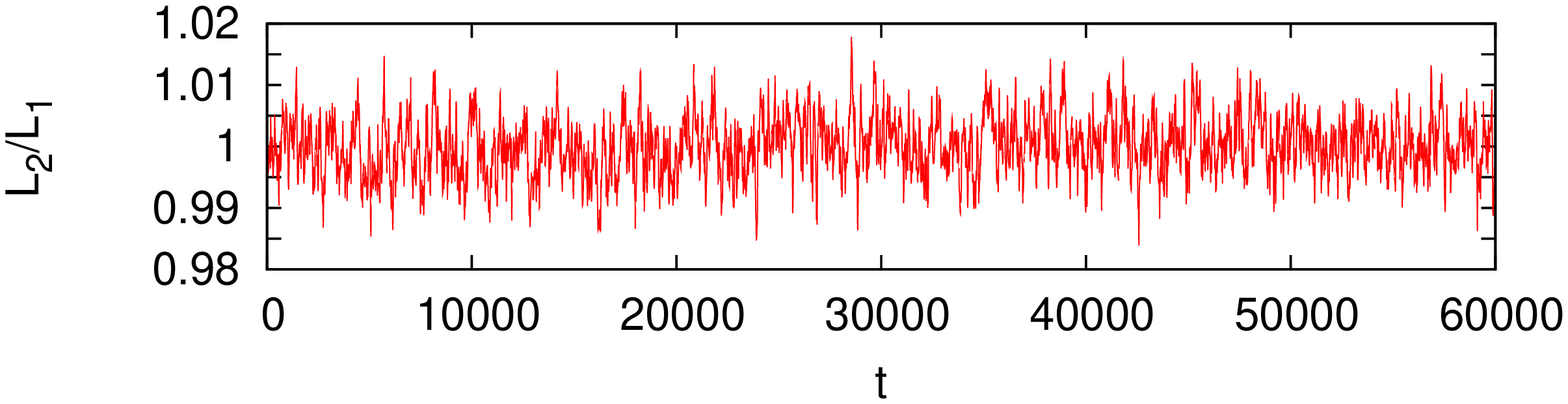} 
\end{tabular}
\vspace{-2truecm}
\caption{Top panel: estimates of the ratio $L_3/(2 L_1)$ as a function
of the number $t$ of iterations 
for the run starting from 
a mixed (fluid-solid) configuration with $n=m=6$ (864 molecules), $c/a = 1$ 
(the solid is a bcc lattice), and $\Phi_p= 0.9$. In the run we fix
$\widetilde{P} = 24$ and keep $L_1 = L_2$. 
Middle and bottom panel: estimates of the ratios $L_3/(2 L_1)$ and 
$L_2/L_1$; the run starts from the final configuration obtained in the 
run considered in the top panel, but now $L_1$, $L_2$, and $L_3$ are 
updated independently. We always set $\widetilde{P} = 24$.
}
\label{bco_SB_P24_suppl}
\end{center}
\end{figure}

\begin{figure}[!t]
\begin{center}
\vspace{-1.5truecm}
\begin{tabular}{c}
\vspace{-1truecm}
\includegraphics[angle=-90,scale=0.33]{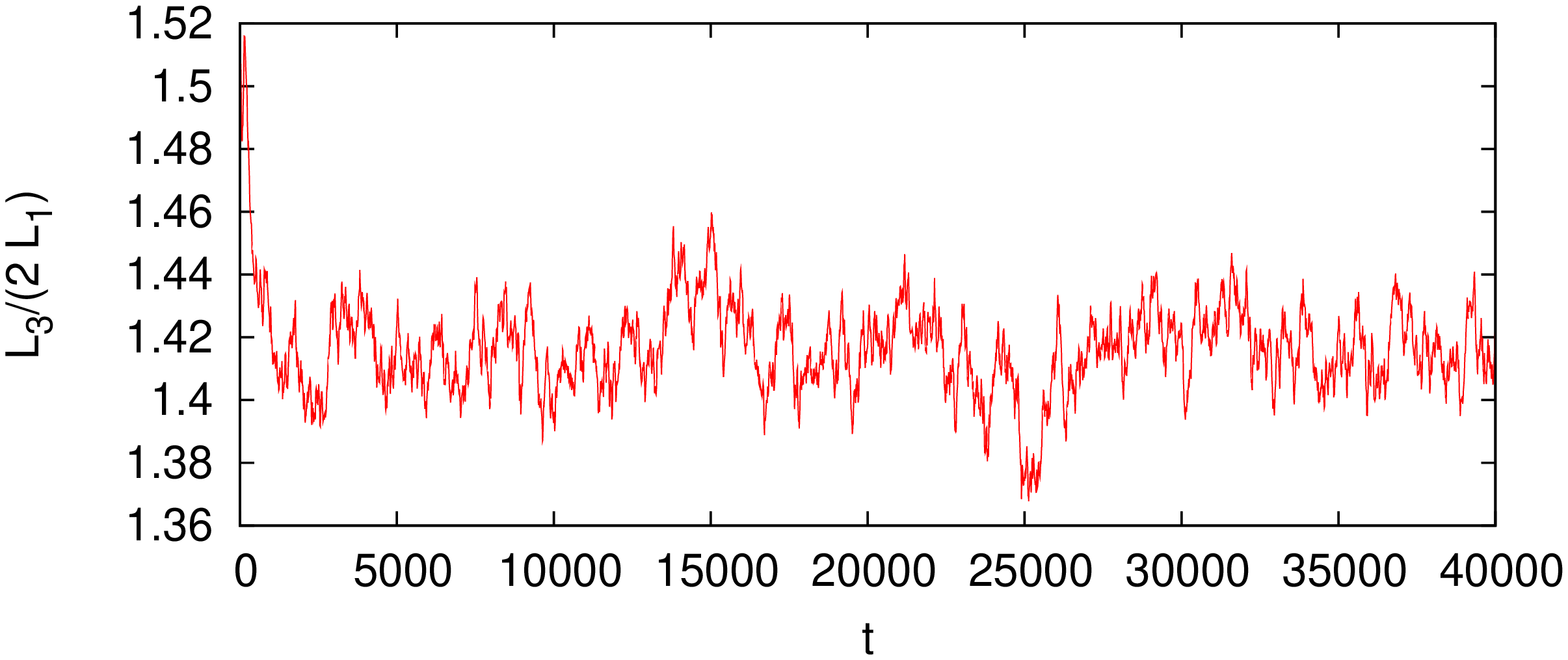} \\[-1.8truecm]
\includegraphics[angle=-90,scale=0.33]{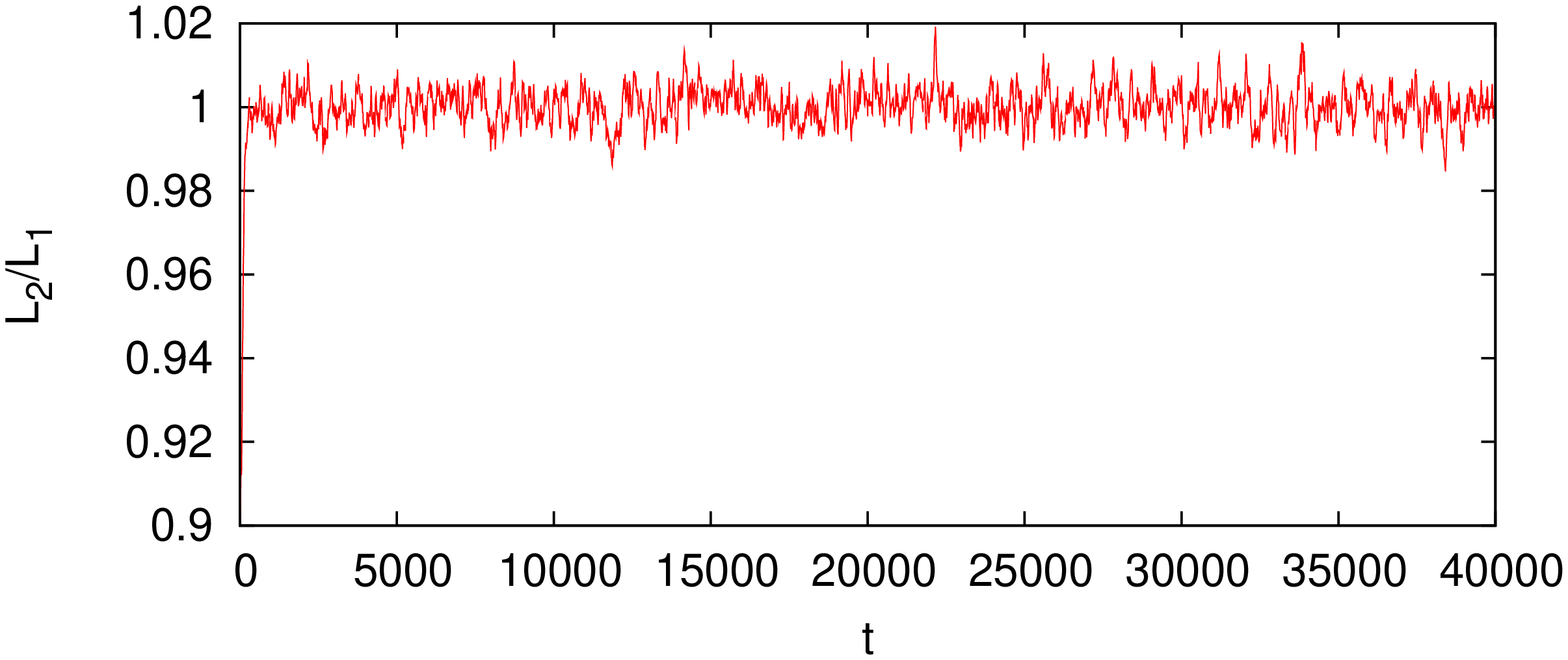} \\[-1.0truecm]
\end{tabular}
\vspace{-0.5truecm}
\caption{Estimates of the ratios $L_3/(2 L_1)$ (top) and 
$L_2/L_1$ (bottom) as a function
of the number $t$ of iterations. 
The simulation starts from a lattice configuration with 1024 molecules 
(8 unit cells in each lattice direction). 
The initial lattice is body-centered orthorombic with unit cell dimensions
$a,b,c$. We fix $L_2/L_1 = b/a = 0.9$, $L_3/L_1 = c/a = 1.5$ and choose $a$ 
so that $\Phi_p = 0.9$.
In the run we set $\widetilde{P} = 24$ and update $L_1$, $L_2$, and 
$L_3$ independently.
}
\label{bco_SB_P24_run2_suppl}
\end{center}
\end{figure}

We also perform simulations at $\widetilde{P} = 10$ using mixed solid-fluid
configurations, taking $c/a = 1.5$ for the starting configuration. In this 
case, we find that the system melts, indicating that the tetragonal phase 
is less stable than the fluid, see Fig.~\ref{NSB2_suppl}. 
This confirms that the stable phase at the 
transition is the bcc lattice structure. However, by increasing the density, 
the relative stability of the different crystal structures can only be 
ascertained by performing a detailed free-energy calculation. 

Simulations at $\widetilde{P} = 24$ (the equilibrium packing fraction is 
$\Phi_p \approx 0.9$) indicate 
that asymmetric crystal structures become the 
stable ones as $\Phi_p$ increases. We have indeed performed a simulation
for such value of $\widetilde{P}$, starting from
a mixed system containing a bcc lattice with $n=m=6$ and 
a disordered configuration (the total number of particles is 864). 
We perform independent updates of $L_3$ and of $L_1$, keeping $L_2 = L_1$. 
At the beginning $L_3/(2 L_1) = c/a$ is 1. In the simulation the 
disordered part of
the system freezes, but at the same time $L_3/(2 L_1)$, which provides 
an estimate of $c/a$, changes, becoming approximately 
1.4 at the end of the simulation,
see Fig.~\ref{bco_SB_P24_suppl}.
We have also analyzed in detail the configurations, looking for lattice 
structures generated by the vectors (\ref{monoclinic-a}) and 
(\ref{monoclinic-b}). We find $\phi \approx 0$, i.e., the stable lattice 
is centered tetragonal. Starting from the final configuration obtained 
in this run, we have performed a second simulation at $\widetilde{P} = 24$, in 
which $L_1$, $L_2$, and $L_3$ are allowed to change independently. 
We find $L_2/L_1 \approx 1$, see the two lower panels in 
Fig.~\ref{bco_SB_P24_suppl}, confirming the (meta)stability of the 
tetragonal structure. As a final check we have performed a new run 
with 1024 particles. We start from a body-centered orthorombic 
lattice with 8 unit cells in each direction and $L_2/L_1 = b/a = 0.9$, 
$L_3/L_1 = c/a = 1.5$ ($a,b,c$ give the size of the unit cell). The ratio 
$L_2/L_1$ as a function of the number of iterations is 
shown in Fig.~\ref{bco_SB_P24_run2_suppl}. After a few iterations, 
we have again $L_2/L_1\approx 1$, indicating the stability of the tetragonal 
structure with respect to orthorombic deformations.

In conclusion, for the SB model, we estimate the pressure $P_{fs}$
at the low-density fluid-solid transition as 
\begin{equation}
     \widetilde{P}_{fs} = 9.2(8).
\end{equation}
If $\Phi_{p,f}$ and $\Phi_{p,s}$ are the boundaries of the fluid and solid
phases, respectively, we obtain 
\begin{equation}
     0.60 \lesssim \Phi_{p,f} < \Phi_{p,s} \lesssim 0.64.
\end{equation}
At the transition the bcc lattice is the stable crystal structure.
The size $a$ of the corresponding unit cell 
is $a = 2.38 \hat{R}_g$, while the distance between the two 
closest lattice points is $a\sqrt{3}/2 \approx 2.06 \hat{R}_g$. 
As $\widetilde{P}$ and $\Phi_p$ increase, the bcc lattice 
becomes unstable. For $\widetilde{P} = 24$ ($\Phi_p = 0.9$) the 
stable lattice structure is apparently centered tetragonal.

\begin{figure}[!t]
\begin{center}
\centering
\includegraphics[angle=-90,scale=0.33]{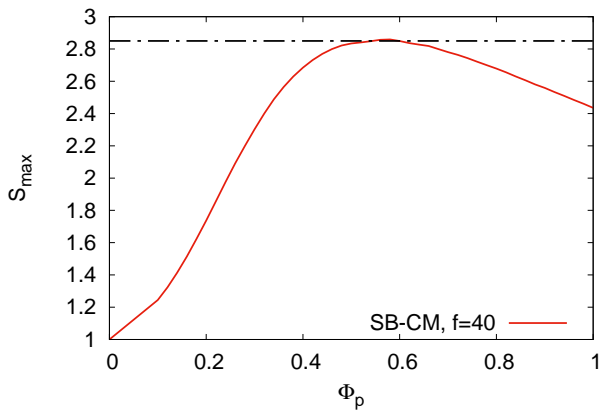}
\caption{Plot of the maximum $S_{\rm max}$ of the structure factor for 
the SB model in the CM representation, as a function of $\Phi_p$.
It has been obtained by using the integral-equation method with 
Rogers-Young closure. The dotted-dashed line at $S_{\rm max}=2.85$ 
corresponds to the Hansen-Verlet criterion for the stability of the 
solid phase.
}
\label{Smax_suppl}
\end{center}
\end{figure}

In Ref.~\onlinecite{MP-13-suppl} we used integral equation methods to estimate 
the location of the fluid-solid transition. In particular, we used 
the Hansen-Verlet criterion:\cite{HV-69-suppl,HS-73-suppl} the phase transition occurs when
the maximum $S_{\rm max}$ of $S(q)$ exceeds 2.85. Unfortunately, the pair
potential used there 
was not accurate, and therefore that result largely overestimates
the correct pressure and density. Here, we have repeated the same 
calculation, using our estimate of the pair potential. The maximum of the 
structure factor $S_{\rm max}$ is reported in Fig.~\ref{Smax_suppl}.
It shows a nonmonotonic behavior with a maximum for $\Phi_p \approx 0.58$, 
where it takes the value 2.86, slightly larger than the Hansen-Verlet value of 
2.85. If we were using the Hansen-Verlet criterion, we would estimate
$\Phi_{p,f} \approx 0.54$, which is not far from the correct estimate 
obtained numerically.

We have also collected some data to estimate the high-density solid-fluid
transition. For this purpose we have performed isobaric simulations with
$\widetilde{P} = 70,$ 100, 110, and 120, starting from mixed
solid-fluid systems. The starting configuration is a centered tetragonal
lattice with $m=n=6$, $c/a = 1.5$, so that the total (solid and fluid) 
number of particles is 864. To verify the stability of 
the tetragonal structure with respect to orthorombic deformations we 
update $L_1$, $L_2$, and $L_3$ independently. For $\widetilde{P} = 70$ 
the system freezes. At the end of the simulation, the lattice has a centered
tetragonal structure with $c/a \approx 1.41$ (the same value
found for $\widetilde{P} = 24$) and $\Phi_p \approx 1.24$. For 
$\widetilde{P} = 100$ and 110, the system is apparently unable to reach
equilibrium. After approximately $10^5$ iterations,
the system settles in an arrested state
and particles no longer diffuse. The presence of arrested states was already 
noted \cite{FSTZLVWDL-03-suppl} in SB models with phenomenological potentials,
using a more realistic dynamics (here we use a Metropolis dynamics with 
local random displacements of the particles). 
Our results show that arrested states
are a generic property of the SB models, which is not very dependent on the 
interaction potentials (the potentials we use are quite different from those
used in Ref.~\onlinecite{FSTZLVWDL-03-suppl}, see 
Ref.~\onlinecite{MP-13-suppl}) 
and on the dynamics.
A visual inspection of the arrested configuration obtained in the run 
with $\widetilde{P} = 100$ shows the molecules belonging to the fluid
half-system have been able to partially order. Thus, we expect the stable 
phase for $\widetilde{P} = 100$ to be solid. To identify the equilibrium
value of $\Phi_p$ for $\widetilde{P} = 100$, we perform a simulation in which 
we start from a solid centered
tetragonal lattice with $n=m=8$ (1024 particles), $b/a = 1$, 
$c/a = 1.5$, $\Phi_p = 1.4$. 
We find at the 
end $b/a\approx 1$, $c/a\approx 1.41$, and $\Phi_p \approx 1.27$. 
For $\widetilde{P} = 110$, there is no indication of partial order.
The arrested configuration consists in a perfectly ordered crystal coexisting 
with a fully disorderd half system.
Finally, for 
$\widetilde{P} = 120$, the solid part of the system melts in a few thousand
iterations. The corresponding volume fraction is $\Phi_p \approx 1.66$.
Therefore, the transition occurs for 
\begin{equation}
   \widetilde{P}_{sf} = 110(10),
\end{equation}
while the coexistence interval $[\Phi_{ps},\Phi_{pf}]$ satisfies 
$1.27 \lesssim \Phi_{ps} <\Phi_{pf} \lesssim 1.66$. At the transition, 
a centered tetragonal crystal with $c/a \approx 1.4$ is 
the stable solid structure.

\subsubsection{The multi-blob model} 

\begin{figure}[!t]
\begin{center}
\begin{tabular}{c}
\includegraphics[angle=-90,scale=0.33]{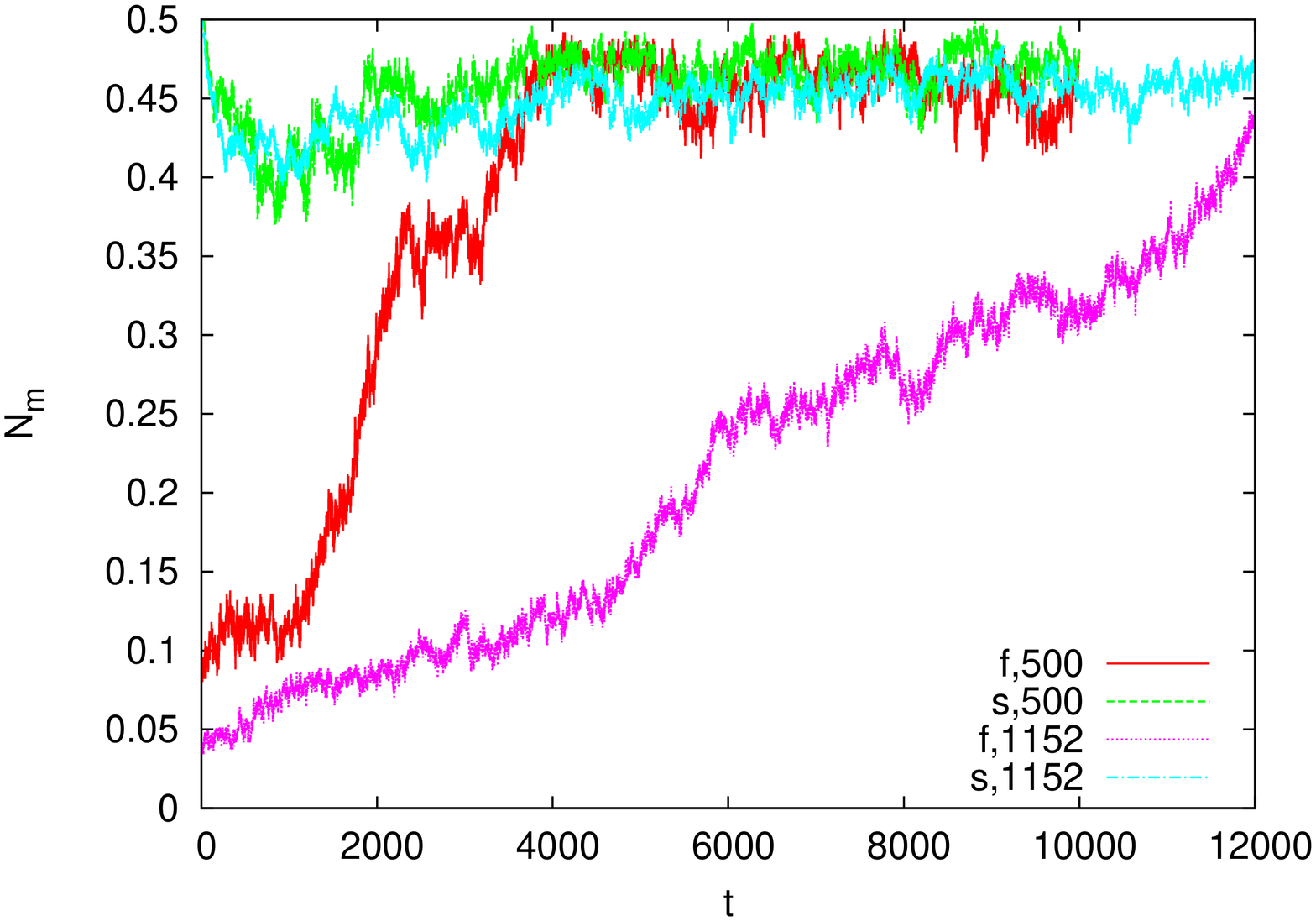} \\
\includegraphics[angle=-90,scale=0.33]{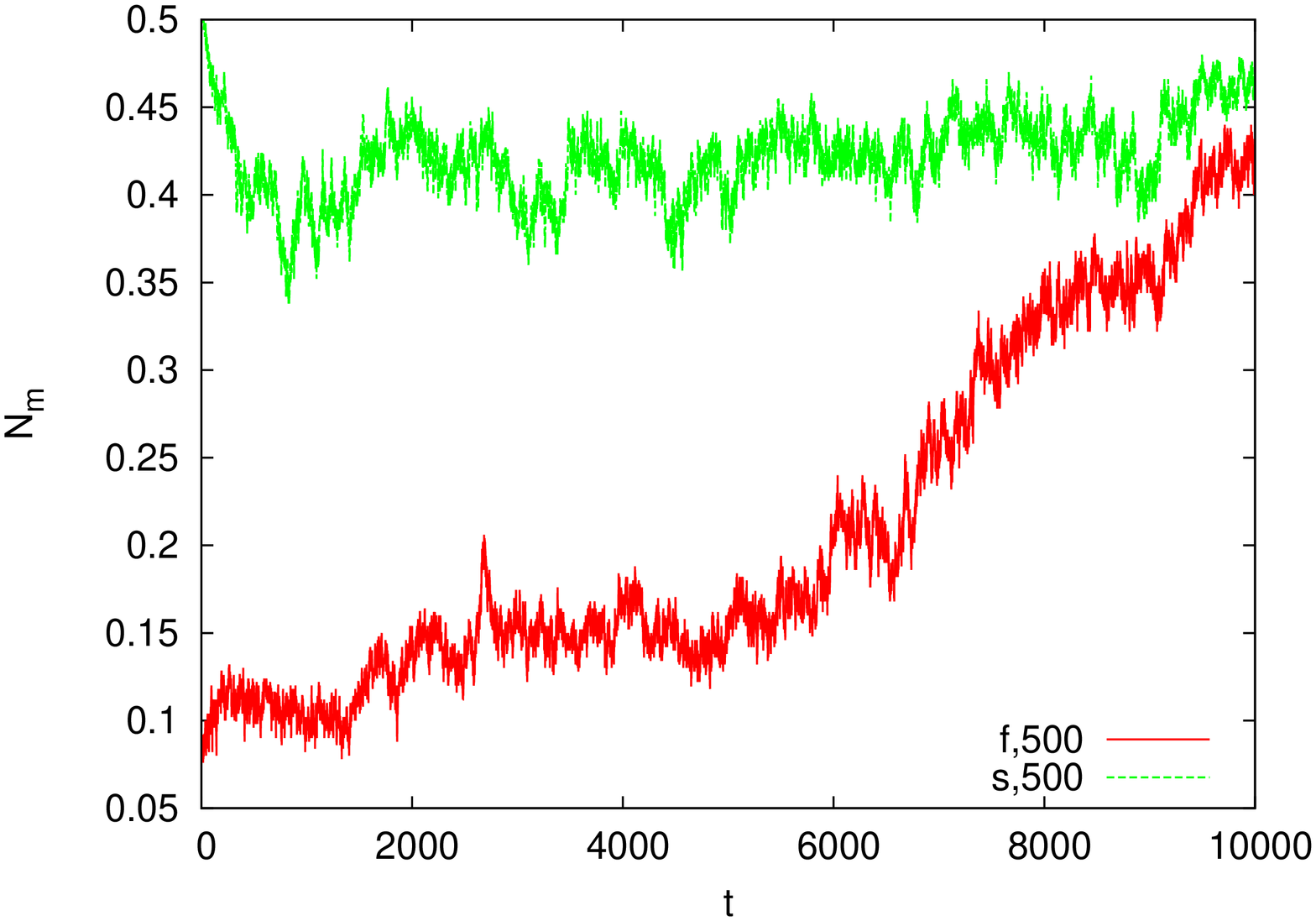} \\
\includegraphics[angle=-90,scale=0.33]{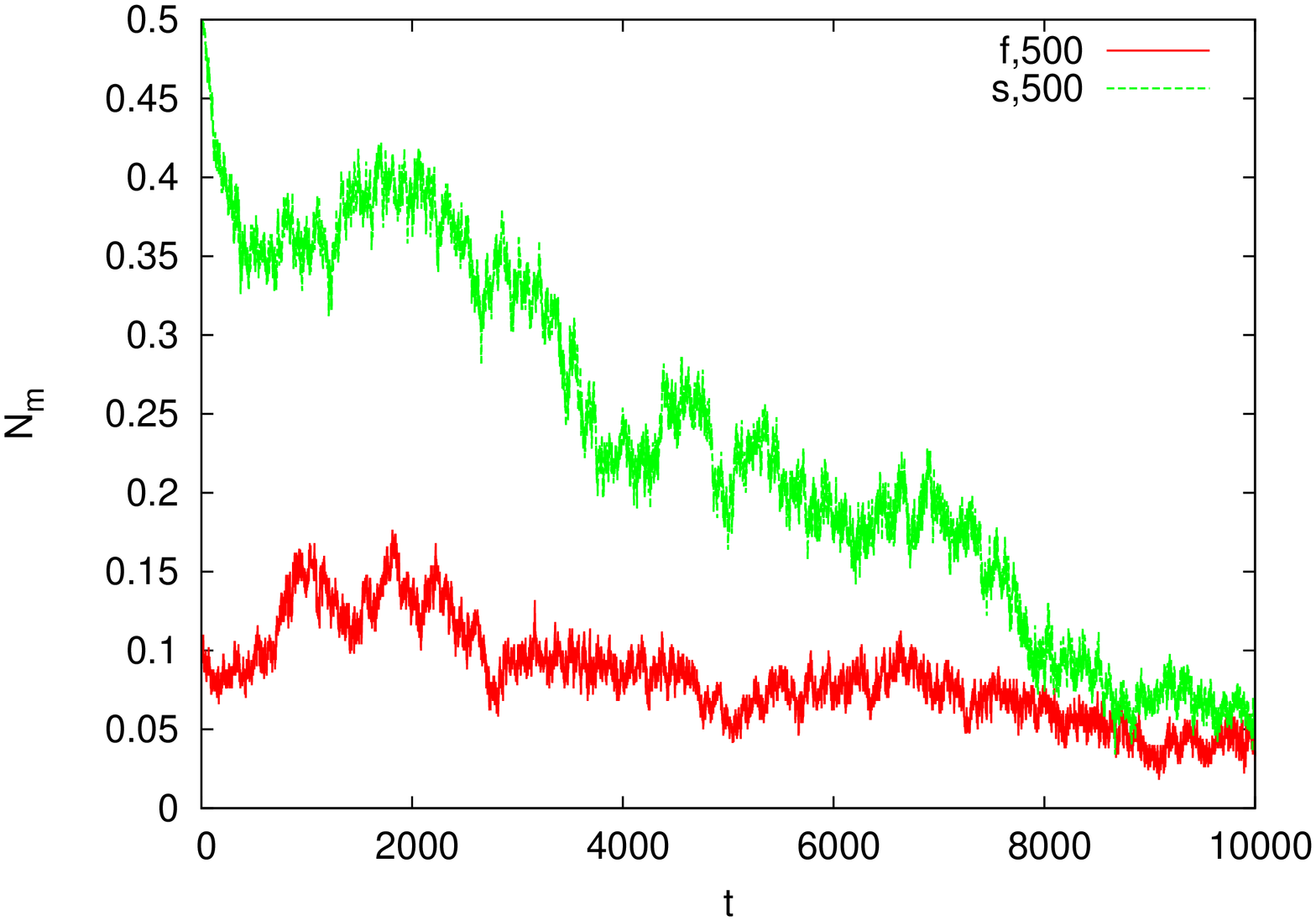} 
\end{tabular}
\caption{Plot of $N_{ms}(\epsilon)$ and of $N_{mf}(\epsilon)$ as 
a function of the number $t$ of iterations. Top: $\widetilde{P} = 6.6$; 
middle: $\widetilde{P} = 5.0$; bottom: $\widetilde{P} = 3.38$.
Here $\epsilon = 0.2$. Simulations with $500$ molecules and, 
for $\widetilde{P} = 6.6$, also with 1152 molecules.  
For $\widetilde{P} = 6.6$ and 5.0, the solid is stable; for 
$\widetilde{P} = 3.38$ the stable phase is fluid.
}
\label{NMB_suppl}
\end{center}
\end{figure}

\begin{figure}[!t]
\begin{center}
\begin{tabular}{c}
\includegraphics[angle=-90,scale=0.33]{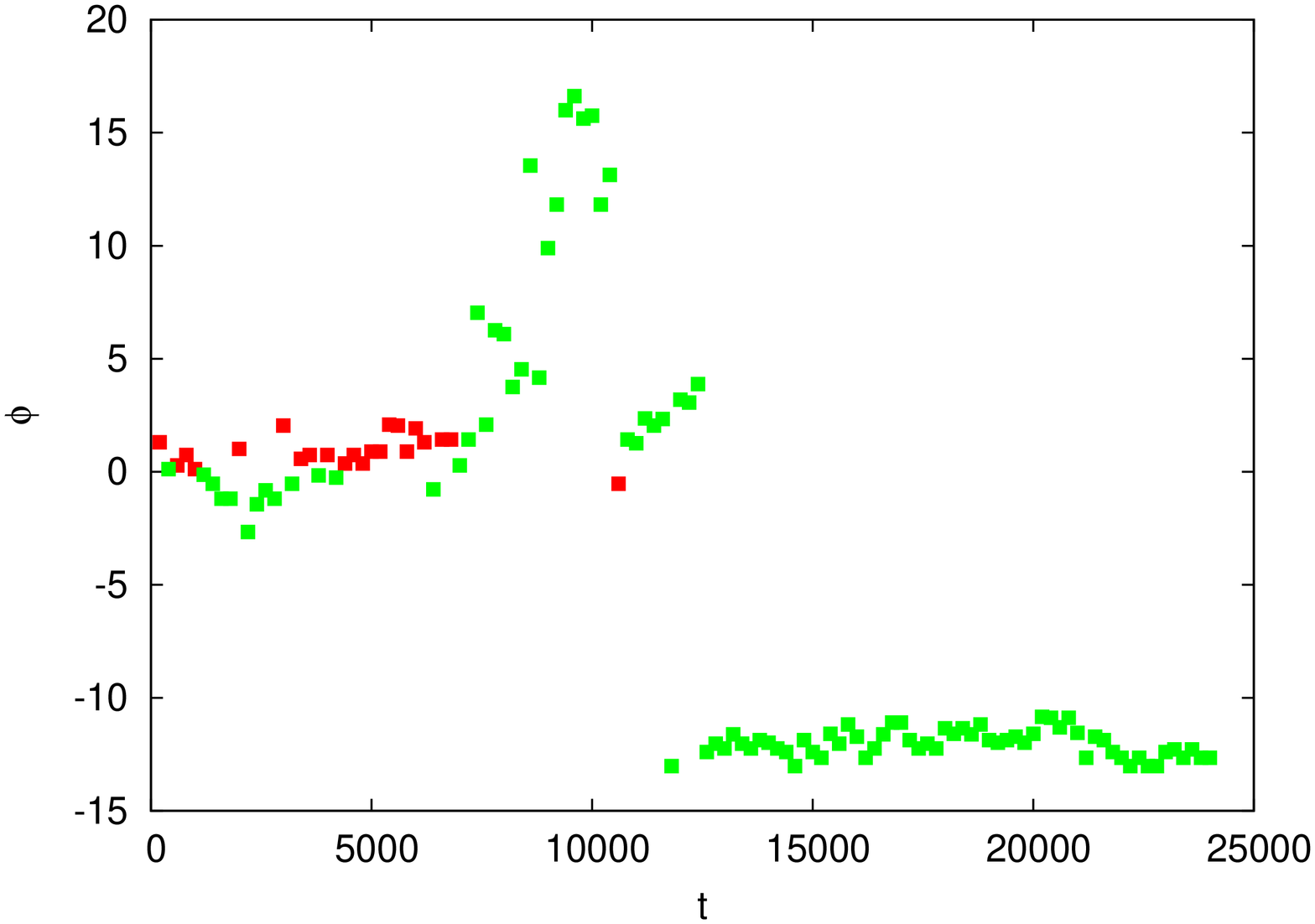} \\ [-1.6truecm]
\includegraphics[angle=-90,scale=0.33]{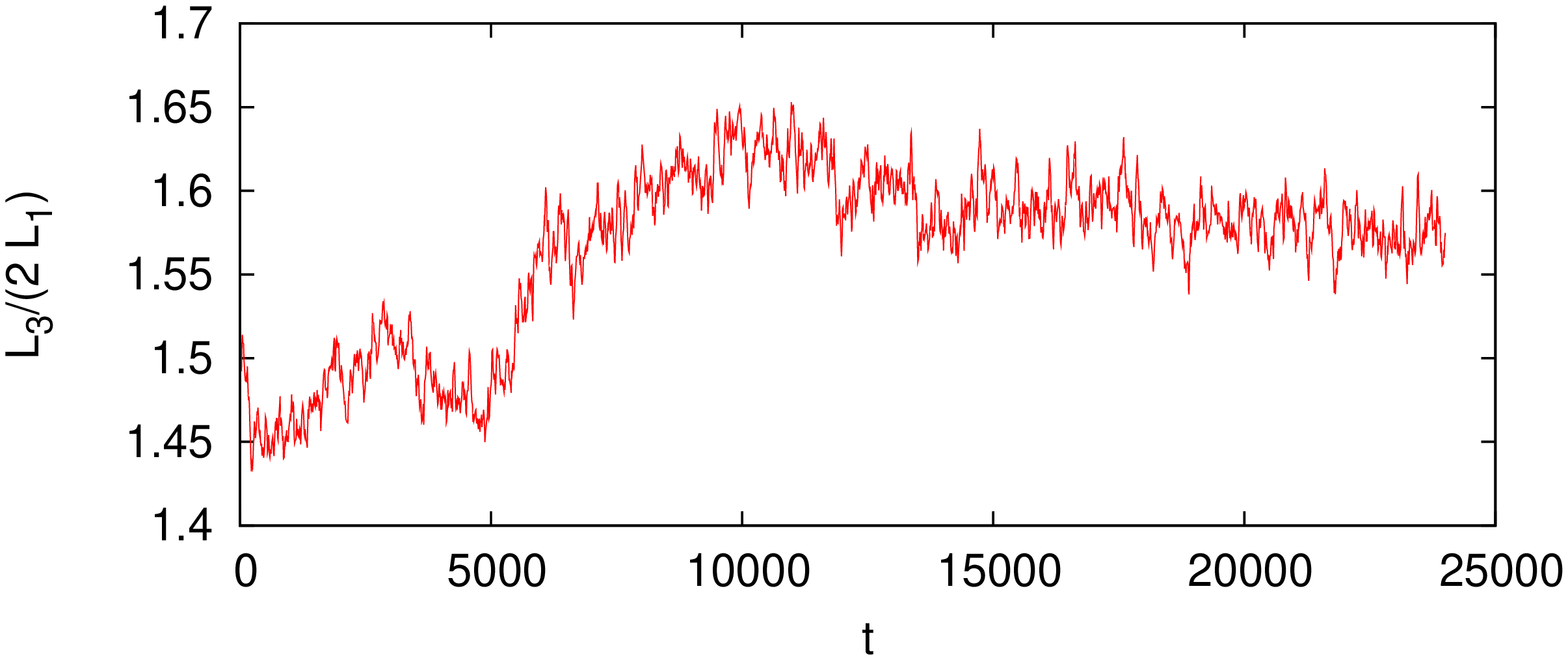} \\[-1.0truecm]
\end{tabular}
\caption{Top: plot of $\phi$ (in degrees) as 
a function of the number $t$ of iterations (we estimate $\phi$ every 200 
iterations): red points correspond to 
systems with unit cell given in Eq.~(\ref{monoclinic-a}), green points
to systems with unit cell given in Eq.~(\ref{monoclinic-b}).
Bottom: plot of $L_3/(2 L_1) = c/a$, as a function of $t$.
Results for a system of 500 molecules
and $\widetilde{P} = 6.6$. We start from a mixed configuration with 
$c/a = 1.5$ and $n=m=5$.
}
\label{PhiMB_suppl}
\end{center}
\end{figure}

The analysis presented in the SB case has been repeated for the MB model. 
Short runs on  very small bcc systems with 128 molecules give indications that 
the solid phase might be stable for $\Phi_p\gtrsim 0.5$. For this reason,
we have performed simulations at $\widetilde{P} = 10.9, 6.6, 3.38$,
which are the values of the pressure obtained in canonical simulations
starting from disordered configurations with $\Phi_p = 0.6, 0.5,$ and
0.4, respectively. For each packing fraction we generated mixed configurations
with $c=a$ and $n=m=5$ (the total number of molecules is 500). They were used
as starting configurations for isobaric runs in which $L_1$ and $L_2$ were kept 
equal. We find that the bcc phase is stable for $\widetilde{P} = 10.9$ and 6.6, 
while for $\widetilde{P} = 3.38$ the solid part of the system melts. This is 
evident from Fig.~\ref{NMB_suppl}, where we report $N_{mf}(\epsilon)$ and 
$N_{ms}(\epsilon)$ for $\epsilon = 0.2$. The results obtained using 
500 molecules are confirmed by simulations with larger systems. In particular,
we considered a starting mixed configuration with $n=6$ and $m=8$ (1152
molecules) at $\widetilde{P} = 6.6$. Again, we observe that the bcc phase
is stable, see Fig.~\ref{NMB_suppl}.
The densities obtained in the isobaric simulations for the solid phase 
do not  differ significantly from those obtained in simulations 
in the (metastable) fluid phase (in these simulations we start 
from a disordered distribution of the molecules). For instance, for 
$\widetilde{P} = 6.6$ we find $\Phi_p = 0.5068(1)$ for the solid phase
and $\Phi_p = 0.5055(1)$ for the metastable fluid phase. 

To improve the estimates of the transition
pressure, we perform a simulation for $\widetilde{P} = 5$. 
We find that the bcc lattice is the stable phase, see
Fig.~\ref{NMB_suppl}, and that the corresponding packing fraction 
is $\Phi_p = 0.46$. 
We thus predict a fluid-solid transition for
\begin{equation}
     \widetilde{P}_{fs} = 4.2 (8).
\end{equation}
If $\Phi_{p,f}$ and $\Phi_{p,s}$ are the boundaries of the fluid and solid
phases, respectively, we obtain 
\begin{equation}
     0.40 \lesssim \Phi_{p,f} < \Phi_{p,s} \lesssim 0.46.
\end{equation}
Note that, for $\Phi_p \approx 0.43$, the size $a$ of the unit cell 
is $a = 2.69 \hat{R}_g$ and that the distance between the two 
closest lattice points is $a\sqrt{3}/2 \approx 2.33 \hat{R}_g$. 

In the MB case, we have not performed a detailed analysis of other possible 
asymmetric lattice structures. We have only performed a single isobaric run 
at $\widetilde{P} = 6.6$, starting with a mixed system with $c/a = 1.5$ 
and $n=m=5$, and keeping $L_1 = L_2$ in the simulation.
After 12000 iterations, the disordered part of the system is completely
frozen. The resulting lattice structure has $c/a \approx 1.6$ and is monoclinic
with $\phi \approx 12$-$13^\circ$, see Fig.~\ref{PhiMB_suppl}.
Note that 
this result is completely analogous to what has been observed in the 
SB simulations (although at a different value of the pressure). Therefore, 
we conclude that the existence of these (meta)stable structures is not due 
the coarse-graining procedure, but that it is a property of the 
underlying star-polymer system.

\begin{figure}[!t]
\begin{center}
\begin{tabular}{c}
\includegraphics[angle=-90,scale=0.33]{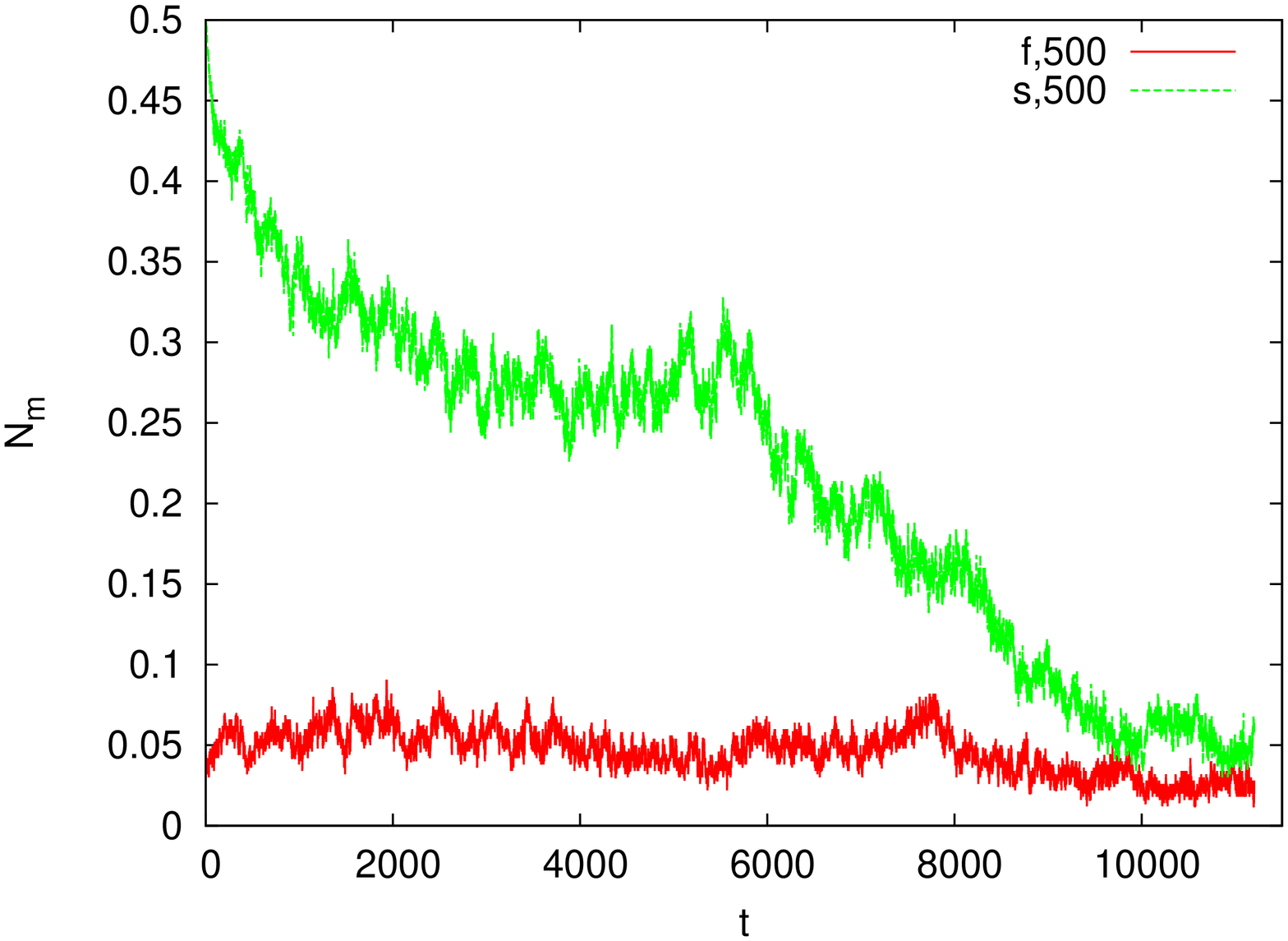} \\
\includegraphics[angle=-90,scale=0.33]{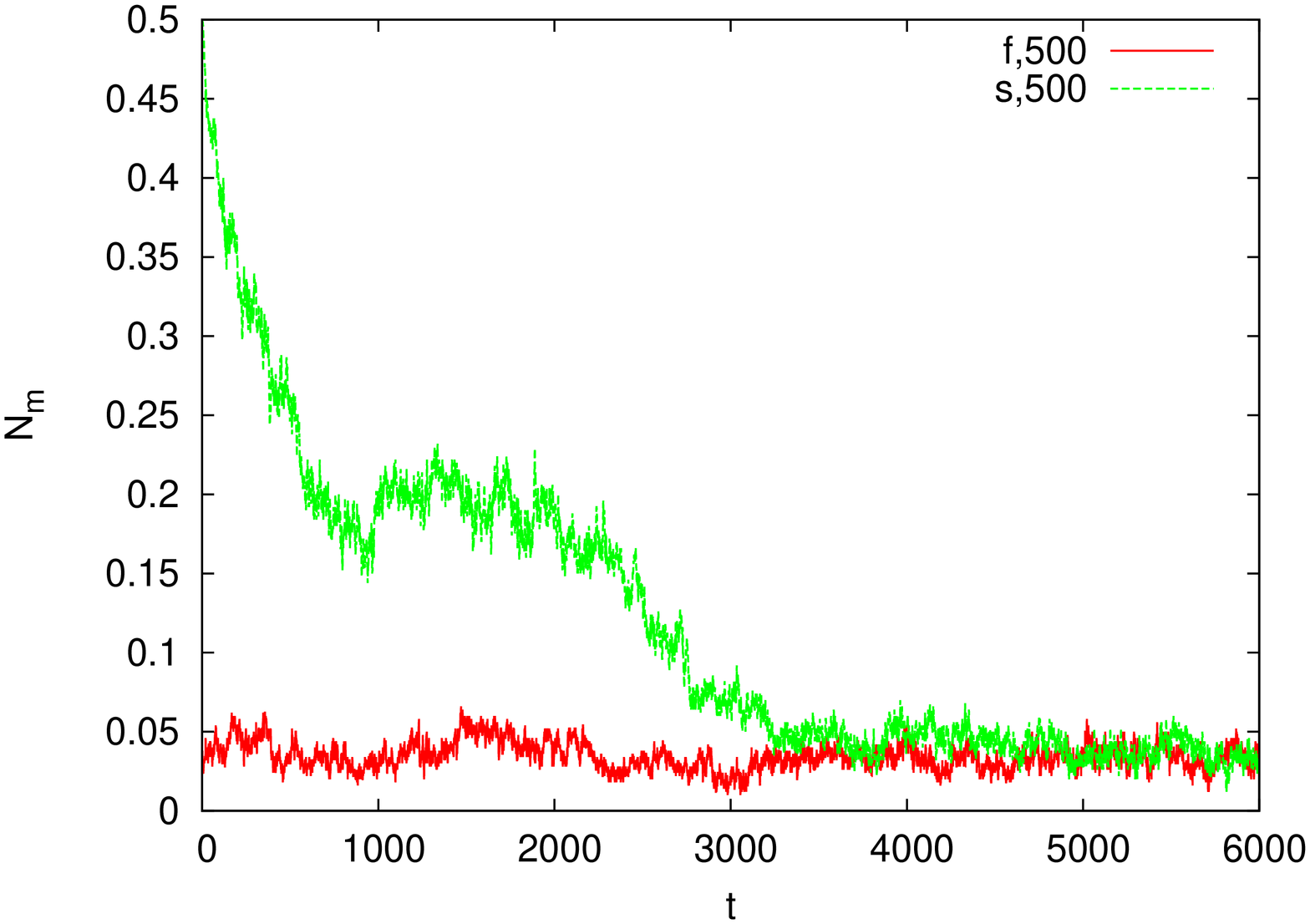} \\
\includegraphics[angle=-90,scale=0.33]{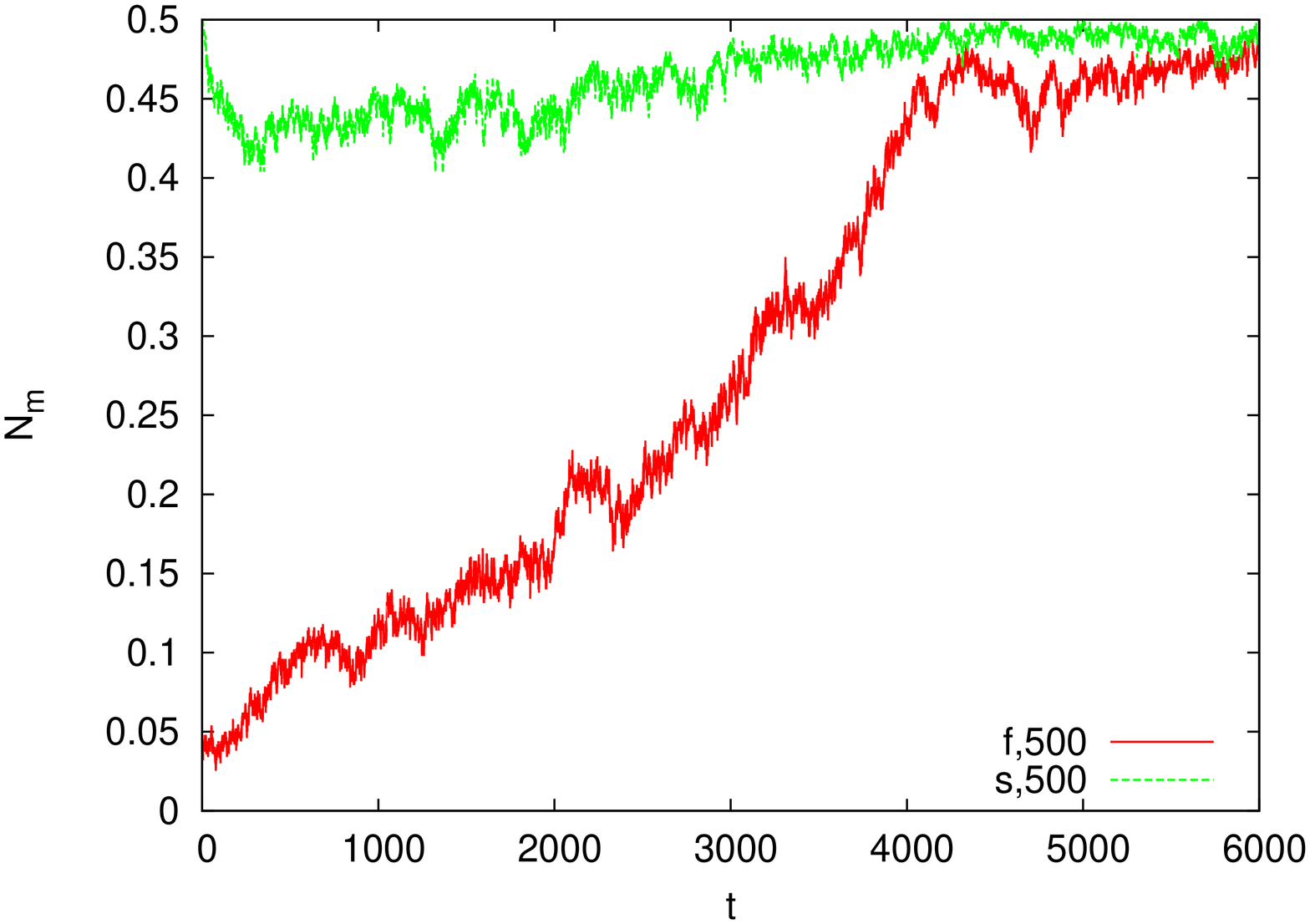} 
\end{tabular}
\caption{Plot of $N_{ms}(\epsilon)$ and of $N_{mf}(\epsilon)$ as 
a function of the number $t$ of iterations. Top: $\widetilde{P} = 70$ and 
starting configuration with $c/a = 1$; 
middle: $\widetilde{P} = 70$ and 
starting configuration with $c/a = 1.5$;
bottom: $\widetilde{P} = 55$ and starting configuration with $c/a = 1$.
Here $\epsilon = 0.2$. Simulations with $500$ molecules.
For $\widetilde{P} = 55$, the solid is stable; for 
$\widetilde{P} = 70$ the stable phase is fluid.
}
\label{NMB2_suppl}
\end{center}
\end{figure}

Finally, we determined the location of the high-density solid-fluid 
transition. We performed simulations for $\widetilde{P} = 24, 55,$ 70, and 85,
starting from mixed solid-fluid configurations with $m=n=5$ (500 particles) and 
$c/a = 1$ (bcc lattice). For $\widetilde{P} = 24$ (corresponding to 
$\Phi_p \approx 0.8$) and $\widetilde{P} = 55$ 
(correspondingly $\Phi_p \approx
1.13$) the system freezes, while for 
$\widetilde{P} = 70, 85$, the solid part of the system melts and 
$\Phi_p \approx 1.26, 1.38$, respectively, at the end of the simulation;
see Fig.~\ref{NMB2_suppl}. 
Since it is possible that the stable phase is not cubic symmetric---it 
might be tetragonal, orthorombic, or
monoclinic---we also performed simulations at $\widetilde{P} = 70, 85$, 
starting from mixed fluid-solid configurations with $c/a = 1.5$ and $m=n=5$. 
In both cases, the crystal melts, confirming that the stable phase is fluid 
for $\widetilde{P} \gtrsim 70$.  
Therefore, the solid-fluid transition occurs at 
\begin{equation}
   \widetilde{P} = 62(8),
\end{equation}
while the coexistence interval $[\Phi_{ps},\Phi_{pf}]$ satisfies 
$1.13 \lesssim \Phi_{ps} <\Phi_{pf} \lesssim 1.26$. Note that we have found no
evidence of structural arrest in the multiblob case.

\onecolumngrid

\subsection{Numerical results}

\begin{table*}[h]
\caption{Estimates of the compressibility factor $Z$ for $f=6$ 
and different values of $\Phi_p$. We report single-blob results in two
different representations (SB-MP and SB-CM), multiblob results (MB),
and estimates obtained by using the Pad\'e extrapolation discussed
in the paper (PADE). }
\begin{tabular}{ c | c c c c c c}
\hline\hline
& $\Phi_p=0.15$ & $\Phi_p=0.25$ & $\Phi_p=0.50$ & $\Phi_p=0.75$ &
$\Phi_p=1$ & $\Phi_p=2$\\
\hline
SB-MP & 1.623 & 2.130 & 3.563 & 5.094 & 6.656 & 12.981 \\
SB-CM & 1.620 & 2.146 & 3.719 & 5.472 & 7.285 & 14.626 \\
MB & 1.638 & 2.182 & 3.927 & 5.919 & 8.132 & 17.782  \\
PADE & 1.618 & 2.141 & 3.776 & 5.797 & 8.132 & 19.74 \\
\hline
\end{tabular}
\end{table*}

\begin{table*}[h]
\caption{Estimates of the compressibility factor $Z$ for $f=12$ 
and different values of $\Phi_p$. We report single-blob results in two
different representations (SB-MP and SB-CM), multiblob results (MB),
and estimates obtained by using the Pad\'e extrapolation discussed
in the paper (PADE). }
\begin{tabular}{ c | c c c c c c}
\hline\hline
$\Phi_p$ & $0.15$ & $0.25$ & $0.50$ & $0.75$ &
$1$ & $2$\\
\hline
SB-MP & 2.255 & 3.532 & 7.587 & 11.998 & 16.440 & 34.116 \\
SB-CM & 2.258 & 3.601 & 8.196 & 13.490 & 18.920 & 40.450 \\
MB & 2.106 & 3.385 & 7.983 & 13.981 & 21.295 & 49.640  \\
PADE & 2.215 & 3.431 & 7.771 &13.725 & 21.295 & 60.051 \\
\hline
\end{tabular}
\end{table*}

\begin{table*}[h]
\caption{Estimates of the compressibility factor $Z$ for $f=40$ 
and different values of $\Phi_p$. We report single-blob results in the
center-of-mass representation (SB-CM)s and multiblob results (MB).
All results refer to the fluid phase, which is metastable for 
$\Phi_p\gtrsim 0.4$ (multiblob model) and for $\Phi_p\gtrsim 0.6$ 
(single blob model).
}
\begin{tabular}{ c | c c c c c c}
\hline\hline
$\Phi_p$ & $0.15$ & $0.25$ & $0.50$ & $0.75$ &
$0.90$\\
\hline
SB-CM & 4.752 & 11.781 & 43.760 & 80.430 & 101.667\\
MB & 4.785 & 12.552 & 54.977 & 112.576 & 147.169\\
\hline
\end{tabular}
\end{table*}

\begin{table*}[h]
\caption{Estimates of the ratio $R_{g,b}(\Phi_p)/\hat{R}_{g,b}$ for 
different values of $\Phi_p$ and $f$. For $f=40$, the results for 
$\Phi_p = 0.50$ and 0.75 are obtained in the metastable fluid phase.
}
\begin{tabular}{ c | c c c c c c}
\hline\hline
$f\backslash\Phi_p$ & $0.15$ & $0.25$ & $0.50$ &
$0.75$ &  $1$ & $2$  \\
\hline
$6$ & 0.989 & 0.982 & 0.961 & 0.941 & 0.924 & 0.883\\
$12$ & 0.992 & 0.984 & 0.954 & 0.924 & 0.894 & 0.820\\
$40$ & 0.995 & 0.985 & 0.936 & 0.888 & --- & --- \\
\hline
\end{tabular}
\end{table*}

\end{document}